\documentclass[11pt]{article}

\usepackage[final]{acl}

\usepackage{times}
\usepackage{latexsym}
\usepackage{tikz}

\usepackage[T1]{fontenc}
\usepackage{tcolorbox}
\usepackage{enumitem}

\usepackage[utf8]{inputenc}
\usepackage{booktabs}

\usepackage{microtype}

\usepackage{inconsolata}

\usepackage{graphicx}
\usepackage{listings}

\usepackage{url}

\usepackage{multirow}
\usepackage{multicol}
\usepackage{subcaption}

\usepackage{lineno}

\usepackage[normalem]{ulem}
\usepackage{listings}
\usepackage{caption}
\usepackage{adjustbox}

\usepackage{hyperref}

 \usepackage{array,multirow}
 \usepackage{tikz}

\usepackage{nowidow}

\definecolor{planningcolor}{HTML}{EF9D65}
\definecolor{implementationcolor}{HTML}{84BCD1}
\definecolor{revisioncolor}{HTML}{9584D1}
\definecolor{BlueGreen}{HTML}{0D98BA}
\usepackage{enumitem} 

\definecolor{darkblue}{rgb}{0, 0, 0.5}
\hypersetup{colorlinks=true, citecolor=darkblue, linkcolor=darkblue, urlcolor=darkblue}

\newcommand{\plan}[1]{\textcolor{planningcolor}{#1}}
\newcommand{\impl}[1]{\textcolor{implementationcolor}{#1}}
\newcommand{\rev}[1]{\textcolor{revisioncolor}{#1}}

\title{\textsc{ScholaWrite}: A Dataset of End-to-End Scholarly Writing Process} 

\author{
 \textbf{Khanh Chi Le}, \textbf{Linghe Wang}, \textbf{Minhwa Lee},
 \textbf{Ross Volkov}, \textbf{Luan Tuyen Chau}, \textbf{Dongyeop Kang}
\\
 University of Minnesota
\\
\texttt{\{le000422,  dongyeop\}@umn.edu}
}

\begin{document}
\maketitle
\begin{abstract}
Writing is a cognitively demanding activity that requires constant decision-making, heavy reliance on working memory, and frequent shifts between tasks of different goals. To build writing assistants that truly align with writers' cognition, it is necessary to capture and analyze the complete thought process behind how writers transform ideas into final texts.
We present \texttt{ScholaWrite}, the first dataset of \textit{end-to-end scholarly writing}, tracing the multi-month journey from initial drafts to final manuscripts. The dataset traces nearly 62K \LaTeX-based edits from five computer science preprints over four months and is enriched with fine-grained annotations of cognitive writing intentions. We demonstrate the value of ScholaWrite through three complementary contributions:
(1) analysis of real-world writing behavior reveals that scholarly writing is highly non-linear and multi-intentional, blending rapid drafting bursts with cognitively sustained writing sessions;
(2) evaluations of current large language models show that they struggle to provide meaningful support throughout the human writing process; and
(3) models finetuned on \textsc{ScholaWrite} demonstrate improved alignment with human writing workflows.
\textsc{ScholaWrite} underscores the value of capturing scientists' cognitive writing process and provides actionable insights and resources for the development of future writing assistants. All data and tools are available on our project page.\footnote{\url{https://minnesotanlp.github.io/scholawrite/}}

\end{abstract}

\section{Introduction}
\begin{figure}[t]
    \centering
    \includegraphics[width=0.45\textwidth]{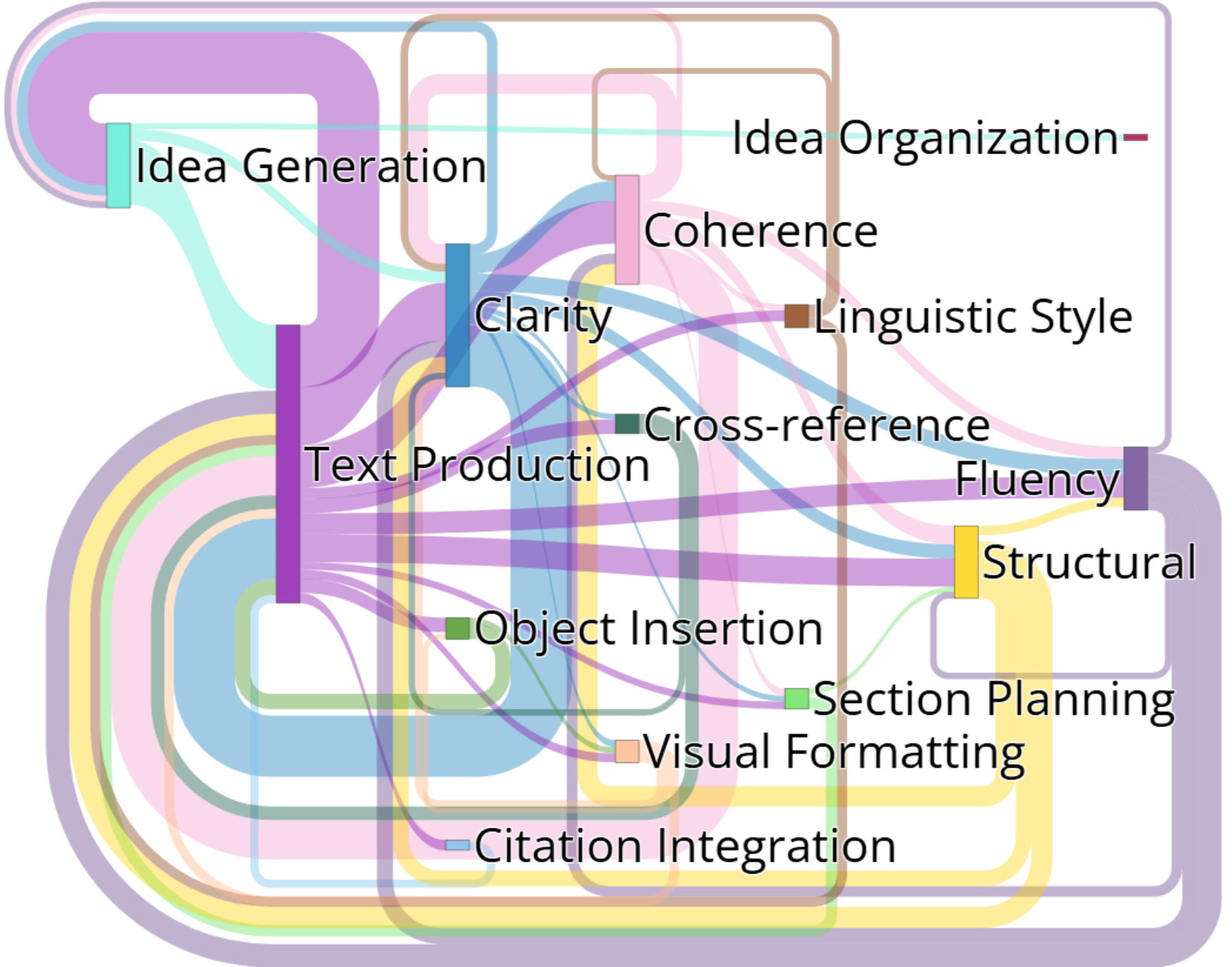}\vspace{-1mm}
    \caption{
    An example scholarly writing process with annotated writing intents in \textsc{ScholaWrite}: it is iterative, non-linear, and switches frequently between multiple activities, tools, and intents over a long range of time. 
    \vspace{-4mm}
    }
    \label{fig:sankey-flow-1}
\end{figure}

Scientific writing is one of the most cognitively demanding forms of human communication. Researchers must articulate novel ideas with precision and structure, iteratively refining arguments, evidence, and phrasing over time \citep{jourdan2023text, kallestinova2011write, bourekkache2022english}. Unlike linear text generation, scholarly writing is a complex cognitive process; writers continually plan, draft, and revise across multiple stages and intentions (see Figure \ref{fig:sankey-flow-1}). Yet, despite the growing integration of large language models (LLMs) into research workflows, our understanding of how human scientists actually write remains limited. Without such understanding, it is difficult to build writing assistants that genuinely align with the cognitive processes of their users.

Recent efforts in writing assistance have leveraged LLMs for tasks such as revision or feedback generation \citep{du-etal-2022-understanding-iterative,liang2024can}. However, these models largely operate autoregressively, producing text from left to right, while human writing unfolds through recursive, non-linear cycles of ideation, organization, and refinement \citep{f508427a-e4c0-3d6a-8abf-03a5d21ec6c4}. This fundamental cognitive gap raises a key question: how can we empirically capture and model the end-to-end scholarly writing process so that future LLMs can support, rather than substitute, human cognition?




Existing work studies writing through revision logs or keystroke analyses \citep{leijten2013keystroke, koo2023decoding}, but it primarily targets short tasks or version level comparisons of completed drafts \citep{du-etal-2022-understanding-iterative, jiang-etal-2022-arxivedits, darcy-etal-2024-aries}, reflecting a broader lack of datasets that capture scholarly writing as a longitudinal and intention driven process and thereby limiting fine grained analysis of writing behavior, evaluation of writing assistants, and training of models aligned with real writers' workflows.

To bridge this gap, we present \textsc{ScholaWrite}, the first dataset capturing the end-to-end scholarly writing process, from initial drafts to final manuscripts, through in-situ keystroke logging and cognitive annotation. Rather than treating the dataset as an endpoint, we demonstrate its value through three complementary uses:

\textbf{(1) Analysis of scholarly writing behavior.} Our analyses reveal that scholarly writing is highly \textit{non-linear} and \textit{multi-intentional}: more than half of all writing sessions involve three or more intertwined intentions. Writers dedicate most time to
producing and revising text for clarity and coherence, while cognitively heavy tasks, such as idea organization or visual structuring, occur in fewer but longer, more focused sessions.

\textbf{(2) Benchmarking cognitively aligned writing assistance.} Evaluations of current LLMs show that they can imitate surface-level edits but \textit{fail to predict next intentions} or \textit{sustain cognitively complex revisions}, highlighting
gaps between human writing processes and
the current capabilities of LLMs in providing meaningful assistance.

\textbf{(3) Training LLMs for cognitive support.} LLMs finetuned on \textsc{ScholaWrite} exhibit consistent gains in next-intention prediction and \textit{intention-conditioned output alignment}. When deployed in iterative writing, these models adopt more human-like patterns -- frequently alternating between Planning, Implementation, and Revision across stages -- despite persistent gaps in surface-level text quality.

\section{Related Work}

\paragraph{Cognitive Theory of Writing Process}
Research on writing has shifted from analyzing final texts to examining cognitive processes across writing phases \citep{diederich1974measuring, Krapels1990SecondLW, macarthur2016writing}. 
Building on this process-oriented approach, \citet{f508427a-e4c0-3d6a-8abf-03a5d21ec6c4} outline three core sub-processes: \textit{planning}, \textit{translating}, and \textit{reviewing}. These dynamic, non-linear stages inform our work, where we extend this model into a finer-grained taxonomy of cognitive writing patterns in scholarly communication.
\citet{koo2023decoding} proposed a taxonomy of scholarly writing based on keystroke data from short, 30-minute research-plan sessions. Expanding on this, we collect larger-scale, longitudinal keystroke data spanning months and culminating in published research. Expert-reviewed and grounded in prior theory, our taxonomy captures end-to-end cognitive trajectories of scholarly writing.

\paragraph{Keystroke Loggers for Scholarly Writing}
Keystroke logging tools (e.g., Inputlog\footnote{\url{https://www.inputlog.net/}}) enable unobtrusive observation of digital writing \citep{chan2017using, johansson2010looking, leijten2013keystroke, lindgren2019observing}. Yet, most operate in closed ecosystems like MS Word and are ill-suited for \LaTeX-based academic writing or extended sessions. 
To address this, we built systems that securely record real-time keystrokes from Overleaf over months while preserving privacy. This workflow supports span-level annotation of writing intentions, providing a natural setting to study long-term cognitive writing activities across document sections.



\paragraph{Writing Datasets} 
Publicly available datasets from previous work tend to track linguistic style changes or grammatical edits during revision \citep{du-etal-2022-understanding-iterative, jiang-etal-2022-arxivedits, ito2019diamonds, mita2022towards, cahill-etal-2013-robust, boyd-2018-using}, while others capture edits based on feedback and peer review \citep{darcy-etal-2024-aries, jourdan2024casimir, kuznetsov2022revise} or citation generation \citep{kobayashi-etal-2022-dataset, narimatsu2021task}. Also, most focus on specific sections of papers (e.g., abstracts, introductions) \citep{du-etal-2022-understanding-iterative, mita2022towards}. In contrast, our dataset covers all cognitive phases of writing over an extended period and all sections.
Furthermore, those existing corpora often compare multiple versions of final manuscripts from preprint databases \citep{jiang-etal-2022-arxivedits, du-etal-2022-understanding-iterative, darcy-etal-2024-aries, jourdan2024casimir}, missing how those manuscripts evolved \citep{jourdan2023text}. To address this, we build a keystroke-based corpus that captures real-time progression of publications.





\section{\textsc{ScholaWrite} Dataset}\label{sec:dataset}

\textsc{ScholaWrite} represents the first large-scale publicly available dataset capturing the cognitive dynamics of end-to-end scholarly writing. It links every keystroke to the author’s underlying writing intention.
Each annotated entry consists of a before-text, after-text, metadata, and intention label, as illustrated below with an example of a Fluency edit in which the writer corrects ``expct'' to ``aspect.''
\begin{table}[ht!]
\centering
\vspace{-2mm}
\begin{minipage}{\linewidth}
\lstset{
    basicstyle=\ttfamily\footnotesize, 
    breaklines=true, 
    frame=single, 
    columns=fullflexible, 
    captionpos=b 
}\vspace{-1mm}
\begin{lstlisting}
{   "Project": 1,
    "timestamp": 1702958491535,
    "author": "author1",
    "before text": "One important expct of studying a LLMs is ..",
    "after text": "One important aspect of studying LLMs is ..",
    "label": "fluency" }
\end{lstlisting}
\vspace{-5mm}
\label{table:single-entry}
\end{minipage}
\end{table}

\subsection{Data Collection \& Annotation Process}\label{sec:collection}

\paragraph{Keystroke Collection in Overleaf} To capture scholarly writing as it naturally unfolds, we developed a Chrome extension that records the end-to-end writing process on Overleaf, from individual keystrokes to their cognitive interpretation, without interrupting authors’ workflow. Please refer to Appendix~\ref{sec:appendix:system} for a description of our extension.

\paragraph{Participant Recruitment}\label{sec:recruitment}
We recruited 10 graduate students in computer science at an R1 university in the United States, each actively writing manuscripts for peer-review AI and NLP conferences using Overleaf. Two participants were native English speakers, and eight reported high proficiency in academic English writing.
The collection period spanned four months (November 2023 - February 2024). Our study has been approved by the IRB institution of the authors. Please refer to Appendix~\ref{sec:appendix:recruitment} for more information.



\paragraph{Annotation Interface}\label{sec:annotation-interface} To decode the nuanced intentions behind each writing action, we annotate each keystroke collected from the Chrome extension using a specialized interactive interface that visualizes writing activity over time and across authors and files.
Please see Appendix \ref{sec:appendix:annotation} for details about the annotation process.

\begin{table*}[h!]
\centering
\footnotesize
\resizebox{\textwidth}{!}{
\begin{tabular}{@{}p{0.5cm}@{\hskip 1mm}@{}p{1.4cm}p{5.8cm}p{6.8cm}@{\hskip 2mm}c@{}}
\toprule
\textbf{1st} & \textbf{Intention} & \textbf{Definition} & \textbf{An example action} & \textbf{Prop.}\\
\midrule

\parbox[t]{2mm}{\multirow{3}{*}{\rotatebox[origin=c]{90}{\textsc{\colorbox{planningcolor}{Planning}}}}}
&  Idea Generation & Formulate and record initial thoughts and concepts. & 
    writing down keywords of a paragraph beforehand (e.g., ``\textit{..\%[Comment out] main point: artifacts lack in human subjectivity..}'') & 7.0\% \\
\cmidrule(r){2-5}
& {Idea Organization} & Select the most useful materials and demarcate those generated ideas in a visually formatted way.  & Linking the generated ideas into a logical sequence and spacing out between ideas (e.g., ``..\% (1) need diff. stress testing...\%\%[Spacing] (2) exp. setup? '') & 0.5\% \\
\cmidrule(r){2-5}
& {Section Planning} & Initially create sections and sub-level structures. & Putting section-related LaTeX commands (e.g., \texttt{\textbackslash section}, \texttt{\textbackslash paragraph}) & 2.2\% \\
\midrule

\parbox[t]{2mm}{\multirow{5}{*}{\rotatebox[origin=c]{90}{\textsc{\colorbox{implementationcolor}{Implementation}}}}}
& Text Production & Translate their ideas into full languages, either from the writers' language or borrowed sentences from an external source. & Generating subsequent sentences with the author's own idea (e.g., ``... GPT-4 (OpenAI, 2023) explains the data ... Our approach is built on top of GPT-4...'') & 57.4\% \\
\cmidrule(r){2-5}
& {Object Insertion} & Insert visual claims of their arguments (e.g., figures, tables, equations, footnotes, lists) & e.g., \texttt{\textbackslash begin\{figure\}[h]} \texttt{\textbackslash centering} \texttt{\textbackslash includegraphics\{figure\_A.pdf\}} \texttt{\textbackslash end\{figure\}} & 4.6\% \\ 
\cmidrule(r){2-5} 
& {Citation Integration} & Incorporate bibliographic references into a document and systematically link these references using citation commands. & Inserting a new BibTeX object in the bibliography file and adding the object name to an existing \texttt{\textbackslash cite\{\}} on the Related Work section & 1.7\% \\
\cmidrule(r){2-5}
& {Cross-reference} & Link sections, figures, tables, or other elements within a document via referencing commands. & Putting a command \texttt{\textbackslash label\{figure-1\}} to a figure and referencing it by calling \texttt{\textbackslash ref\{figure-1\}} & 1.1\% \\
\cmidrule(r){2-5}
& {Macro Insertion} & Incorporate predefined commands or packages into a LaTeX document to alter its formatting. & Putting a \texttt{\textbackslash usepackage\{minted\}} for formatting a LLM prompt & 0.2\% \\
\midrule

\parbox[t]{2mm}{\multirow{7}{*}{\rotatebox[origin=c]{90}{\textsc{\colorbox{revisioncolor}{Revision}}}}}
& {Fluency} & Fix grammatical or syntactic errors in the text or LaTeX commands. & ``We desig{\textcolor{red}{\sout{ining}}}{\textcolor{teal}{ned}} several experiment setups for 
{\textcolor{teal}{the}} LLM evaluations as described..'' & 1.4\%  \\
\cmidrule(r){2-5} 
& {Coherence} & Logically link (1) multiple sentences within the same paragraph; (2) any two subsequent paragraphs; or (3) objects to be consistent.  & ``Each comment was annotated by three different annotators \textcolor{red}{\sout{, which}} \textcolor{teal}{and we} achieved high inter-annotator agreement.'' & 3.3\% \\
\cmidrule(r){2-5} 
& {Clarity} & Improve the semantic relationships between texts to be more straightforward and concise. & ``..relevant studies have examined \textcolor{red}{\sout{one of the several textual styles}} \textcolor{teal}{one aspect of texts}, the formality, ....'' & 11.5\% \\
\cmidrule(r){2-5}
& {Structural} & Improve the flow of information by modifying the location of texts and objects. & ``..human alignment \textcolor{teal}{to compare the alignment between lexicon-based preferences and humans' original preferences.} First, we calculated the score from each human participant \textcolor{red}{\sout{to compare the alignment between..}}'' & 3.7\% \\
\cmidrule(r){2-5}
& {Linguistic Style} & Modify texts with the writer’s writing preferences regarding styles and word choices, etc. & ``We \textcolor{red}{\sout{believe}} \textcolor{teal}{posit} that ...'' & 1.6\% \\
\cmidrule(r){2-5}
& {Scientific Accuracy} & Update or correct scientific evidence (e.g., numbers, equations) for more accurate claims. & ``..Pearson's $r$ correlation (\textcolor{red}{\sout{0.78}}\textcolor{teal}{0.68}; \textcolor{teal}{p < 0.01})'' & 0.7\% \\
\cmidrule(r){2-5}
& {Visual Formatting} & Modify the stylistic formatting of texts, objects, and citations & \texttt{\textbackslash cite} $\rightarrow$ \texttt{\textbackslash citet}, \texttt{\textbackslash textbf} $\rightarrow$ \texttt{\textbackslash textsc}, etc. & 3.2\% \\
\bottomrule
\end{tabular}
}\vspace{-1mm}
\caption{The developed taxonomy of scholarly writing process in \textsc{ScholaWrite}. }\vspace{-3mm}
\label{table:taxonomy-full}
\end{table*}

\subsection{Intention Taxonomy}\label{sec:taxonomy}
Building on cognitive theories of writing \citep{f508427a-e4c0-3d6a-8abf-03a5d21ec6c4} and recent empirical studies on scholarly revision \citep{du-etal-2022-understanding-iterative, koo2023decoding}, we developed a taxonomy of 15 distinct writing intentions grouped into three overarching categories (Table~\ref{table:taxonomy-full}).
Following \citet{pustejovsky2017designing}, two annotators conducted iterative open coding on initial subsets of keystrokes to inductively identify recurring cognitive patterns. Each span of edits was treated as a \textit{meaningful writing unit}, such as composing a sentence or improving phrasing, and assigned an appropriate intention label. Disagreements were resolved through multiple rounds of discussion with a cognitive linguist, who refined label definitions and boundaries to ensure coherence.

To validate the taxonomy, we applied two principles from taxonomy design \citep{nickerson2013method, kundisch2021update}: mutual exclusivity (each edit corresponds to one primary intention) and collective comprehensiveness (taxonomy covers all observed behaviors). After iterative refinement, we achieved a weighted F1 inter-annotator agreement of $0.71$ on a 1K-keystroke subset, confirming strong reliability. 
See details in Appendix~\ref{sec:appendix:annotation}.

\paragraph{Post-processing}
After annotation, we post-processed the data to improve usability and ensure privacy. All personal identifiers were removed, and keystrokes labeled as non-informative artifacts were filtered out. Please refer to Appendix \ref{sec:appendix:postprocess} for more information. 

\subsection{Overview}
\begin{figure}[h]
    \centering
                \includegraphics[width=0.8\columnwidth]{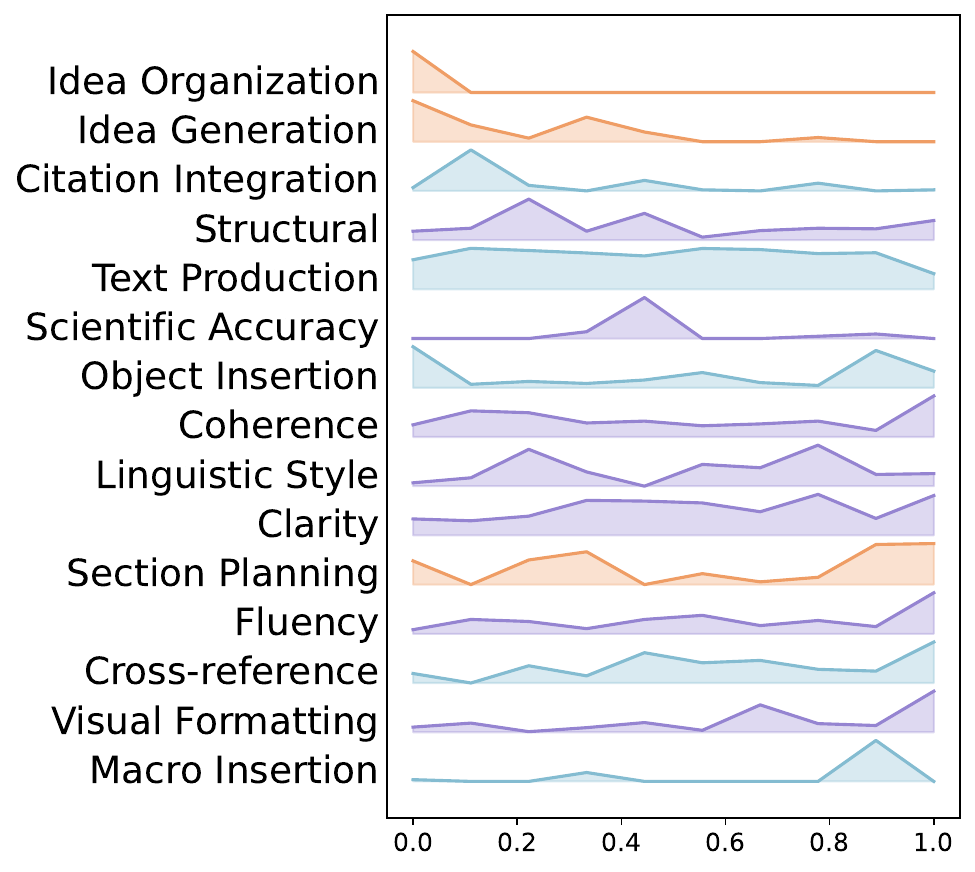}
    \caption{Distribution of intent writing activities over time. See more in Figure \ref{fig:writing-step-detailed-all} and \ref{fig:label-dist-all} in Appendix}\vspace{-2mm}
    \label{fig:writing-detailed-proj4}
\end{figure}

The final \textsc{ScholaWrite} dataset contains 61,504 keystroke-based writing actions collected across five Overleaf projects, each of which culminated in an arXiv preprint authored by graduate researchers. Every action is annotated with a cognitive writing intention drawn from the 15-category taxonomy. 
Table~\ref{tab:data-stat} summarizes key statistics across projects. In total, the dataset captures over 62K text edits, encompassing more than 118K words added or deleted across multi-month writing trajectories. Each project contains thousands of temporally ordered keystrokes, providing a detailed chronicle of the writing process at the sentence, paragraph, and document levels (See Fig \ref{fig:writing-detailed-proj4}) for one of projects). 
\subsection{Analytical Units: Capturing Writing Flow}

\textsc{ScholaWrite} records writing activity every millisecond. For the sake of analysis, we organize this continuous stream into cognitively coherent units. We define two complementary units: \textit{intention sessions} and \textit{sessions}.

\textbf{Intention session.}\label{sec:intention-session-def}
An intention session is a continuous period during which the writer focuses on a \textit{single cognitive goal}, such as producing text, before switching to another intention. These sessions typically last from a few seconds to several minutes, capturing a concentrated burst of focused cognitive activity.

\textbf{Session.} \label{sec:session-def}
A session \textit{aggregates all writing actions within a continuous working period}, regardless of how many intentions are involved. It represents the natural rhythm of writing, how authors weave together planning, drafting, and revising while working on a particular section or idea.

These two units capture both micro-level cognitive focus and macro-level writing flow, forming the backbone of our analyses in \S\ref{sec:analysis}.

\begin{table}[h]
\centering
\footnotesize
\resizebox{\columnwidth}{!}{%
\begin{tabular}{@{}l|c|c|c|c|c@{}}
\toprule
\textbf{Project} & \textbf{1} & \textbf{2} & \textbf{3} & \textbf{4} & \textbf{5} \\ 
\midrule \midrule
\# Authors & 1 (3) & 1 (4) & 1 (3) & 1 (4) & 9 (18) \\ 
\# Keystrokes & 14{,}217 & 5{,}059 & 6{,}641 & 8{,}348 & 27{,}239 \\ 
\# Words added & 17{,}387 & 23{,}835 & 7{,}779 & 12{,}448 & 57{,}511 \\ 
\# Words deleted & 11{,}739 & 15{,}158 & 2{,}308 & 7{,}621 & 25{,}853 \\ 
\bottomrule
\end{tabular}%
}
\caption{Statistics of writing actions per Overleaf project in \textsc{ScholaWrite}. “\# Authors” indicates the number of participants who contributed data, with parentheses showing the total number of actual authors.}\vspace{-3mm}
\label{tab:data-stat}\vspace{-4mm}
\end{table}

\section{Analysis of Scholarly Writing Behavior}\label{sec:analysis}
\textsc{ScholaWrite} offers an unprecedented window into the dynamic, cognitive process of scholarly writing. Our analysis pursues two complementary goals:
(1) to uncover how scholarly writing intentions evolve, interact, and shift across time, and
(2) to inform the design of cognitively aligned writing assistants capable of supporting real-world research workflows.
We focus on three research questions:
\begin{itemize}[noitemsep, topsep=0pt, leftmargin=1em]
    \item (\S\ref{sec:dominate}) What writing intentions dominate the scholarly writing process?  
    \item  (\S\ref{sec:intertwined}) How are different intentions intertwined and sequenced in practice?
    \item (\S\ref{sec:stable}) How do these patterns change across phases of manuscript development?
\end{itemize}

\subsection{Which Writing Intentions Dominate?}\label{sec:dominate}
We examine which writing intentions account for the largest share of time and cognitive effort during scholarly writing.

\paragraph{Time Distribution across Intentions}

Overall, writing activity is dominated by \textcolor{implementationcolor}{Text Production} and \textcolor{revisioncolor}{Revision}, while \textcolor{planningcolor}{Planning} activities play a comparatively smaller role (See Appendix Table~\ref{table:intention-time-avg} for the average share of time spent per intention).
These results suggest that scholars devote most of their time to articulating and refining text rather than explicit planning. The considerable effort spent on revision and formatting underscores that much of scholarly writing involves clarifying and visually shaping ideas, an aspect often overlooked in computational models of writing.




\paragraph{Session Duration and Cognitive Effort}
We analyze the duration of each \textit{intention session}. Intention sessions associated with planning and structural work -- such as \textcolor{planningcolor}{Idea Organization}, \textcolor{implementationcolor}{Object Insertion}, and \textcolor{revisioncolor}{Visual Formatting} -- tend to last longer and exhibit greater variability, reflecting the sustained cognitive effort required for reorganizing content or managing visual elements. (See Appendix Table~\ref{table:intention-session-stats} for intention session level time statistics).

In contrast, \textcolor{implementationcolor}{Text Production}, while accounting for the largest share of overall writing time, typically unfolds in short, frequent bursts. Together, these patterns reveal two complementary cognitive modes in scholarly writing: (1) rapid, iterative drafting episodes, and (2) longer, focused periods of planning or visual composition that demand sustained attention.

\subsection{How Are Writing Intentions Intertwined?}\label{sec:intertwined}
Scholarly writing rarely unfolds linearly. Instead, it involves recursive alternations between complementary intentions, writers generate ideas, structure them, produce text, and immediately refine it.
As seen in Figure~\ref{fig:sankey-flow-1} (and other projects in Appendix Figure \ref{fig:writing-sankey-all}), the intertwined processes of writing reveals dense bidirectional flows across intentions.



\paragraph{Transition Patterns}
The transition probability matrix (Appendix Figure~\ref{fig:trans-prob-all}) shows that \textcolor{planningcolor}{Planning} and \textcolor{implementationcolor}{Implementation} are tightly coupled. Planning activities, such as \textit{Idea Generation}, are frequently followed by \textit{Text Production}. Within \textcolor{planningcolor}{Implementation}, transitions predominantly return to \textcolor{implementationcolor}{Text Production}, whereas \textcolor{revisioncolor}{Revision} intentions exhibit strong recursive loops -- particularly among \textit{Clarity}, \textit{Fluency}, and \textit{Coherence} -- reflecting iterative refinement of language and argumentation.
Overall, the results indicate that planning, implementation, and revision are not distinct stages but interdependent cycles continually revisited throughout the writing process.

\paragraph{Multitasking and Alternation Patterns}

\begin{figure}[ht!]
    \centering
    \includegraphics[width=0.45\textwidth,trim={0.5cm 0 0.5cm 0cm}, clip]{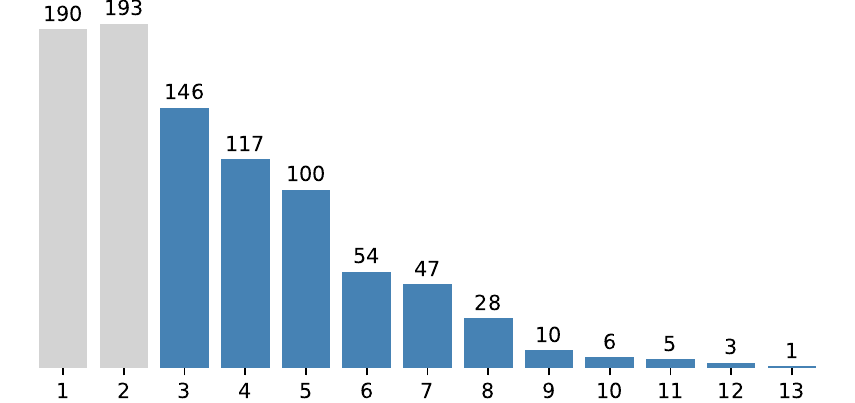}
    \hfill
    \caption{Distribution of the number of intentions per writing session where x-axis represents the number of intentions and the y-axis represents the session count.}
    \label{fig:count_label_session}
\end{figure}

Writing sessions are highly multitasking in nature. As shown in Figure~\ref{fig:count_label_session}, 57\% of all sessions involve three or more distinct intentions, reflecting frequent switching between cognitive modes within a single sitting.
This dynamic interplay reveals how writers manage multiple goals in parallel.

\begin{figure*}[t!]
    \centering
    \begin{subfigure}[t]{0.56\textwidth}
        \centering
        \includegraphics[width=\textwidth,trim=0cm 0cm 0 0cm, clip]{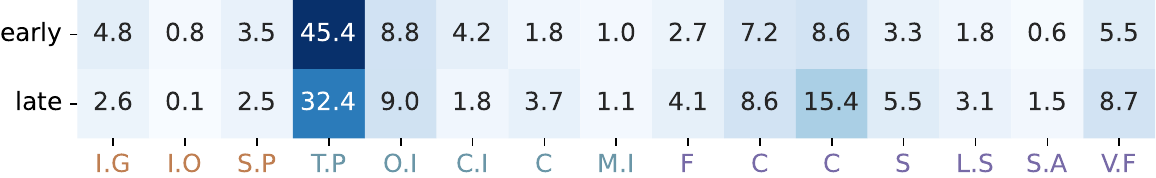}
        \caption{Share of time by intention across early and late phases.}
        \label{fig:time-share-across}
    \end{subfigure}
    \hfill
    \begin{subfigure}[t]{0.4\textwidth}
        \centering
        \includegraphics[width=\textwidth,trim=1.5cm 1.2cm 5cm 0cm,clip]{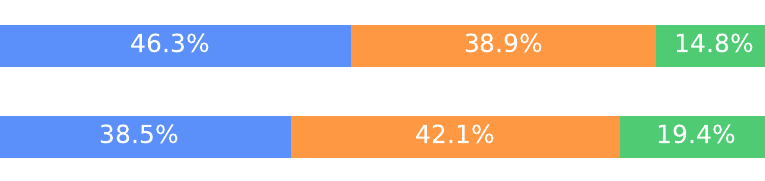}
        \caption{Distribution of number of intentions per session.}
        \label{fig:session-intentions-across}
    \end{subfigure}
    \vspace{-2mm}
    \caption{Dynamics across early and late phases of writing. (a) The share of time devoted to each intention shifts from planning to revision as writing progresses. (b) Later sessions involve more overlapping intentions (blue for 1-2, orange for 3-5, and green for >5 intentions), reflecting higher cognitive integration. }
    \label{fig:intention-temporal-dynamics}\vspace{-2mm}
\end{figure*}

\begin{table}[t!]
\centering
\footnotesize
\begin{tabular}{@{}lccc@{}}
\toprule
\textbf{Intention} & \textbf{Early} & \textbf{Late} & \textbf{Trend} \\ 
\midrule
\plan{Idea Gen.} & 1.6 (23) & \textbf{2.5 (11)} & longer, fewer \\
\plan{Idea Org.} & \textbf{2.6 (2)} & --- & early-only \\
\plan{Sect. Plan.} & 1.6 (13) & \textbf{1.0 (17)} & more, shorter \\ 
\midrule
\impl{Text Prod.} & 2.1 (221) & 2.0 (227) & stable \\
\impl{Obj. Insert.} & 2.4 (28) & \textbf{3.4 (42)} & up both \\
\impl{Cite Int.} & 1.4 (16) & 1.8 (6) & fewer later \\
\impl{Cross-ref.} & 0.7 (8) & \textbf{1.2 (18)} & more later \\ 
\midrule
\rev{Clarity} & 1.5 (43) & \textbf{1.3 (99)} & more, quicker \\
\rev{Structural} & 2.4 (8) & \textbf{1.4 (29)} & more later \\
\rev{Ling. Style} & 3.5 (3) & \textbf{1.7 (12)} & concise edits \\
\rev{Vis. Form.} & 3.6 (16) & \textbf{1.7 (43)} & quick fixes \\
\bottomrule
\end{tabular}
\caption{Early-late phase shift in writing intentions (mean duration [min]; counts in parentheses). Late writing involves more frequent, shorter revision and formatting sessions.}
\label{table:intention-early-late}\vspace{-3mm}
\end{table}

\subsection{Are writing patterns stable throughout the process, or do they change across phases?}\label{sec:stable}
Scholarly writing evolves as authors move from idea formation to refinement. To examine how writing behavior changes across the writing process, we compare the \emph{early phase} (first third of project duration) with the \emph{late phase} (final third). 
\paragraph{Phase-dependent intention patterns}

Writing intentions exhibit clear stage-dependent shifts (Figure~\ref{fig:time-share-across}).
\textcolor{planningcolor}{Planning} intentions are more frequent early, reflecting front-loaded structuring. Within \textcolor{implementationcolor}{Implementation}, \textit{Citation Integration} is higher early, while \textit{Cross-referencing} increases late as authors add visuals and refine internal links.  
\textcolor{revisioncolor}{Revision} grows sharply in the late phase, consistent with end-stage polishing for coherence and presentation.

Table~\ref{table:intention-early-late} summarizes session-level differences. 
In \textcolor{planningcolor}{Planning}, \textit{Idea Generation} and \textit{Idea Organization} sessions decrease in count but lengthen, suggesting deeper cognitive engagement later. \textit{Section Planning} rises as writers reorganize material for submission. For \textcolor{revisioncolor}{Revision}  session counts increase, though durations shorten, indicating rapid, localized edits characteristic of final revisions.

\paragraph{How intention interactions change over time?}
Figure~\ref{fig:transition-prob-accross-time} in the Appendix compares transition probabilities across phases.  
Early writing features strong flows into \textcolor{implementationcolor}{Implementation}, whereas late writing shifts toward \textcolor{revisioncolor}{Revision}, marking the transition from building content to refining it. 
Some intention paths also evolve -- for example, transitions from \textcolor{implementationcolor}{Macro Insertion} lead to \textcolor{planningcolor}{Idea Generation} early but to \textcolor{planningcolor}{Section Planning} late, as layout decisions become more prominent.
Writers also juggle more intentions later in the process as shown in Figure~\ref{fig:session-intentions-across},  reflecting increased multitasking and integration of planning, drafting, and revising.


\section{Benchmarking and Improving AI  Writing Assistants}

\begin{figure*}[t!]
\centering
\scalebox{0.8}{
\begin{tikzpicture}
  \node[anchor=south west] (img) at (0,0) {
    \includegraphics[
      width=\linewidth,
      trim={0.2cm 4cm 4cm 3.5cm},
      clip
    ]{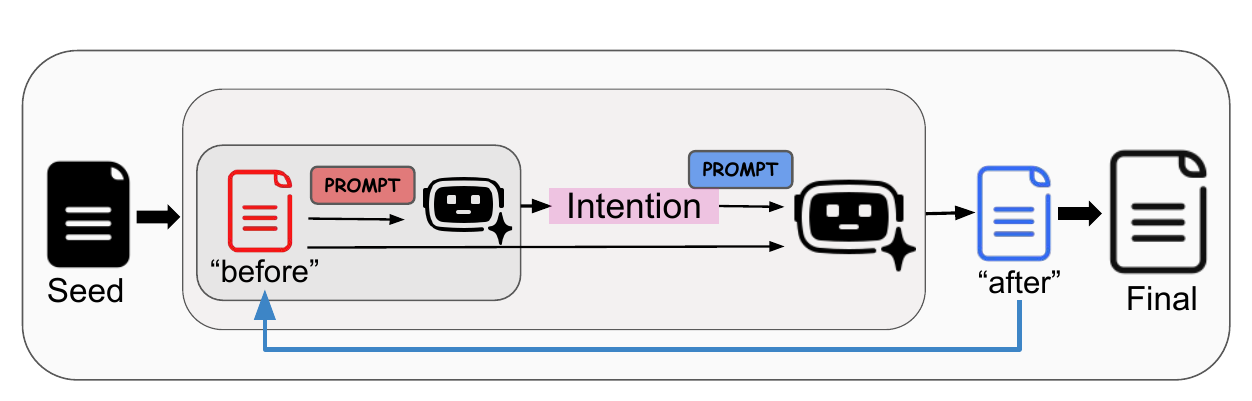}
  };

  \begin{scope}[x={(img.south east)}, y={(img.north west)}]

    \node[
      fill=gray!50,
      minimum width=0.15\linewidth,
      minimum height=0.05\linewidth,
      align=center,
      font=\large
    ] at (0.3, 0.67)
    {\hyperref[sec:next_predict]{Prediction}};

    \node[
      fill=gray!50,
      minimum width=0.15\linewidth,
      minimum height=0.05\linewidth,
      align=center,
      font=\large
    ] at (0.50, 0.75)
    {\hyperref[sec:alignment]{Alignment}};

    \node[
      fill=gray!50,
      minimum width=0.15\linewidth,
      minimum height=0.05\linewidth,
      align=center,
      font=\large
    ] at (0.82, 0.85)
    {\hyperref[sec:iterative]{Generation}};

  \end{scope}
\end{tikzpicture}
}
\vspace{-2mm}
\caption{Overview of our AI writing assistant evaluation setup: intention prediction (\S\ref{sec:next_predict}), output alignment (\S\ref{sec:alignment}), and iterative writing (\S\ref{sec:iterative}).}
\label{fig:all-exp}
\vspace{-3mm}
\end{figure*}

Scientific writing is a cognitively demanding, non-linear, and phase-dependent process in which writers continuously shift between intentions over time. We use \textsc{ScholaWrite} to evaluate and improve cognitively aligned writing assistance along three complementary dimensions: a model’s ability to predict writers’ next intentions (\S\ref{sec:next_predict}), to align its outputs with human writing actions at different temporal granularity (\S\ref{sec:alignment}), and to test how intention prediction and output alignment interact during iterative writing (\S\ref{sec:iterative}) (Figure~\ref{fig:all-exp}). 

\begin{table*}[t!]
\centering
\small
\begin{minipage}[t]{0.55\linewidth}
\centering
\resizebox{\linewidth}{!}{%
\begin{tabular}{@{}lccccc@{}}
\toprule
 & BERT & RoBERTa & Llama-8B & GPT-5 & Qwen-14B \\ 
\midrule
Base & 0.04 & 0.02 & 0.12 & 0.41 & 0.08 \\
+ SW & \textbf{0.64} & \textbf{0.64} & \textbf{0.13} & -- & -- \\
\bottomrule
\end{tabular}%
}
\captionof{table}{Weighted F1 scores for next-intention prediction. “+SW” = fine-tuned on \textsc{ScholaWrite}.}
\label{table:intention-prediction}
\end{minipage}
\hspace{0.02\linewidth}
\begin{minipage}[t]{0.4\linewidth}
\centering
\begin{tabular}{@{}lcc@{}}
\toprule
 & BERTScore-F1 & Levenshtein \\
\midrule
\textcolor{teal}{Llama-8B-Zero} & 0.91 & 0.64 \\
\textcolor{magenta}{Llama-8B-SW}   & \textbf{0.99} & \textbf{0.98} \\
\bottomrule
\end{tabular}
\captionof{table}{Keystroke-level output alignment results}
\label{table:keystroke-output-alignment}
\end{minipage}
\vspace{1em}

\begin{minipage}[t]{\linewidth}
\centering
\resizebox{\linewidth}{!}{%
\begin{tabular}{@{}lcccccc@{}}
\toprule
 & \multicolumn{3}{c}{\textbf{Automatic Evaluation}}
 & \multicolumn{3}{c}{\textbf{Human Evaluation}} \\
\cmidrule(lr){2-4}
\cmidrule(lr){5-7}
 & Lexical Diversity & Topic Consistency & Intention Coverage
 & Accuracy & Alignment / Fluency / Coherence & Relevance \\
\midrule
\textcolor{teal}{Llama-8B-Zero} & 0.20 & 0.66 & 6.75 & \textbf{38.4} & \textbf{3} & \textbf{3} \\
\textcolor{magenta}{Llama-8B-SW} & \textbf{0.43} & \textbf{0.67} & \textbf{9.75} & 16.2 & 0 & 1.75 \\
\bottomrule
\end{tabular}%
}
\captionof{table}{Automatic and human evaluation results for iterative writing across four seed documents. Accuracy reports the average number of generated keystrokes with correctly inferred intentions. Other metrics report the number of human evaluators who agreed on the model’s performance.}
\label{table:iterative-result}
\end{minipage}
\end{table*}

\subsection{Can models anticipate what writers do next?}\label{sec:next_predict}

Because writing involves frequent task-switching, an effective assistant should anticipate the writer’s next move.
Using \textsc{ScholaWrite}, we evaluate whether state-of-the-art language models can infer writers’ upcoming intentions from ongoing writing context, and we further examine the extent to which this capability improves through finetuning.

\paragraph{Setup \& Metrics} 

This task takes the ``before-text'' from a keystroke pair and context prompt, and predicts the writing intention to apply for the subsequent actions.

We evaluate GPT-5 \citep{gpt5} and Qwen2.5-14B-Instruct \citep{qwen2} in a zero-shot setting for next intention prediction.

To assess the impact of finetuning on \textsc{ScholaWrite}, we use BERT \citep{devlin2019bert}, RoBERTa \citep{Liu2019RoBERTaAR}, and Llama-3.1-8B-Instruct \citep{dubey2024llama} as trainable baselines, fine-tuning each on the \textsc{ScholaWrite} training split (80\%) and evaluating on the held-out test split (20\%). See Appendix \ref{sec:appendix:pred-intent} for finetuning details.

To evaluate model performance, we used a weighted F-1 score\footnote{Weighted F-1 was chosen to address skewed label distribution.}.

\paragraph{Findings} \emph{Finding 1: Current state-of-the-art language models struggle to accurately anticipate writers’ intentions during the writing process.} When evaluated on next-intention prediction, state-of-the-art models such as GPT-5 and Qwen-2.5-14B-Instruct still perform poorly on this task, achieving weighted F1 scores below 0.5 (Table~\ref{table:intention-prediction}). 

\emph{Finding 2: Finetuning on \textsc{ScholaWrite} substantially improves models’ ability to predict writers’ intentions.} 
Regardless of the intricate nature of the task itself\footnote{Each model predicts the next intention using only the ``before'' text, while human annotators consider multiple "before and after" edits. Moreover, the chosen next intention is not necessarily the only correct one.}, all models finetuned on \textsc{ScholaWrite} show an improved performance compared to their baselines (Table \ref{table:intention-prediction}). BERT and RoBERTa achieved the most improvement, while Llama-8B-Instruct showed a modest improvement after fine-tuning. Furthermore, a per-intention breakdown (Appendix Table \ref{tab:per-class-f1-bert}) reveals that BERT finetuned on \textsc{ScholaWrite} achieves improved performance across all 15 intention categories, with 9 intentions reaching an F1 score above 0.40.

Those results demonstrate the effectiveness of our \textsc{ScholaWrite} dataset to align language models with writers' intentions.

\subsection{Can models align their output with human writing actions?}\label{sec:alignment}

To assist writers effectively, a model must not only infer their intentions but also generate text aligned with those goals. We evaluate output alignment at two temporal granularities: at the \emph{session level}, which assesses whether model-generated text is consistent with the overall writing actions observed in a human session, and at the \emph{keystroke level}, which examines alignment at the level of fine-grained, moment-to-moment writing actions.

\paragraph{Setup \& Metrics}
For each session, models receive the “before-text” and either the sequence of intentions observed in that session (session-level) or just a single intention (keystroke-level), then generate the corresponding “after-text.” Output alignment is evaluated using two complementary metrics: semantic similarity, measured by BERTScore-F1 \citep{zhang2019bertscore}, and operational similarity, measured by Levenshtein distance.

\subsubsection{Coarse-grained alignment (session-level)}

Session-level alignment is necessary to test whether models can align with human writing goals at the level of overall writing outcomes.

We report main results using GPT-5 as the representative model and evaluate alignment with human writing using BERTScore-F1. Results for other model (Qwen2.5-14B-Instruct) and metric (Levenshtein distance) appear in the Appendix \ref{sec:appendix:coarse-fig} and show consistent patterns.

\begin{figure}[t!]
    \begin{subfigure}[t]{0.49\columnwidth}
        \centering
        \includegraphics[width=\linewidth, trim={0cm 0.4cm 0cm 0cm}, clip]{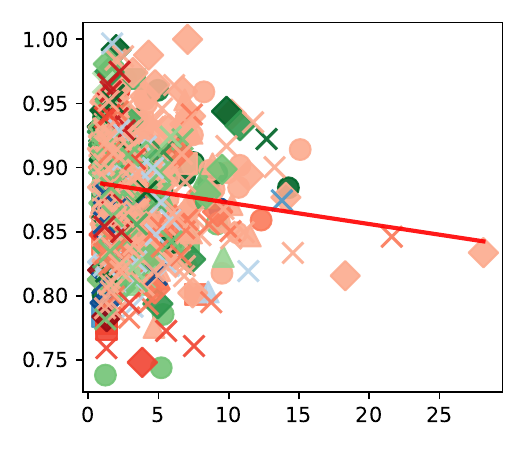}
        \caption{Duration vs.\ alignment}
        \label{fig:duration_gpt5}
    \end{subfigure}
    \begin{subfigure}[t]{0.5\columnwidth}
        \centering
        \includegraphics[width=\linewidth, trim={0.8cm 0.9cm 3.5cm 0cm}, clip]{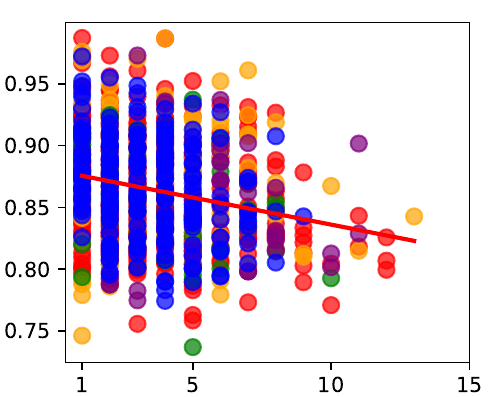}
        \caption{Intents vs.\ alignment}
        \label{fig:intentions_gpt5}
    \end{subfigure}\vspace{-2mm}
    \caption{
    \textbf{Model alignment patterns.} 
    (a; minutes-vs-alignment) Longer writing sessions show lower alignment, indicating higher cognitive complexity. 
    (b; \#-intents-vs-alignment) Alignment decreases as more intentions intertwine within a session. 
    }
    \label{fig:intention-session-summary}\vspace{-3mm}
\end{figure}

\begin{figure}[ht!]
  \centering
        \includegraphics[width=\linewidth, trim={0.8cm 1cm 0cm 0cm}, clip]{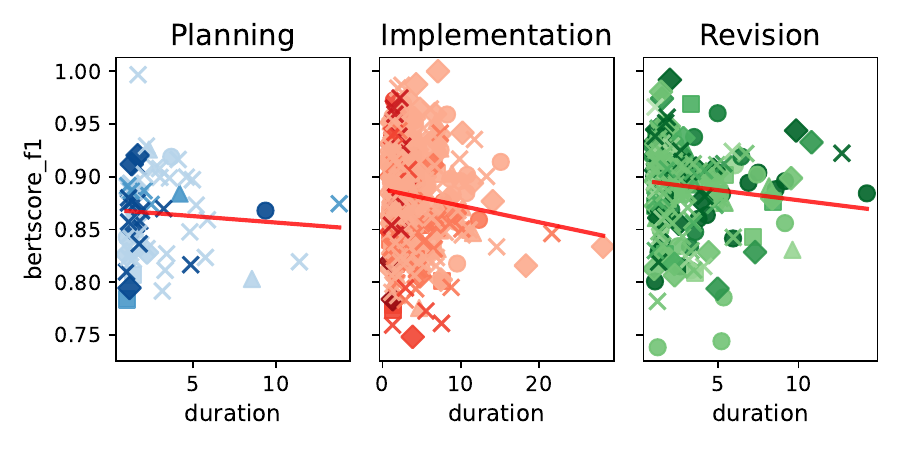}\vspace{-2mm}
    \caption{Writing duration (minutes) vs.\ alignment across different intention categories}
    \label{fig:intention-session-category-only-bertscore}\vspace{-3mm}
\end{figure}

\paragraph{Finding} \emph{Model-human alignment degrades as writing sessions become longer or cognitively more complex} (Figure~\ref{fig:intention-session-summary}). As shown in Figure~\ref{fig:intention-session-category-only-bertscore}, GPT-5 (and Qwen) perform best on \textcolor{revisioncolor}{Revision} intentions (such as \textit{Clarity}, \textit{Fluency}) but struggle with \textcolor{planningcolor}{Planning} and \textcolor{implementationcolor}{Implementation}, which demand deeper reasoning and structural organization.
Alignment further drops when multiple intentions co-occur within a session, indicating models’ limited ability to integrate overlapping cognitive goals.
Overall, current LLMs excel at surface refinement but remain weak at sustaining the intertwined reasoning processes that drive human scholarly writing.


\subsubsection{Fine-grained alignment (keystroke-level)}\label{sec:alignment:keystroke}

While session-level alignment evaluates overall writing outcomes, keystroke-level alignment allows us to study how models can be fine-tuned to align with the fine-grained writing actions that give rise to those outcomes.

We fine-tune Llama-3.1-8B-Instruct (\textcolor{magenta}{Llama-8B-SW}) and compare it against the vanilla Llama-3.1-8B-Instruct (\textcolor{teal}{Llama-8B-Zero}). All models use the same train-test split (80\%, 20\%) across experiments. Training details are provided in Appendix~\ref{sec:appendix:fine-alignment}.

\paragraph{Finding} \emph{Models finetuned on \textsc{ScholaWrite} generate outputs that are more accurately aligned with human writing actions and better positioned within the writing workflow.} Table~\ref{table:keystroke-output-alignment} shows that (\textcolor{magenta}{Llama-8B-SW}) achieves higher BERTScore-F1 and Levenshtein ratio than the base model, indicating improved semantic alignment and closer correspondence to human editing actions. This suggests that \textsc{ScholaWrite} fine-tuning helps models generate outputs that better match both the content and the placement of human revisions.

\begin{table}[t!]
\vspace{-0.5em}
\centering
\footnotesize
\resizebox{\columnwidth}{!}{
\begin{tabular}{@{}c@{\hskip 1mm}@{}|p{4.4cm}p{4.4cm}@{}}
\toprule
\textbf{Iter.} & \textbf{\textcolor{magenta}{Llama-8B-SW}} & \textbf{\textcolor{teal}{Llama-8B-Zero}}\\
\midrule
25 & [\textit{..Editing Abstract..}] but rather should be used to improve the flow of information to avoid information overload\textcolor{red}{\sout{,}}\textcolor{teal}{.} (\textbf{Text Production}) & [\textit{..Editing Table}]
Acceptance rate (\%) \& 75 \& 8\textcolor{red}{\sout{7.5}}\textcolor{teal}{8.2} //
Revision time (minutes) \& 45 \& 2\textcolor{red}{\sout{9}}\textcolor{teal}{8.5}
Human effort (minutes) \& 60 \& 4\textcolor{red}{\sout{3}}\textcolor{teal}{2} 
... (\textbf{Scientific Accuracy})\\
\hline 
51 & [\textit{..Editing Abstract..}] but rather should be used to improve the flow of information, offering \textcolor{red}{\sout{teach}}\textcolor{teal}{previously trained} to \textcolor{teal}{a} load more related information over the\textcolor{teal}{-} load. (\textbf{Clarity}) & \textcolor{teal}{\textbackslash section\{Impact of the Proposed System\}
The proposed system, $\mathcal{R}3$, has the potential to impact the writing process in several ways.}... \textcolor{teal}{\textbackslash section\{Future Research Direction\}}... (\textbf{Structural})\\
\hline
100 & [..\textit{Same as the 99th iteration}] \textcolor{teal}{\textbackslash end\{document\}} (\textbf{Macro Insertion}) & [..\textit{Same as the 99th }] \textcolor{teal}{\textbackslash usepackage[margin=1in]\{geometry\} \texttt{\% Customizes page margins}} \textcolor{teal}{\textbackslash usepackage\{hyperref\} \texttt{\% Enables hyperlinks}} (\textbf{Fluency}) \\ 
\bottomrule
\end{tabular}
}
\caption{Example model outputs at different iterations from the seed document \cite{du-etal-2022-read} (Appendix Listing \ref{table:seed-entry-read})}\vspace{-4mm}
\label{table:iteration_text_example}
\end{table}

\subsection{Does alignment persist across iterative writing?}\label{sec:iterative}

Iterative writing requires jointly predicting writers’ intentions and generating text that is aligned with those intentions across multiple steps of revision. We use this setting to test whether our \textsc{ScholaWrite}-fine-tuned model can sustain intention-aware, aligned generation over extended writing trajectories.

\paragraph{Setup \& Metrics}
During iterative self-writing, models process a LaTeX-formatted seed document (as ``before-text'') with a context prompt to predict the next writing intention, then revises the text (``after-text'') accordingly given prompt. The revised document then serves as the new seed for the next iteration. This process repeats until the iteration limit of 100 is reached.

We compare the same finetuned \textcolor{magenta}{Llama-8B-SW} model (from \ref{sec:alignment:keystroke}) to \textcolor{teal}{Llama-8B-Zero}. Seed documents were derived from LaTeX-formatted abstracts of four award-winning NLP papers on diverse topics \citep{zeng-etal-2024-johnny, lu-etal-2024-semisupervised, du-etal-2022-read, etxaniz-etal-2024-latxa}, as shown in Appendix Listings \ref{table:seed-entry-johnny}-\ref{table:seed-entry-latxa}. Table \ref{table:iteration_text_example} illustrates example outputs from both models across iterations.

We evaluated \textit{lexical diversity} (unique-to-total token ratio), \textit{topic consistency} (cosine similarity between seed and final output), and \textit{intention coverage} (unique writing intentions used over 100 iterations).
For \textbf{human evaluation}, three native English speakers\footnote{The IAA scores are 0.84 (\textcolor{magenta}{SW}) and 0.76 (\textcolor{teal}{Zero}) for accuracy, all 100\% for alignment, fluency, coherence, and 49.8\% (\textcolor{magenta}{SW}) and 100\% (\textcolor{teal}{Zero}) for relevance.} with LaTeX expertise assessed outputs from \textcolor{teal}{Llama-8B-Zero} and \textcolor{magenta}{Llama-8B-SW} on \textit{accuracy} (alignment with predicted intention), \textit{alignment} (similarity to human writing), \textit{fluency} (grammatical correctness), \textit{coherence} (logical structure), and \textit{relevance} (connection to the seed document). Please refer to the Appendix \ref{sec:appendix:iterative-writing} for details about the iterative writing experiments setup.

\begin{figure}[h!]
    \centering
    \begin{subfigure}[t]{\columnwidth}
        \centering
        \includegraphics[width=\columnwidth,trim=0.3cm 1cm 0 0.2cm, clip]{figures/llama8_sw_output_broad_seed3.pdf}
        \caption{\textcolor{magenta}{Llama-8B-SW}}
    \end{subfigure}
    \vspace{2mm}
    \begin{subfigure}[t]{\columnwidth}
        \centering
        \includegraphics[width=\columnwidth,trim=0.3cm 1cm 0cm 0.5cm,clip]{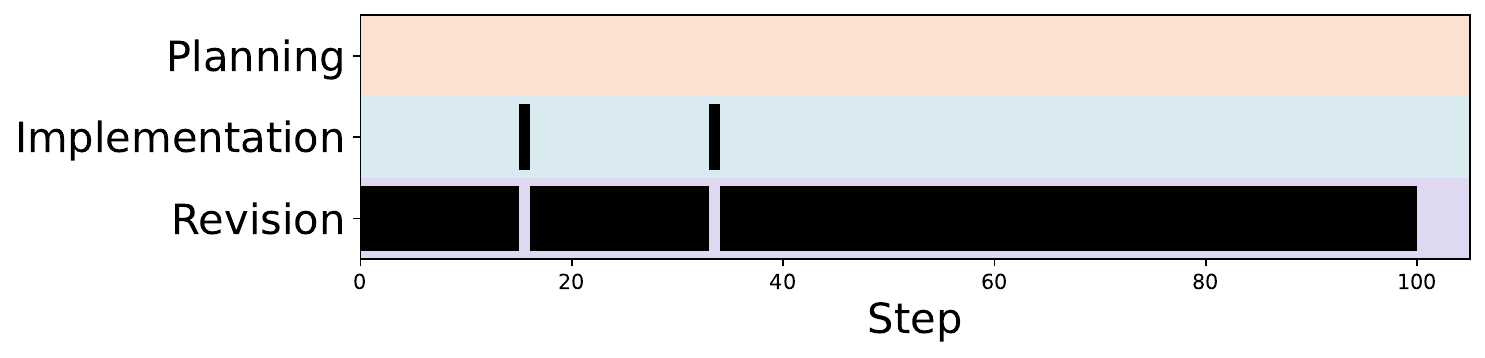}
        \caption{\textcolor{teal}{Llama-8B-Zero}}
    \end{subfigure}
    \vspace{-5mm}
    \caption{Distribution of writing intents on seed \ref{table:seed-entry-semi} over time \textcolor{magenta}{with} and \textcolor{teal}{without} finetuning with ScholaWrite}\vspace{-2mm}
    \label{fig:distribution-models}
\end{figure}

\paragraph{Finding} \emph{Finetuning improves alignment with human writing behavior in iterative settings, even though surface-level text quality remains limited.} \textcolor{magenta}{Llama-8B-SW} shows consistent gains on automatic measures (Table ~\ref{table:iterative-result}). At the process level, it also displays writing activity patterns that more closely resemble human behavior, frequently alternating between implementation and revision and engaging all three high-level writing processes over time (Figure~\ref{fig:distribution-models}). In contrast, \textcolor{teal}{Llama-8B-Zero} tend to remain within a single dominant stage throughout iterative writing.

Human evaluation reveals that, despite these process-level improvements, \textcolor{magenta}{Llama-8B-SW} still lags in surface-level fluency, logical coherence, and consistent realization of predicted intentions (Appendix Figure~\ref{fig:sec5-human-all}). Nevertheless, it produces more relevant content in some settings, aligning with observed gains in topic consistency. 

Together, these results suggest that fine-tuning on \textsc{ScholaWrite} enables models to better emulate the process of human scholarly writing, even as challenges in text quality remain.


\section{Conclusion and Future Work} 

This work introduces \textsc{ScholaWrite}, the first dataset to capture the \emph{end-to-end cognitive process} of scholarly writing through unobtrusive, in-situ logging of real writing activity paired with a cognitively grounded annotation taxonomy. By focusing on how text is produced, rather than solely on the final output, \textsc{ScholaWrite} provides a foundation for studying, evaluating, and training writing assistants grounded in real human writing workflows.

Using \textsc{ScholaWrite}, we demonstrate that scholarly writing is fundamentally non-linear and multi-intentional, and phase-depedent. We further show that, despite strong text generation capabilities, current large language models remain misaligned with human writing workflows, struggling to support writing at the process level. Finally, we find that models trained on \textsc{ScholaWrite} exhibit improved alignment with human writing behavior, suggesting that exposure to process-level supervision enables models to better adapt to evolving writing goals.

Together, \textsc{ScholaWrite} contributes: (1) a \textbf{fine-grained lens} on the cognitive dynamics of scholarly writing; (2) a \textbf{benchmark} for evaluating LLMs’ alignment with human writing intentions; and (3) a \textbf{training resource} for models that adapt to evolving, multi-intent writing behavior.

Future work will extend beyond keystroke logs to include pre-writing ideation, collaboration, and multimodal composition (e.g., figure and data integration). By tracing the full scholarly writing lifecycle, we aim to deepen our understanding of human cognitive patterns and to design AI writing assistants that not only complete text, but truly \textit{complement} how humans think and write.

\section*{Limitations and Ethical Considerations}

We acknowledge several limitations in our study. First, the \textsc{ScholaWrite} dataset is \textbf{currently limited to the computer science domain}, as LaTeX is predominantly used in computer science journals and conferences. This domain-specific focus may restrict the dataset's generalizability to other scientific disciplines. Future work could address this limitation by collecting keystroke data from a broader range of fields with diverse writing conventions and tools, such as the humanities or biological sciences. For example, students in humanities usually write book-length papers and integrate more sources, so it could affect cognitive complexities.

Additionally, the \textsc{ScholaWrite} dataset \textbf{captures only the observable editing activity within the LaTeX environment}, and therefore does not reflect the full writing process of each author. Many researchers engage in pre-writing activities outside of LaTeX -- such as outlining or planning in separate tools, drafting initial content in other documents before transferring text, or conducting ideation and brainstorming offline. As a result, the dataset may underrepresent the early cognitive stages of writing, such as planning and idea generation. Future work could address this gap by combining keystroke logging with complementary data collection methods, such as screen recording, think-aloud protocols, or self-reported writing journals, to capture a more holistic view of the writing process.

Second, our dataset includes \textbf{contributions from only 10 participants, resulting in five final preprints on arXiv}. This small-to-medium sample size is partly due to privacy concerns, as the dataset captures raw keystrokes that transparently reflect real-time human reasoning. To mitigate these concerns, we removed all personally identifiable information (PII) during post-processing and obtained full IRB approval for the study's procedures. However, the highly transparent nature of keystroke data may still have discouraged broader participation. Future studies could explore more robust data collection protocols, such as advanced anonymization or de-identification techniques, to better address privacy concerns and enable larger-scale participation.
We also call for community-wise collaboration and participation for our next version of our dataset, \textsc{ScholaWrite 2.0} and encourage researchers to contact authors for future participation.

Furthermore, \textbf{all participants were early-career researchers} (e.g., PhD students) at an R1 university in the United States. Expanding the dataset to include senior researchers, such as post-doctoral fellows and professors, could offer valuable insights into how writing strategies and revision behaviors evolve with research experience and expertise.
Despite these limitations, our study captured an end-to-end writing process for 10 unique authors, resulting in a diverse range of writing styles and revision patterns. The dataset contains approximately 62,000 keystrokes, offering fine-grained insights into the human writing process, including detailed editing and drafting actions over time. While the number of articles is limited, the granularity and volume of the data provide a rich resource for understanding writing behaviors. Prior research has shown that detailed keystroke logs, even from small datasets, can effectively model writing processes \citep{leijten2013keystroke, guo2018modeling, vandermeulen2023writing}. Unlike studies focused on final outputs, our dataset enables a process-oriented analysis, emphasizing the cognitive and behavioral patterns underlying scholarly writing.

Third, \textbf{collaborative writing is underrepresented} in our dataset, as only one Overleaf project involved multiple authors. This limits our ability to analyze co-authorship dynamics and collaborative writing practices, which are common in scientific writing. Future work should prioritize collecting multi-author projects to better capture these dynamics. Additionally, the dataset is \textbf{exclusive to English-language writing}, which restricts its applicability to multilingual or non-English writing contexts. Expanding to multilingual settings could reveal unique cognitive and linguistic insights into writing across languages.


Finally, the human evaluation process in Section \ref{sec:appendix:human-eval} was determined as exempt from IRB review by the authors' primary institution, while the data collection using our Chrome extension program was fully approved by the IRB at our institution. Importantly, no LLMs were used during any stage of the study, except for grammatical error correction in this manuscript. 

\section*{Acknowledgements}
This work was supported by Grammarly. We also thank members of the Minnesota NLP group for their valuable feedback and comments on initial drafts of this work.

\bibliography{custom}

@inproceedings{cahill-etal-2013-robust,
    title = "Robust Systems for Preposition Error Correction Using {W}ikipedia Revisions",
    author = "Cahill, Aoife  and
      Madnani, Nitin  and
      Tetreault, Joel  and
      Napolitano, Diane",
    editor = "Vanderwende, Lucy  and
      Daum{\'e} III, Hal  and
      Kirchhoff, Katrin",
    booktitle = "Proceedings of the 2013 Conference of the North {A}merican Chapter of the Association for Computational Linguistics: Human Language Technologies",
    month = jun,
    year = "2013",
    address = "Atlanta, Georgia",
    publisher = "Association for Computational Linguistics",
    url = "https://aclanthology.org/N13-1055/",
    pages = "507--517"
}

@inproceedings{boyd-2018-using,
    title = "Using {W}ikipedia Edits in Low Resource Grammatical Error Correction",
    author = "Boyd, Adriane",
    editor = "Xu, Wei  and
      Ritter, Alan  and
      Baldwin, Tim  and
      Rahimi, Afshin",
    booktitle = "Proceedings of the 2018 {EMNLP} Workshop W-{NUT}: The 4th Workshop on Noisy User-generated Text",
    month = nov,
    year = "2018",
    address = "Brussels, Belgium",
    publisher = "Association for Computational Linguistics",
    url = "https://aclanthology.org/W18-6111/",
    doi = "10.18653/v1/W18-6111",
    pages = "79--84"
}

@article{qwen2,
      title={Qwen2 Technical Report}, 
      author={An Yang and Baosong Yang and Binyuan Hui and Bo Zheng and Bowen Yu and Chang Zhou and Chengpeng Li and Chengyuan Li and Dayiheng Liu and Fei Huang and Guanting Dong and Haoran Wei and Huan Lin and Jialong Tang and Jialin Wang and Jian Yang and Jianhong Tu and Jianwei Zhang and Jianxin Ma and Jin Xu and Jingren Zhou and Jinze Bai and Jinzheng He and Junyang Lin and Kai Dang and Keming Lu and Keqin Chen and Kexin Yang and Mei Li and Mingfeng Xue and Na Ni and Pei Zhang and Peng Wang and Ru Peng and Rui Men and Ruize Gao and Runji Lin and Shijie Wang and Shuai Bai and Sinan Tan and Tianhang Zhu and Tianhao Li and Tianyu Liu and Wenbin Ge and Xiaodong Deng and Xiaohuan Zhou and Xingzhang Ren and Xinyu Zhang and Xipin Wei and Xuancheng Ren and Yang Fan and Yang Yao and Yichang Zhang and Yu Wan and Yunfei Chu and Yuqiong Liu and Zeyu Cui and Zhenru Zhang and Zhihao Fan},
      journal={arXiv preprint arXiv:2407.10671},
      year={2024}
}

@article{zhang2019bertscore,
  title={Bertscore: Evaluating text generation with bert},
  author={Zhang, Tianyi and Kishore, Varsha and Wu, Felix and Weinberger, Kilian Q and Artzi, Yoav},
  journal={arXiv preprint arXiv:1904.09675},
  year={2019}
}

@article{pustejovsky2017designing,
  title={Designing annotation schemes: From theory to model},
  author={Pustejovsky, James and Bunt, Harry and Zaenen, Annie},
  journal={Handbook of Linguistic Annotation},
  pages={21--72},
  year={2017},
  publisher={Springer}
}

@article{chan2017using,
  title={Using keystroke logging to understand writers’ processes on a reading-into-writing test},
  author={Chan, Sathena},
  journal={Language Testing in Asia},
  volume={7},
  pages={1--27},
  year={2017},
  publisher={Springer}
}

@article{johansson2010looking,
  title={Looking at the keyboard or the monitor: relationship with text production processes},
  author={Johansson, Roger and Wengelin, {\AA}sa and Johansson, Victoria and Holmqvist, Kenneth},
  journal={Reading and writing},
  volume={23},
  pages={835--851},
  year={2010},
  publisher={Springer}
}

@article{leijten2013keystroke,
  title={Keystroke logging in writing research: Using Inputlog to analyze and visualize writing processes},
  author={Leijten, Mari{\"e}lle and Van Waes, Luuk},
  journal={Written Communication},
  volume={30},
  number={3},
  pages={358--392},
  year={2013},
  publisher={Sage Publications Sage CA: Los Angeles, CA}
}

@book{lindgren2019observing,
  title={Observing writing: Insights from keystroke logging and handwriting},
  author={Lindgren, Eva and Sullivan, Kirk},
  volume={38},
  year={2019},
  publisher={Brill}
}

@article{f508427a-e4c0-3d6a-8abf-03a5d21ec6c4,
 ISSN = {0010096X},
 URL = {http://www.jstor.org/stable/356600},
 author = {Linda Flower and John R. Hayes},
 journal = {College Composition and Communication},
 number = {4},
 pages = {365--387},
 publisher = {National Council of Teachers of English},
 title = {A Cognitive Process Theory of Writing},
 urldate = {2024-08-20},
 volume = {32},
 year = {1981}
}

@article{koo2023decoding,
  title={Decoding the End-to-end Writing Trajectory in Scholarly Manuscripts},
  author={Koo, Ryan and Martin, Anna and Wang, Linghe and Kang, Dongyeop},
  journal={arXiv preprint arXiv:2304.00121},
  year={2023}
}

@inproceedings{du-etal-2022-understanding-iterative,
    title = "Understanding Iterative Revision from Human-Written Text",
    author = "Du, Wanyu  and
      Raheja, Vipul  and
      Kumar, Dhruv  and
      Kim, Zae Myung  and
      Lopez, Melissa  and
      Kang, Dongyeop",
    editor = "Muresan, Smaranda  and
      Nakov, Preslav  and
      Villavicencio, Aline",
    booktitle = "Proceedings of the 60th Annual Meeting of the Association for Computational Linguistics (Volume 1: Long Papers)",
    month = may,
    year = "2022",
    address = "Dublin, Ireland",
    publisher = "Association for Computational Linguistics",
    url = "https://aclanthology.org/2022.acl-long.250",
    doi = "10.18653/v1/2022.acl-long.250",
    pages = "3573--3590",
}

@inproceedings{Krapels1990SecondLW,
  title={Second Language Writing: An overview of second language writing process research},
  author={Alexandra Rowe Krapels},
  year={1990},
  url={https://api.semanticscholar.org/CorpusID:60659113}
}

@article{nickerson2013method,
  title={A method for taxonomy development and its application in information systems},
  author={Nickerson, Robert C and Varshney, Upkar and Muntermann, Jan},
  journal={European Journal of Information Systems},
  volume={22},
  number={3},
  pages={336--359},
  year={2013},
  publisher={Taylor \& Francis}
}

@article{kundisch2021update,
  title={An update for taxonomy designers: methodological guidance from information systems research},
  author={Kundisch, Dennis and Muntermann, Jan and Oberl{\"a}nder, Anna Maria and Rau, Daniel and R{\"o}glinger, Maximilian and Schoormann, Thorsten and Szopinski, Daniel},
  journal={Business \& Information Systems Engineering},
  pages={1--19},
  year={2021},
  publisher={Springer}
}

@inproceedings{jiang-etal-2022-arxivedits,
    title = "ar{X}iv{E}dits: Understanding the Human Revision Process in Scientific Writing",
    author = "Jiang, Chao  and
      Xu, Wei  and
      Stevens, Samuel",
    editor = "Goldberg, Yoav  and
      Kozareva, Zornitsa  and
      Zhang, Yue",
    booktitle = "Proceedings of the 2022 Conference on Empirical Methods in Natural Language Processing",
    month = dec,
    year = "2022",
    address = "Abu Dhabi, United Arab Emirates",
    publisher = "Association for Computational Linguistics",
    url = "https://aclanthology.org/2022.emnlp-main.641",
    doi = "10.18653/v1/2022.emnlp-main.641",
    pages = "9420--9435",
}

@article{kuznetsov2022revise,
  title={Revise and resubmit: An intertextual model of text-based collaboration in peer review},
  author={Kuznetsov, Ilia and Buchmann, Jan and Eichler, Max and Gurevych, Iryna},
  journal={Computational Linguistics},
  volume={48},
  number={4},
  pages={949--986},
  year={2022},
  publisher={MIT Press One Broadway, 12th Floor, Cambridge, Massachusetts 02142, USA~…}
}

@inproceedings{darcy-etal-2024-aries,
    title = "{ARIES}: A Corpus of Scientific Paper Edits Made in Response to Peer Reviews",
    author = "D{'}Arcy, Mike  and
      Ross, Alexis  and
      Bransom, Erin  and
      Kuehl, Bailey  and
      Bragg, Jonathan  and
      Hope, Tom  and
      Downey, Doug",
    editor = "Ku, Lun-Wei  and
      Martins, Andre  and
      Srikumar, Vivek",
    booktitle = "Proceedings of the 62nd Annual Meeting of the Association for Computational Linguistics (Volume 1: Long Papers)",
    month = aug,
    year = "2024",
    address = "Bangkok, Thailand",
    publisher = "Association for Computational Linguistics",
    url = "https://aclanthology.org/2024.acl-long.377",
    doi = "10.18653/v1/2024.acl-long.377",
    pages = "6985--7001",
}

@article{ito2019diamonds,
  title={Diamonds in the rough: Generating fluent sentences from early-stage drafts for academic writing assistance},
  author={Ito, Takumi and Kuribayashi, Tatsuki and Kobayashi, Hayato and Brassard, Ana and Hagiwara, Masato and Suzuki, Jun and Inui, Kentaro},
  journal={arXiv preprint arXiv:1910.09180},
  year={2019}
}

@article{jourdan2024casimir,
  title={CASIMIR: A Corpus of Scientific Articles enhanced with Multiple Author-Integrated Revisions},
  author={Jourdan, L{\'e}ane and Boudin, Florian and Hernandez, Nicolas and Dufour, Richard},
  journal={arXiv preprint arXiv:2403.00241},
  year={2024}
}

@article{mita2022towards,
  title={Towards automated document revision: Grammatical error correction, fluency edits, and beyond},
  author={Mita, Masato and Sakaguchi, Keisuke and Hagiwara, Masato and Mizumoto, Tomoya and Suzuki, Jun and Inui, Kentaro},
  journal={arXiv preprint arXiv:2205.11484},
  year={2022}
}

@inproceedings{narimatsu2021task,
  title={Task definition and integration for scientific-document writing support},
  author={Narimatsu, Hiromi and Koyama, Kohei and Dohsaka, Kohji and Higashinaka, Ryuichiro and Minami, Yasuhiro and Taira, Hirotoshi},
  booktitle={Proceedings of the Second Workshop on Scholarly Document Processing},
  pages={18--26},
  year={2021}
}

@inproceedings{kobayashi-etal-2022-dataset,
    title = "Dataset Construction for Scientific-Document Writing Support by Extracting Related Work Section and Citations from {PDF} Papers",
    author = "Kobayashi, Keita  and
      Koyama, Kohei  and
      Narimatsu, Hiromi  and
      Minami, Yasuhiro",
    editor = "Calzolari, Nicoletta  and
      B{\'e}chet, Fr{\'e}d{\'e}ric  and
      Blache, Philippe  and
      Choukri, Khalid  and
      Cieri, Christopher  and
      Declerck, Thierry  and
      Goggi, Sara  and
      Isahara, Hitoshi  and
      Maegaard, Bente  and
      Mariani, Joseph  and
      Mazo, H{\'e}l{\`e}ne  and
      Odijk, Jan  and
      Piperidis, Stelios",
    booktitle = "Proceedings of the Thirteenth Language Resources and Evaluation Conference",
    month = jun,
    year = "2022",
    address = "Marseille, France",
    publisher = "European Language Resources Association",
    url = "https://aclanthology.org/2022.lrec-1.609",
    pages = "5673--5682",
}

@article{jourdan2023text,
  title={Text revision in scientific writing assistance: An overview},
  author={Jourdan, L{\'e}ane and Boudin, Florian and Dufour, Richard and Hernandez, Nicolas},
  journal={arXiv preprint arXiv:2303.16726},
  year={2023}
}

@article{kallestinova2011write,
  title={How to write your first research paper},
  author={Kallestinova, Elena D},
  journal={The Yale journal of biology and medicine},
  volume={84},
  number={3},
  pages={181},
  year={2011},
  publisher={Yale Journal of Biology and Medicine}
}

@article{bourekkache2022english,
  title={English for specific purposes: writing scientific research papers. case study: Phd students in the computer science department},
  author={Bourekkache, Samir},
  year={2022}
}

@article{liang2024can,
  title={Can large language models provide useful feedback on research papers? A large-scale empirical analysis},
  author={Liang, Weixin and Zhang, Yuhui and Cao, Hancheng and Wang, Binglu and Ding, Daisy Yi and Yang, Xinyu and Vodrahalli, Kailas and He, Siyu and Smith, Daniel Scott and Yin, Yian and others},
  journal={NEJM AI},
  volume={1},
  number={8},
  pages={AIoa2400196},
  year={2024},
  publisher={Massachusetts Medical Society}
}

@article{dubey2024llama,
  title={The llama 3 herd of models},
  author={Dubey, Abhimanyu and Jauhri, Abhinav and Pandey, Abhinav and Kadian, Abhishek and Al-Dahle, Ahmad and Letman, Aiesha and Mathur, Akhil and Schelten, Alan and Yang, Amy and Fan, Angela and others},
  journal={arXiv preprint arXiv:2407.21783},
  year={2024}
}

@misc{gpt5,
    author = "OpenAI",
    title={GPT-5 is here},
    URL = {https://openai.com/gpt-5/}, 
    year={2025},
    note = {Accessed: 2025-10-3}
}

@inproceedings{devlin2019bert,
  title={BERT: Pre-training of Deep Bidirectional Transformers for Language Understanding},
  author={Devlin, Jacob and Chang, Ming-Wei and Lee, Kenton and Toutanova, Kristina},
  booktitle={Proceedings of the 2019 Conference of the North American Chapter of the Association for Computational Linguistics: Human Language Technologies, Volume 1 (Long and Short Papers)},
  pages={4171--4186},
  year={2019}
}

@article{Liu2019RoBERTaAR,
  title={RoBERTa: A Robustly Optimized BERT Pretraining Approach},
  author={Yinhan Liu and Myle Ott and Naman Goyal and Jingfei Du and Mandar Joshi and Danqi Chen and Omer Levy and Mike Lewis and Luke Zettlemoyer and Veselin Stoyanov},
  journal={ArXiv},
  year={2019},
  volume={abs/1907.11692},
  url={https://api.semanticscholar.org/CorpusID:198953378}
}

@inproceedings{zeng-etal-2024-johnny,
    title = "How Johnny Can Persuade {LLM}s to Jailbreak Them: Rethinking Persuasion to Challenge {AI} Safety by Humanizing {LLM}s",
    author = "Zeng, Yi  and
      Lin, Hongpeng  and
      Zhang, Jingwen  and
      Yang, Diyi  and
      Jia, Ruoxi  and
      Shi, Weiyan",
    editor = "Ku, Lun-Wei  and
      Martins, Andre  and
      Srikumar, Vivek",
    booktitle = "Proceedings of the 62nd Annual Meeting of the Association for Computational Linguistics (Volume 1: Long Papers)",
    month = aug,
    year = "2024",
    address = "Bangkok, Thailand",
    publisher = "Association for Computational Linguistics",
    url = "https://aclanthology.org/2024.acl-long.773",
    doi = "10.18653/v1/2024.acl-long.773",
    pages = "14322--14350",
}

@inproceedings{du-etal-2022-read,
    title = "Read, Revise, Repeat: A System Demonstration for Human-in-the-loop Iterative Text Revision",
    author = "Du, Wanyu  and
      Kim, Zae Myung  and
      Raheja, Vipul  and
      Kumar, Dhruv  and
      Kang, Dongyeop",
    editor = "Huang, Ting-Hao 'Kenneth'  and
      Raheja, Vipul  and
      Kang, Dongyeop  and
      Chung, John Joon Young  and
      Gissin, Daniel  and
      Lee, Mina  and
      Gero, Katy Ilonka",
    booktitle = "Proceedings of the First Workshop on Intelligent and Interactive Writing Assistants (In2Writing 2022)",
    month = may,
    year = "2022",
    address = "Dublin, Ireland",
    publisher = "Association for Computational Linguistics",
    url = "https://aclanthology.org/2022.in2writing-1.14",
    doi = "10.18653/v1/2022.in2writing-1.14",
    pages = "96--108",
}

@inproceedings{lu-etal-2024-semisupervised,
    title = "Semisupervised Neural Proto-Language Reconstruction",
    author = "Lu, Liang  and
      Xie, Peirong  and
      Mortensen, David",
    editor = "Ku, Lun-Wei  and
      Martins, Andre  and
      Srikumar, Vivek",
    booktitle = "Proceedings of the 62nd Annual Meeting of the Association for Computational Linguistics (Volume 1: Long Papers)",
    month = aug,
    year = "2024",
    address = "Bangkok, Thailand",
    publisher = "Association for Computational Linguistics",
    url = "https://aclanthology.org/2024.acl-long.788",
    doi = "10.18653/v1/2024.acl-long.788",
    pages = "14715--14759",
}

@inproceedings{etxaniz-etal-2024-latxa,
    title = "Latxa: An Open Language Model and Evaluation Suite for {B}asque",
    author = "Etxaniz, Julen  and
      Sainz, Oscar  and
      Miguel, Naiara  and
      Aldabe, Itziar  and
      Rigau, German  and
      Agirre, Eneko  and
      Ormazabal, Aitor  and
      Artetxe, Mikel  and
      Soroa, Aitor",
    editor = "Ku, Lun-Wei  and
      Martins, Andre  and
      Srikumar, Vivek",
    booktitle = "Proceedings of the 62nd Annual Meeting of the Association for Computational Linguistics (Volume 1: Long Papers)",
    month = aug,
    year = "2024",
    address = "Bangkok, Thailand",
    publisher = "Association for Computational Linguistics",
    url = "https://aclanthology.org/2024.acl-long.799",
    doi = "10.18653/v1/2024.acl-long.799",
    pages = "14952--14972",
}

@article{guo2018modeling,
  title={Modeling basic writing processes from keystroke logs},
  author={Guo, Hongwen and Deane, Paul D and van Rijn, Peter W and Zhang, Mo and Bennett, Randy E},
  journal={Journal of Educational Measurement},
  volume={55},
  number={2},
  pages={194--216},
  year={2018},
  publisher={Wiley Online Library}
}

@article{vandermeulen2023writing,
  title={Writing process feedback based on keystroke logging and comparison with exemplars: Effects on the quality and process of synthesis texts},
  author={Vandermeulen, Nina and Van Steendam, Elke and De Maeyer, Sven and Rijlaarsdam, Gert},
  journal={Written Communication},
  volume={40},
  number={1},
  pages={90--144},
  year={2023},
  publisher={Sage Publications Sage CA: Los Angeles, CA}
}

@article{dettmers2024qlora,
  title={Qlora: Efficient finetuning of quantized llms},
  author={Dettmers, Tim and Pagnoni, Artidoro and Holtzman, Ari and Zettlemoyer, Luke},
  journal={Advances in Neural Information Processing Systems},
  volume={36},
  year={2024}
}

@article{diederich1974measuring,
  title={Measuring growth in English.},
  author={Diederich, Paul B},
  year={1974},
  publisher={ERIC}
}

@article{macarthur2016writing,
  title={Writing research from a cognitive perspective.},
  author={MacArthur, Charles A and Graham, Steve},
  year={2016},
  publisher={The Guilford Press}
}

\appendix

\newpage
\clearpage



\section{More about the Data Collection Process}
\subsection{Participant Recruitment \& Demographics} \label{sec:appendix:recruitment}

We recruited ten graduate students in the computer science department who actively prepared their manuscripts in Overleaf, an online \LaTeX{} editor, and who aimed to submit their manuscripts to peer-reviewed conferences. We held a consent procedure with each participant through a 30-minute virtual meeting remotely.  After the consent process, we installed on their computers a Chrome extension program that we designed and implemented only for this study and asked for the ID number of only the Overleaf projects that the participants agreed to share as their manuscripts. 
We provided all participants with a \$100 Amazon gift card per project, which could be divided among the authors involved in the project. This compensation level was determined based on the expected duration of longitudinal participation and the sustained engagement required for collecting personalized writing workflow data. Our data collection process is approved by the IRB of the authors' primary institution. 

All ten participants of our study are graduate students who currently study computer science domain at an accredited university in the United States. Out of the ten, two of them identified themselves as native English speakers, and the remaining participants identified themselves as proficient in English in terms of writing. Also, two of the ten participants attained a Master of Science degree in Computer Science with several publication experiences, and the remaining eight of them are currently PhD students with extensive research experiences. 

\subsection{Technical Details of System Implementations} \label{sec:appendix:system} 
We built a custom Chrome extension (Appendix Figure \ref{fig:chrome-extension}) that unobtrusively logs real-time keystroke trajectories from Overleaf. The extension operates in the background once participants consent and authenticate via unique credentials. Every time a key-up event occurs, the system captures the visible text within the user’s Overleaf editor, compares it with the previous version using Google’s diff\_match\_patch algorithm, and stores the resulting text differences alongside metadata such as timestamp, file name, and author ID.
To protect privacy, only project IDs pre-approved through participant consent were collected; any non-consented Overleaf projects were automatically filtered out. All data were stored on a secure institutional server and anonymized before analysis.

\begin{figure}[ht!]
    \centering
    {\includegraphics[width=0.7\columnwidth]{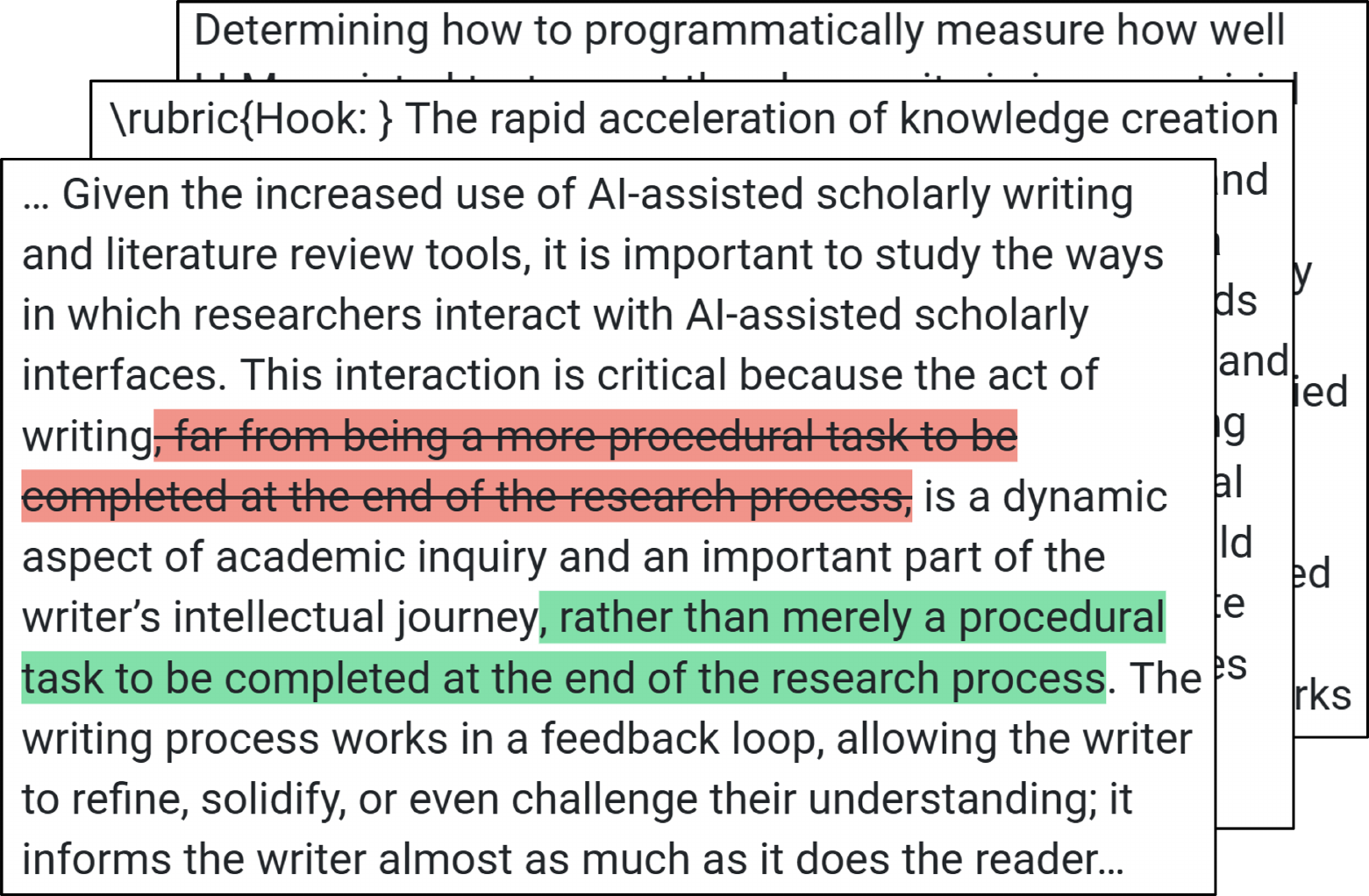}}
    \caption{The array of differences between two subsequent texts, generated by \texttt{diff\_match\_patch}}
    \label{fig:system-diff}
\end{figure}

When a key-up event fires in a browser, the extension collects the writer's viewable texts in the code editor panel\footnote{To prevent privacy concerns, the extension filters out keystroke data from any unauthorized Overleaf projects. Please see Appendix \ref{sec:appendix:system} for more details.}. When each of these actions\footnote{Example actions are (1) inserting a space/newline; (2) copy/paste; (3) undo/redo; (4) switching files and (5) scrolling a page.} occurs, the extension uses `\texttt{diff\_match\_patch}' package\footnote{https://github.com/google/diff-match-patch} to generate an array of differences between two subsequent texts (i.e., Figure \ref{fig:system-diff}). Then, the extension will send the array along with metadata (e.g., time stamp, author ID, etc.) to the backend server. 

For any Overleaf project that consists of multiple LaTeX files, we also collected all keystrokes from subfiles associated with the main LaTeX file. Our comprehensive data collection process captures the end-to-end writing processes of the participating authors across all parts of Overleaf projects. This approach ensures that our dataset reflects the full scope of scholarly writing including edits made in auxiliary files such as files of each section, appendix, bibliography, etc. 

We explain the technical implementation details of the two systems for the data collection process. For the Chrome extension, we implemented a backend application using Flask and Python and stored all keystroke data in the MongoDB database. 

For the annotation interface, we used HTML/CSS and JavaScript for the client side and Flask for the backend. All data for the annotation interface was retrieved from the MongoDB database used in the Chrome extension system. 

\paragraph{Privacy Concerns}
To prevent any issue of private data collection, we designed the backend of our Chrome extension to fetch only the IDs of the Overleaf projects that participants consented to share during the recruitment process and filter out participants' keystroke data from any unauthorized projects. We used Google Sheet API to retrieve ID information that we collected during the recruitment process.

\subsection{Data Post-Processing}
\label{sec:appendix:postprocess}

For use during the annotation phase, each keystroke entry from the raw collection includes the following fields: (1) a valid file name; (2) a valid writing action that triggered keystroke logging (e.g., copy, paste, typing, etc.); (3) a valid array of differences to enable visualization of writing trajectories; and (4) the line numbers in the Overleaf editor. Data entries annotated with a valid intention label (i.e., labels except `artifact') and having a difference array length of fewer than or equal to $300$ are then used for model training.

Regarding the additional postprocessing for public use, we took the following steps to post-process our data with the annotations to promote the usability of our dataset and prevent any privacy issues. 
For the annotation data, we only include the keystroke changes, anonymized project ID, timeframe information, and anonymized author's name (e.g., 0, 1, 2, etc.) from the metadata. Then, we extract the before and after texts from the differences array. We also include the annotated intention label for each entry.

Then, we analyzed any `artifact' generated due to natural keyboard/mouse activities or user switching files, and we discarded them as they are not informative to any writing intention in our taxonomy. 
Lastly, to prevent any privacy issues we removed keystrokes containing any private author information such as names, affiliations,  contact information, and any personally identifiable information (PID) from the collected keystrokes. Instead,  we replaced those with an arbitrary command (e.g., `\texttt{\textbackslash anonymous}').


\section{More about the Annotation} \label{sec:appendix:annotation}

\subsection{Annotator Recruitment}

Due to privacy concerns, we did not hire external freelancers with expertise, rather the two corresponding authors of this paper annotated the data are graduate students who possess extensive scholarly writing experiences in natural language processing and data annotation skills. The raw keystroke data collected by our Chrome extension could potentially contain personally identifiable information, such as specific content edits or metadata that could reveal the identity of the authors. To ensure the confidentiality and ethical handling of sensitive information, we restricted access to the data to the authors only. This annotation process was also authorized by the IRB of the authors’ institution. Please note that the final dataset which will be released publicly is ensured not to contain any PII information through several sophisticated post-processing steps.

\subsection{Annotation Interface}
The annotation process transforms raw writing traces into interpretable cognitive data, forming the foundation of the \textsc{ScholaWrite} taxonomy and subsequent analyses.
Each keystroke record contains metadata, such as file name, type of action, text differences between two states, and line number in the Overleaf editor, which allows precise reconstruction of how writers iteratively develop and refine their manuscripts. 
This process is crucial for developing a detailed taxonomy of the end-to-end scholarly writing process, which serves as the basis for annotating the collected keystrokes. To perform annotations, we processed those raw keystroke entries by file name, type of writing actions, an array of differences between two subsequent texts, and line number in an Overleaf editor (Figure~\ref{fig:annotation-interface}). Annotators can navigate a timeline of keystrokes, examine spans of related edits (e.g., drafting a sentence or clarifying an argument), and assign one or more intention labels from a predefined taxonomy via a dropdown menu.

\subsection{Detailed Annotation Process}

    

\begin{figure}[t!]
    \centering
    {\includegraphics[width=\columnwidth]{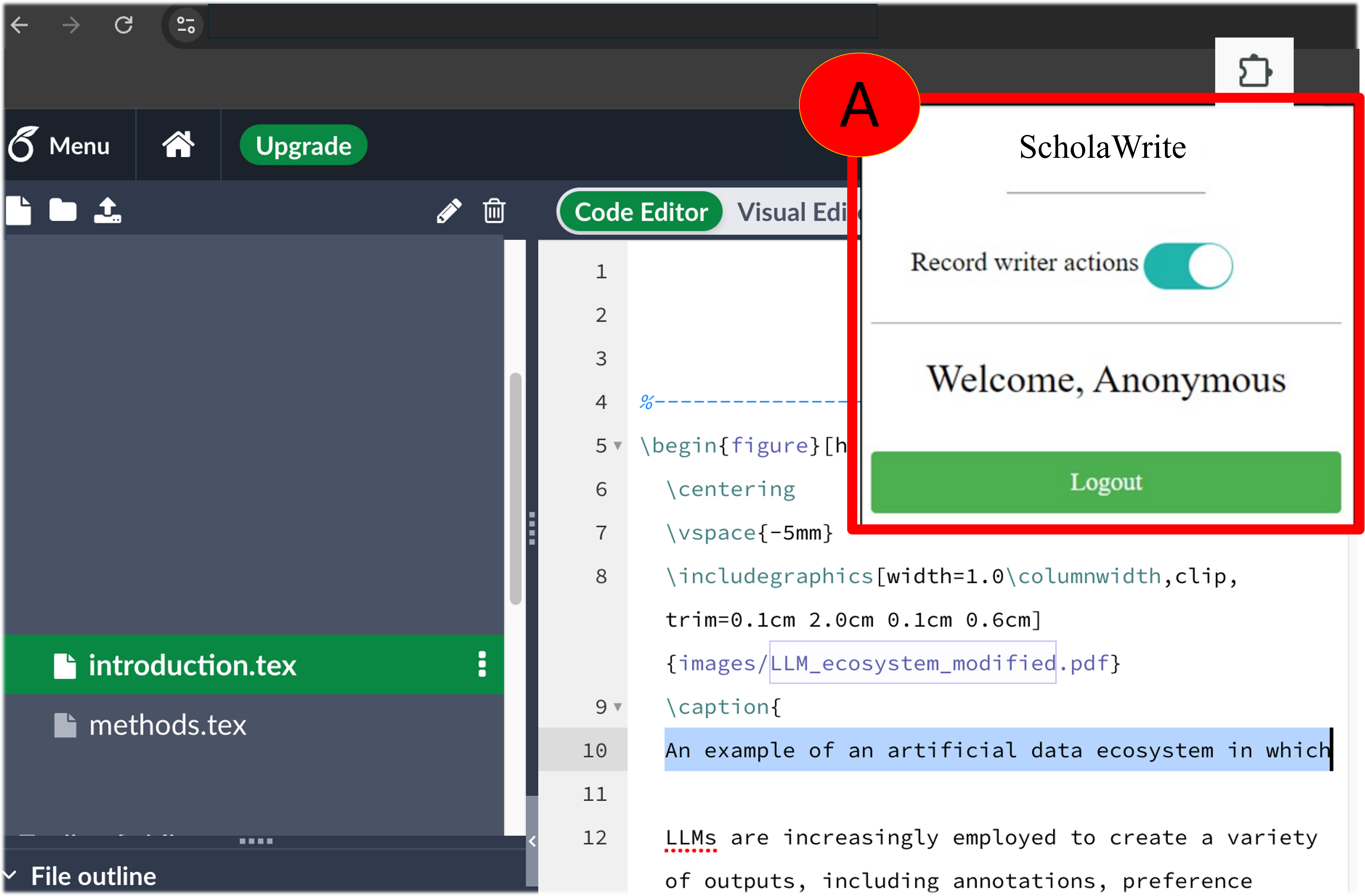}}
    \caption{The Chrome extension interface (A) on the Overleaf project, where it collects real-time keystrokes in the Overleaf editor (highlighted). \vspace{-4mm}}
    \label{fig:chrome-extension}
\end{figure}

\begin{figure}[ht!]
    \centering
    {\includegraphics[width=\columnwidth]{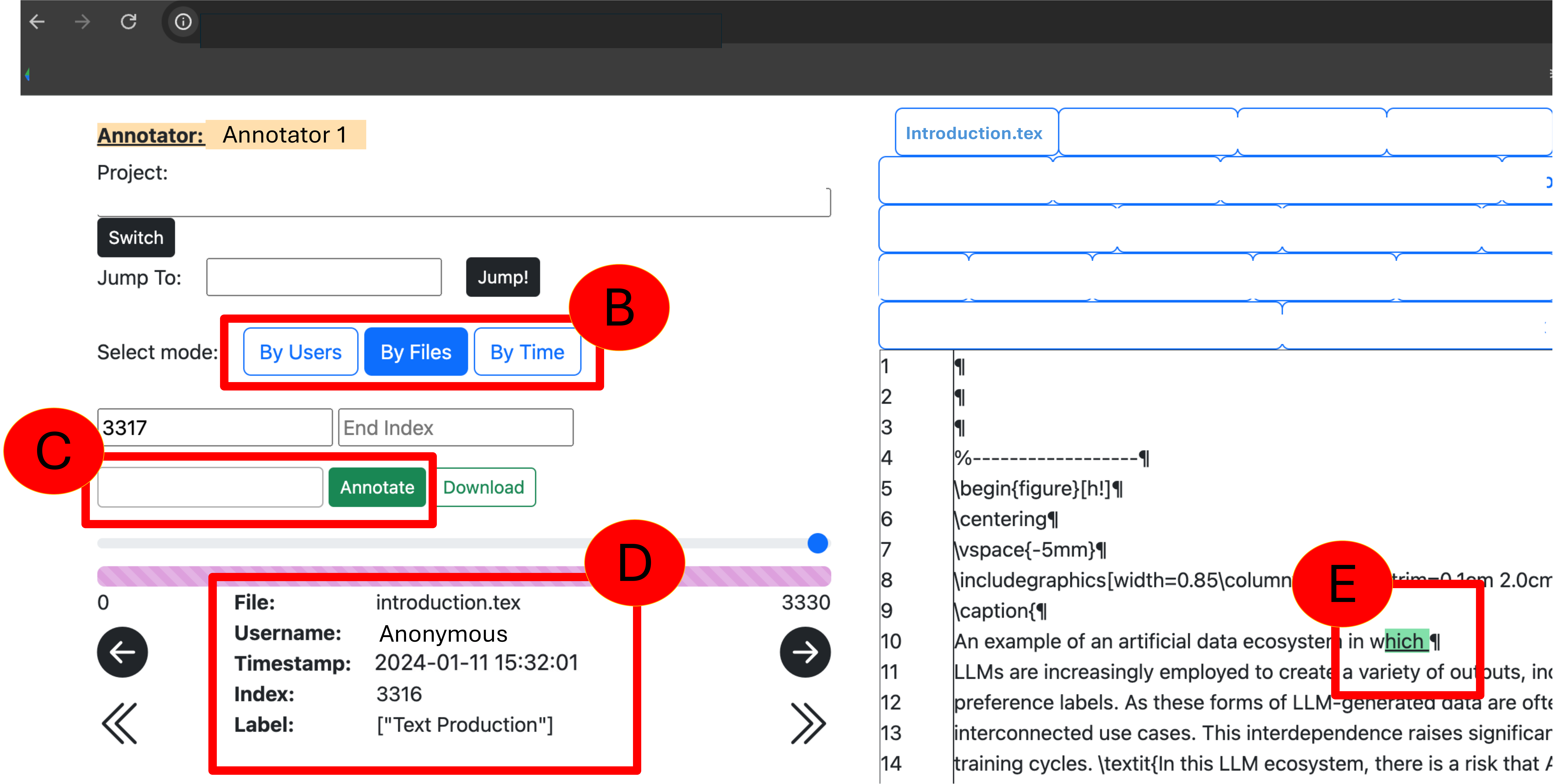}}
    \caption{Annotation interface. During the annotation stage, annotators can click a viewing mode of the collected keystroke data (B). By right-clicking to navigate the timeline of keystroke trace in the interactive panel on the right side (E), annotators can choose an intention label under the drop-down menu (C). They can also view the meta-information of each annotated keystroke (D). }
    \label{fig:annotation-interface}
\end{figure}


Two members of the research team served as annotators, collaborating with a cognitive linguist to develop a codebook and review the annotation results applied to the participants' keystroke data. Also, the two annotators conducted an iterative open coding approach to identify several unique writing intentions from keystrokes and developed a codebook of intention labels (“ground-truth labels”) within each high-level process (Planning, Implementation, and Revision) based on the findings from \citet{f508427a-e4c0-3d6a-8abf-03a5d21ec6c4, koo2023decoding}. Using this codebook, those annotators re-labeled each span of keystrokes with the corresponding label during the annotation process. 

The annotators were fully informed about all the labels and had complete access to them when annotating each data point. The annotation process for all the labels is the same: First, they view through multiple consecutive data points and identify which high-level label occurs (e.g., Planning, Implementation, or Revision). Once annotators have identified the current high-level label, attempting to identify where it ends. Then, they decide on the low-level label within the high-level label (e.g., idea generation or organization under the Planning stage, etc.). Finally, they identify the interval for low-level labels and annotate data points in the interval with the identified low-level label. If a keystroke does not deliver any insight, then label it as an `artifact.' 

We calculated \textbf{inter-annotator agreement} using the weighted F1 score in a multi-label, multi-class setting, which is suitable for our complex annotation schema involving multiple labels per instance. The weighted F1 score achieved was 0.71, indicating a high level of agreement between the annotators.

\section{More About the Taxonomy} \label{sec:appendix:taxonomy}

During the \textbf{planning} stage, the writers engage in a process of generating and organizing raw ideas, arguments, or content structures that were not introduced in the previous trajectory. Based on the plan, the writers \textbf{implement} their plan by drafting full sentences and paragraphs and structuring the contents tangibly. At the same time, the writers enter the \textbf{revision} stage by improving the quality of their implemented sentences and LaTeX objects in terms of linguistic styles, format, or information accuracy. Particularly, spans of keystrokes whose intentions involved any changes but did not change the meaning of original texts are classified as \textbf{Revision}. For those edits that show changes in the meaning, we considered them as \textbf{Implementation}. 
Furthermore, if an author repeatedly adds, removes, and revises text back and forth until a sentence is completed, we consider this process as part of \textbf{text production}. Any subsequent changes made to the sentence after it is finished are considered \textbf{revision}. Table \ref{table:taxonomy-full} presents the comprehensive, complete definitions of each intention of end-to-end scholarly writing process, identified from \textsc{ScholaWrite} dataset.





\section{\textsc{ScholaWrite} dataset statistics}

Table \ref{table:label_distribution_all} shows the distribution of intention labels per Overleaf project from \textsc{ScholaWrite}.


\begin{table}[ht!]
\footnotesize
\centering
\begin{tabular}{@{}lp{1pt}rr@{}}
\toprule
Label & & Subsequent label & Probability  \\
\midrule
\colorbox{planningcolor}{Idea Generation} & $\rightarrow$ & \colorbox{implementationcolor}{Text Production} & 0.52 \\
\colorbox{planningcolor}{Idea Organization} & $\rightarrow$ & \colorbox{planningcolor}{Idea Generation} & 0.34  \\
\colorbox{planningcolor}{Section Planning} & $\rightarrow$ & \colorbox{implementationcolor}{Text Production} & 0.33  \\
\midrule
\colorbox{implementationcolor}{Text Production} & $\rightarrow$ & \colorbox{revisioncolor}{Clarity} & 0.20\\
\colorbox{implementationcolor}{Object Insertion} & $\rightarrow$ & \colorbox{implementationcolor}{Text Production} & 0.32 \\
\colorbox{implementationcolor}{Citation Integration} & $\rightarrow$ & \colorbox{implementationcolor}{Text Production} & 0.37 \\
\colorbox{implementationcolor}{Cross-reference} & $\rightarrow$ & \colorbox{implementationcolor}{Text Production} & 0.36  \\
\colorbox{implementationcolor}{Macro Insertion} & $\rightarrow$ & \colorbox{planningcolor}{Idea Generation} & 0.29 \\
\midrule
\colorbox{revisioncolor}{Fluency} & $\rightarrow$ & \colorbox{implementationcolor}{Text Production}  & 0.30 \\
\colorbox{revisioncolor}{Coherence} & $\rightarrow$ & \colorbox{implementationcolor}{Text Production}  & 0.34  \\
\colorbox{revisioncolor}{Clarity} & $\rightarrow$ & \colorbox{implementationcolor}{Text Production}  & 0.35  \\
\colorbox{revisioncolor}{Structural} & $\rightarrow$ & \colorbox{implementationcolor}{Text Production}  & 0.27  \\
\colorbox{revisioncolor}{Linguistic Style} & $\rightarrow$ & \colorbox{implementationcolor}{Text Production}  & 0.29 \\
\colorbox{revisioncolor}{Scientific Accuracy} & $\rightarrow$ & \colorbox{implementationcolor}{Text Production}  & 0.34 \\
\colorbox{revisioncolor}{Visual Formatting} & $\rightarrow$ & \colorbox{implementationcolor}{Text Production} & 0.25 \\
\bottomrule
\end{tabular}
\caption{Probability of inter-connections between writing intentions in \textsc{ScholaWrite}. For example, in 34\% of instances where an author engaged in ``Idea Organization,'' the subsequent intention was ``Idea Generation.'' }
\label{table:flow-intention-full}
\end{table}

\begin{table}[ht!]
    \centering
    \footnotesize
    \begin{tabular}{@{}l|ccccc@{}}
        \toprule
         & 1 & 2 & 3 & 4 & 5 \\
        \midrule \midrule
        Idea Generation & 515 & 130 & 116 & 309 & 3255 \\
        Idea Organization & 0 & 45 & 25 & 9 & 231 \\
        Section Planning & 182 & 57 & 111 & 201 & 773 \\
        \midrule
        Text Production & 9267 & 2438 & 5109 & 4478 & 14031 \\
        Object Insertion & 583 & 383 & 62 & 486 & 1300 \\
        Cross-reference & 141 & 112 & 13 & 292 & 458 \\
        Citation Integration & 75 & 151 & 69 & 127 & 245 \\
        Macro Insertion & 16 & 7 & 51 & 29 & 33 \\
        \midrule 
        Linguistic Style & 233 & 75 & 42 & 201 & 411 \\
        Coherence & 422 & 242 & 126 & 193 & 1021 \\
        Clarity & 1249 & 645 & 721 & 1180 & 3301 \\
        Scientific Accuracy & 307 & 15 & 2 & 24 & 95 \\
        Structural & 359 & 506 & 105 & 257 & 1042 \\
        Fluency & 116 & 90 & 46 & 135 & 476 \\
        Visual Formatting & 752 & 163 & 43 & 427 & 567 \\
        \bottomrule
        \end{tabular}
    \caption{Distribution of intention labels annotated across all five Overleaf projects.} \label{table:label_distribution_all}
\end{table}

\begin{figure*}
    \centering
    \begin{subfigure}{0.32\textwidth}
        \includegraphics[width=\textwidth]{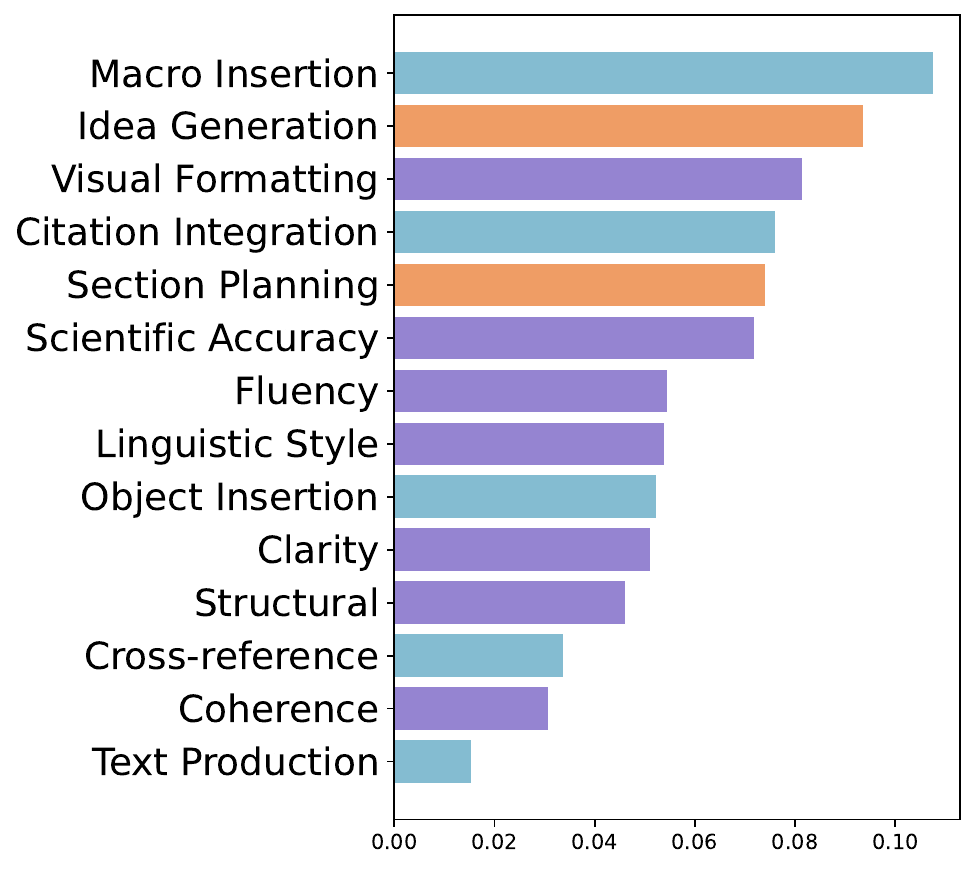}
        \caption{Project 1}
    \end{subfigure}
    \begin{subfigure}{0.32\textwidth}
        \includegraphics[width=\textwidth]{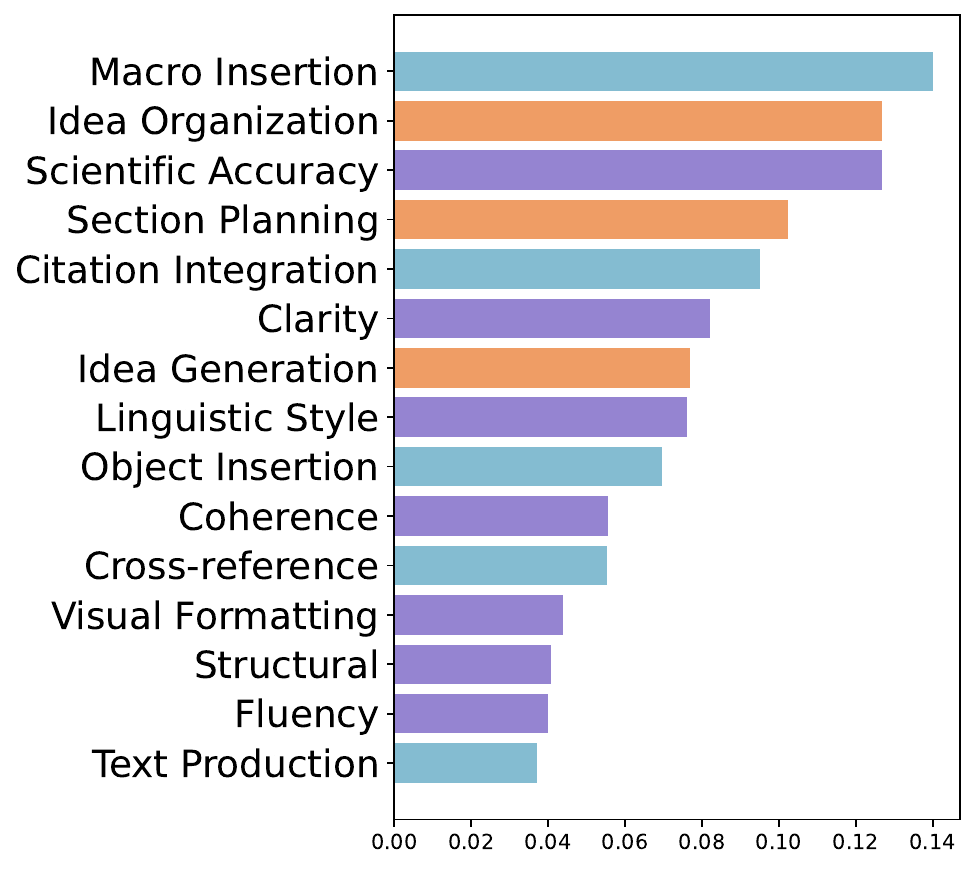}
        \caption{Project 2}
    \end{subfigure}
    \begin{subfigure}{0.32\textwidth}
        \includegraphics[width=\textwidth]{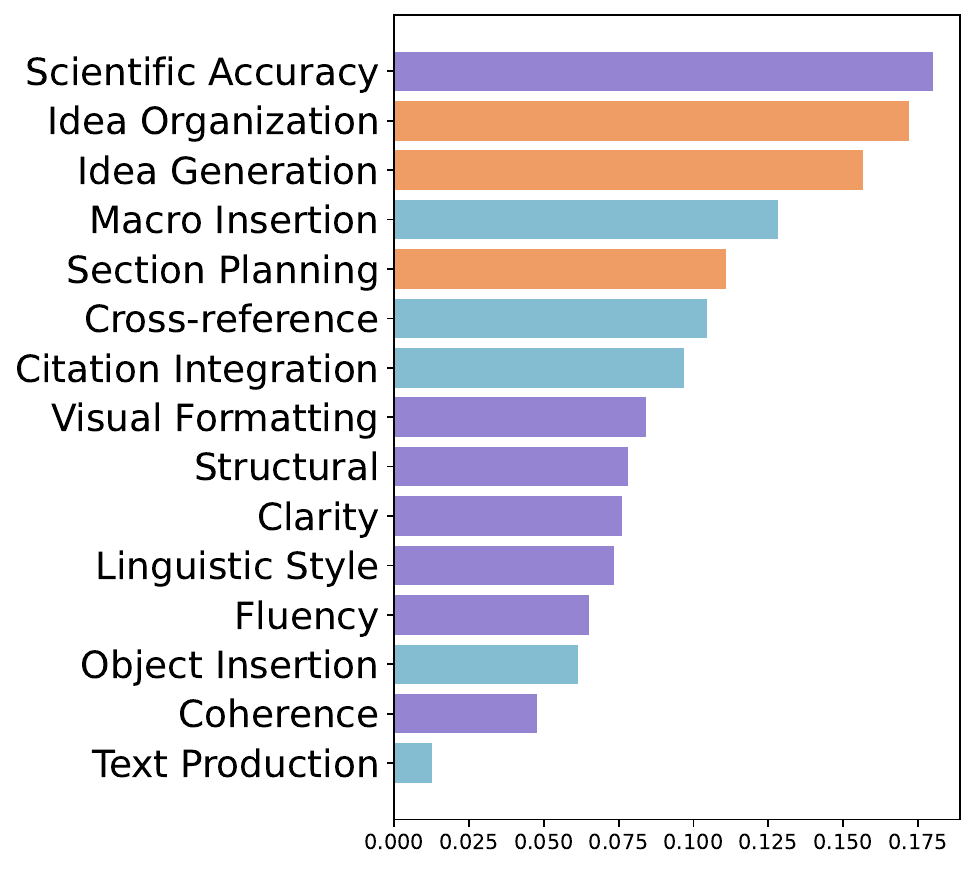}
        \caption{Project 3}
    \end{subfigure}
    \begin{subfigure}{0.32\textwidth}
        \includegraphics[width=\textwidth]{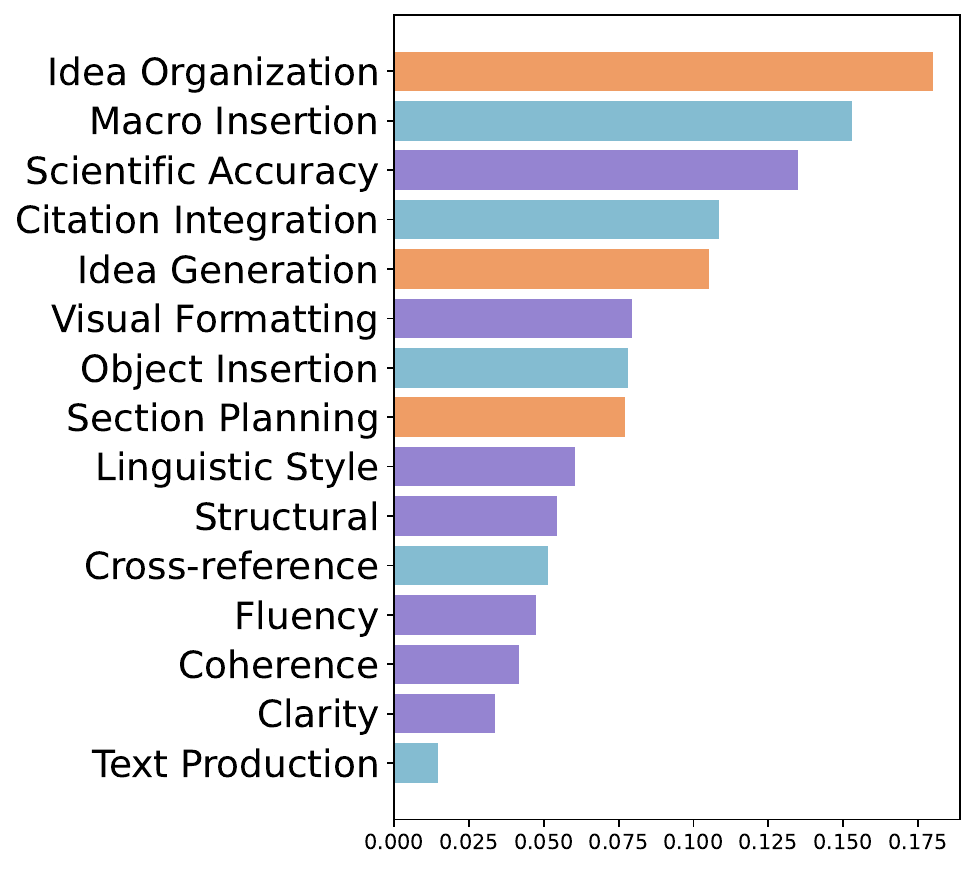}
        \caption{Project 4}
    \end{subfigure}
    \begin{subfigure}{0.32\textwidth}
        \includegraphics[width=\textwidth]{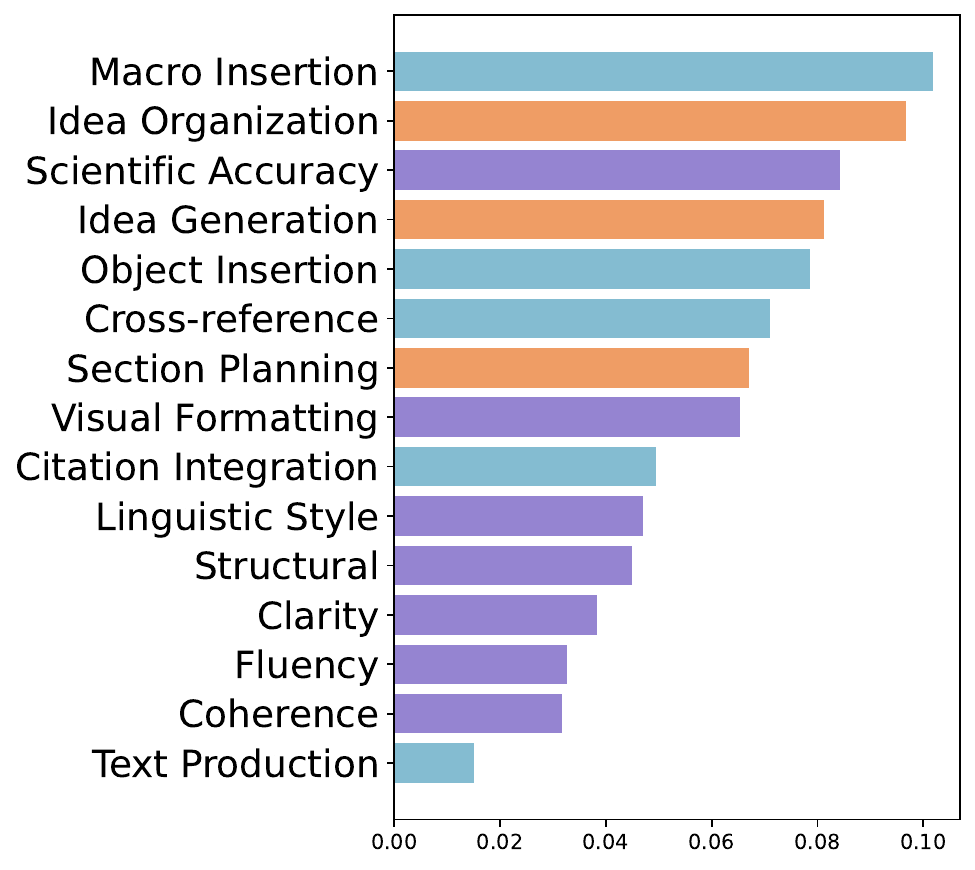}
        \caption{Project 5}
    \end{subfigure}
    \caption{Wasserstein distance to uniform distribution for each distribution of writing intentions. Orange, Blue, and Purple represent Planning, Implementation, and Revision writing actions, respectively.}
    \label{fig:dist-to-uni-all}
\end{figure*}

\begin{figure*}
    \centering
    \begin{subfigure}{0.32\textwidth}
        \includegraphics[width=\textwidth]{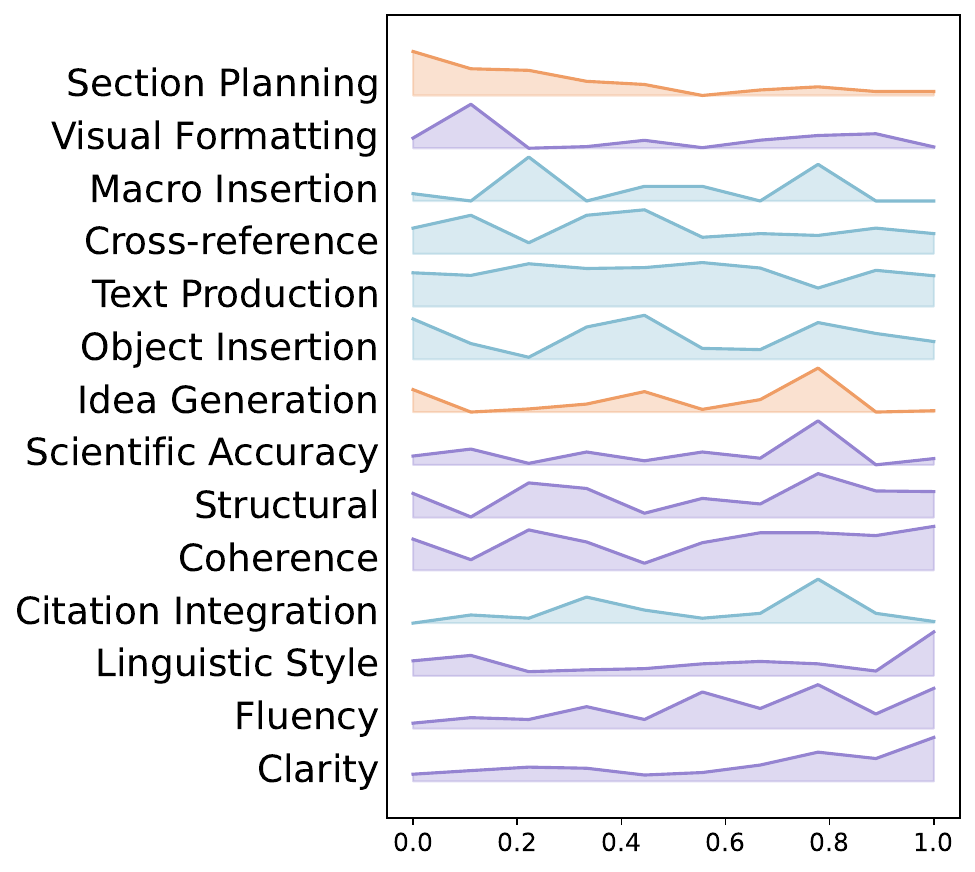}
        \caption{Project 1}
    \end{subfigure}
    \begin{subfigure}{0.32\textwidth}
        \includegraphics[width=\textwidth]{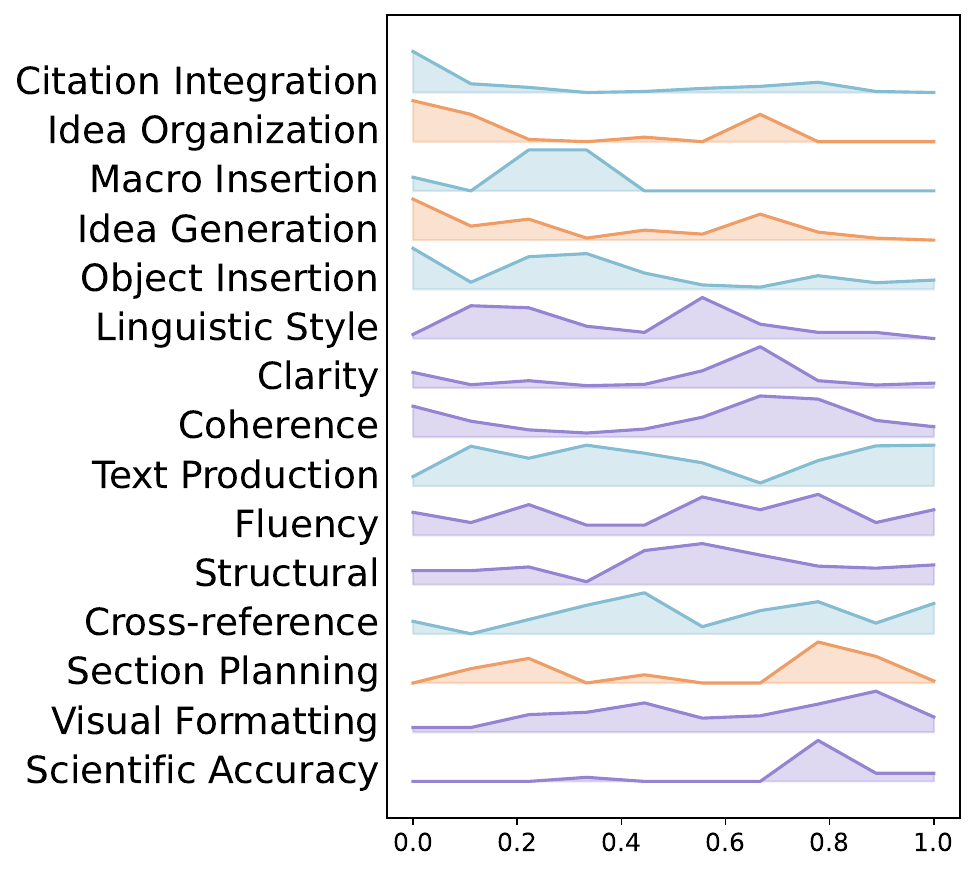}
        \caption{Project 2}
    \end{subfigure}
    \begin{subfigure}{0.32\textwidth}
        \includegraphics[width=\textwidth]{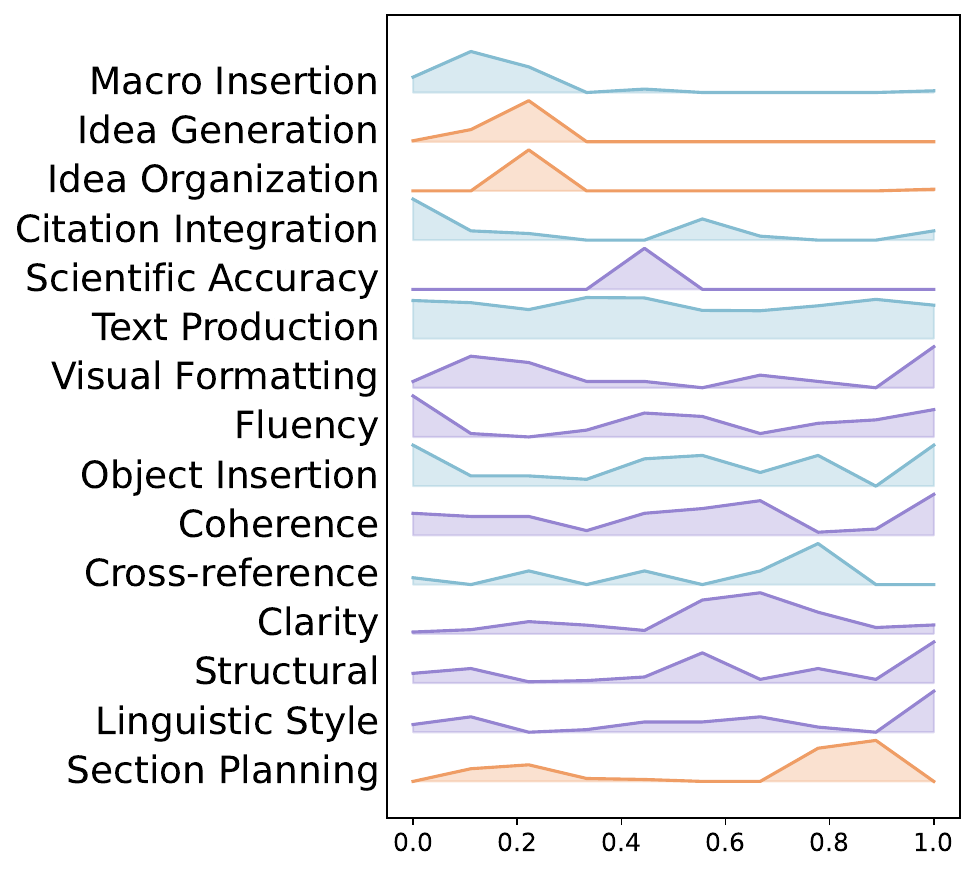}
        \caption{Project 3}
    \end{subfigure}
    \begin{subfigure}{0.32\textwidth}
        \includegraphics[width=\textwidth]{figures/project_label_distributions/project_4_distributions.pdf}
        \caption{Project 4}
    \end{subfigure}
    \begin{subfigure}{0.32\textwidth}
        \includegraphics[width=\textwidth]{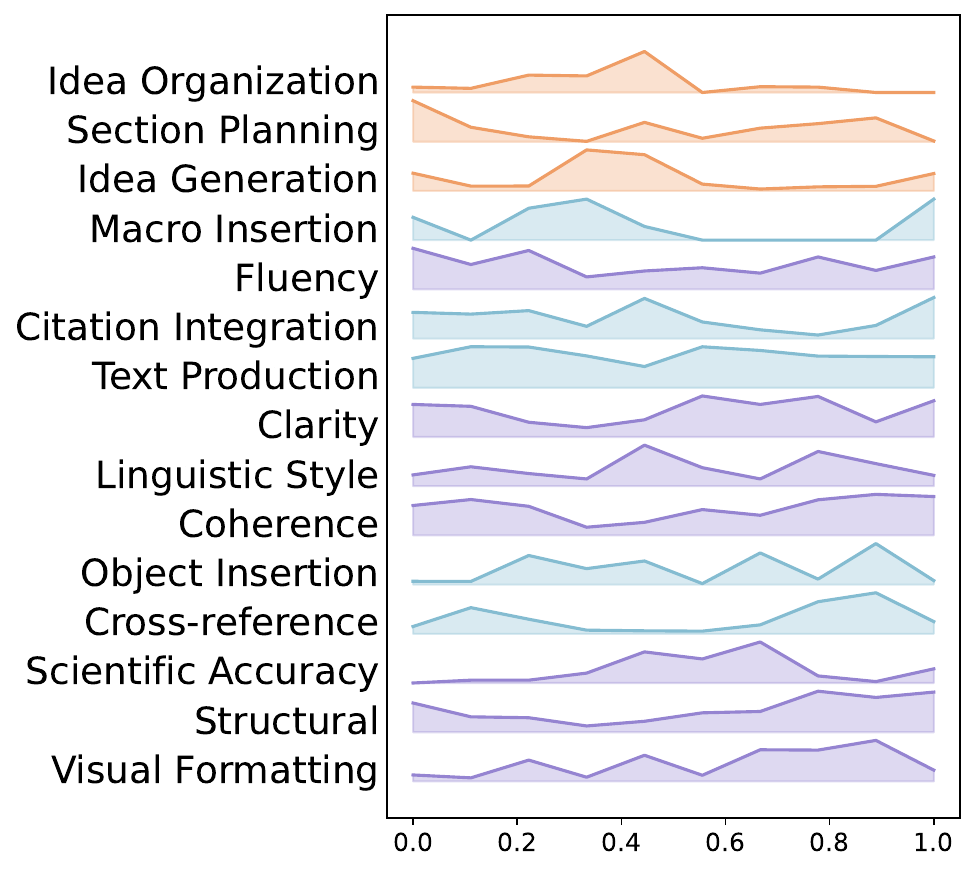}
        \caption{Project 5}
    \end{subfigure}
    \caption{Distribution of labels over time across projects. Orange, Blue and Purple represent Planning, Implementation, and Revision writing actions respectively. The writing actions are sorted in ascending order, top to bottom, according to their distribution mean.}
    \label{fig:label-dist-all}
\end{figure*}

\begin{figure*}
    \centering
    \begin{subfigure}{0.32\textwidth}
        \includegraphics[width=\textwidth]{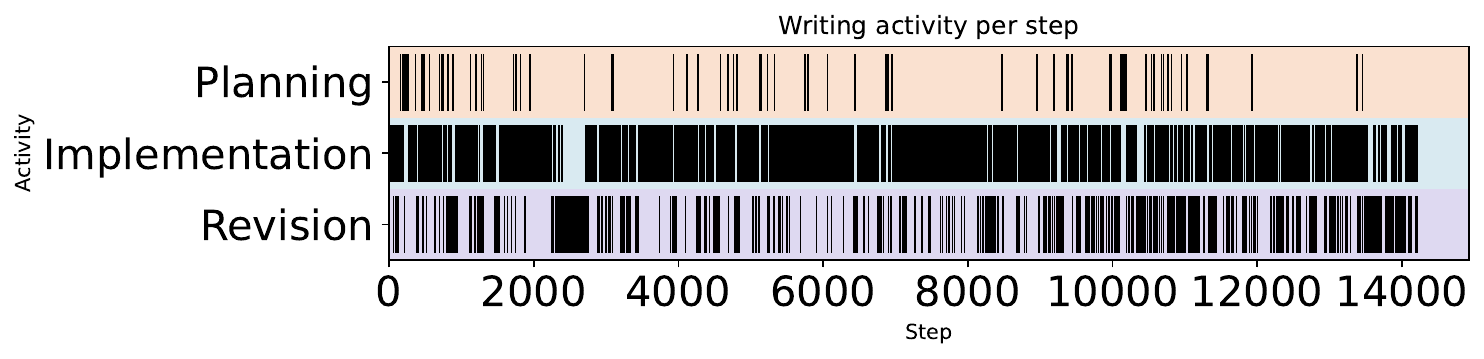}
        \caption{Project 1}
    \end{subfigure}
    \begin{subfigure}{0.32\textwidth}
        \includegraphics[width=\textwidth]{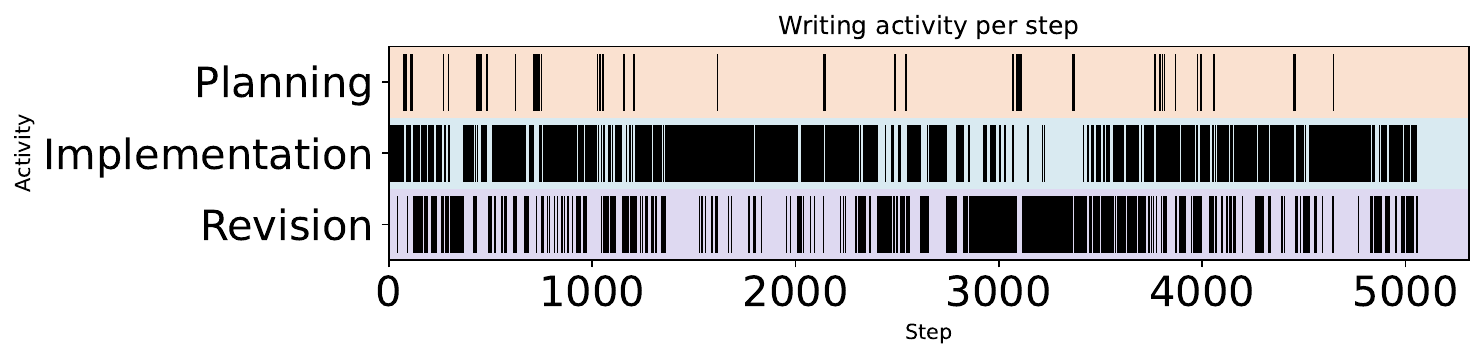}
        \caption{Project 2}
    \end{subfigure}
    \begin{subfigure}{0.32\textwidth}
        \includegraphics[width=\textwidth]{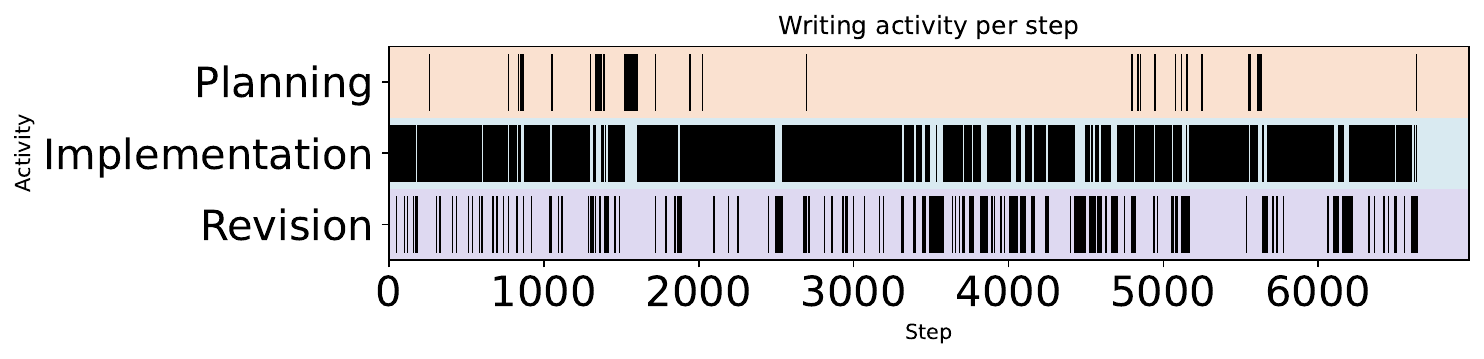}
        \caption{Project 3}
    \end{subfigure}
    \begin{subfigure}{0.32\textwidth}
        \includegraphics[width=\textwidth]{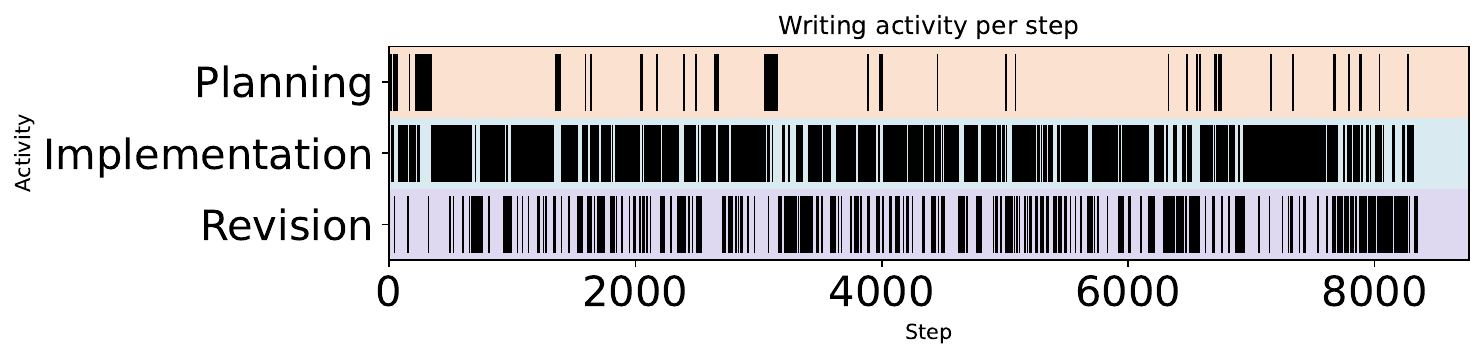}
        \caption{Project 4}
    \end{subfigure}
    \begin{subfigure}{0.32\textwidth}
        \includegraphics[width=\textwidth]{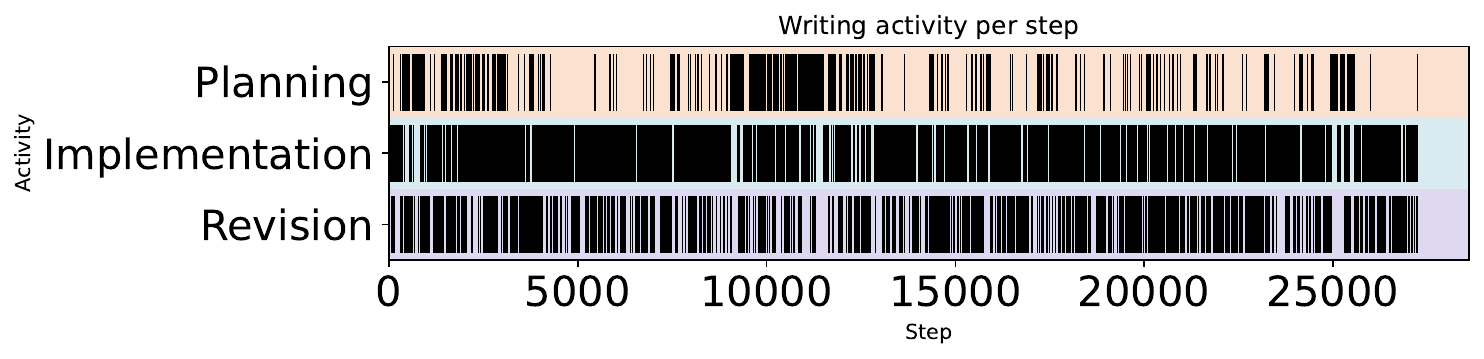}
        \caption{Project 5}
    \end{subfigure}
    \caption{Distribution of high-level intention activities over time. Orange, Blue and Purple represent Planning, Implementation, and Revision writing actions respectively.}
    \label{fig:writing-step-broad-all}
\end{figure*}

\begin{figure*}
    \centering
    \begin{subfigure}{0.32\textwidth}
        \includegraphics[width=\textwidth]{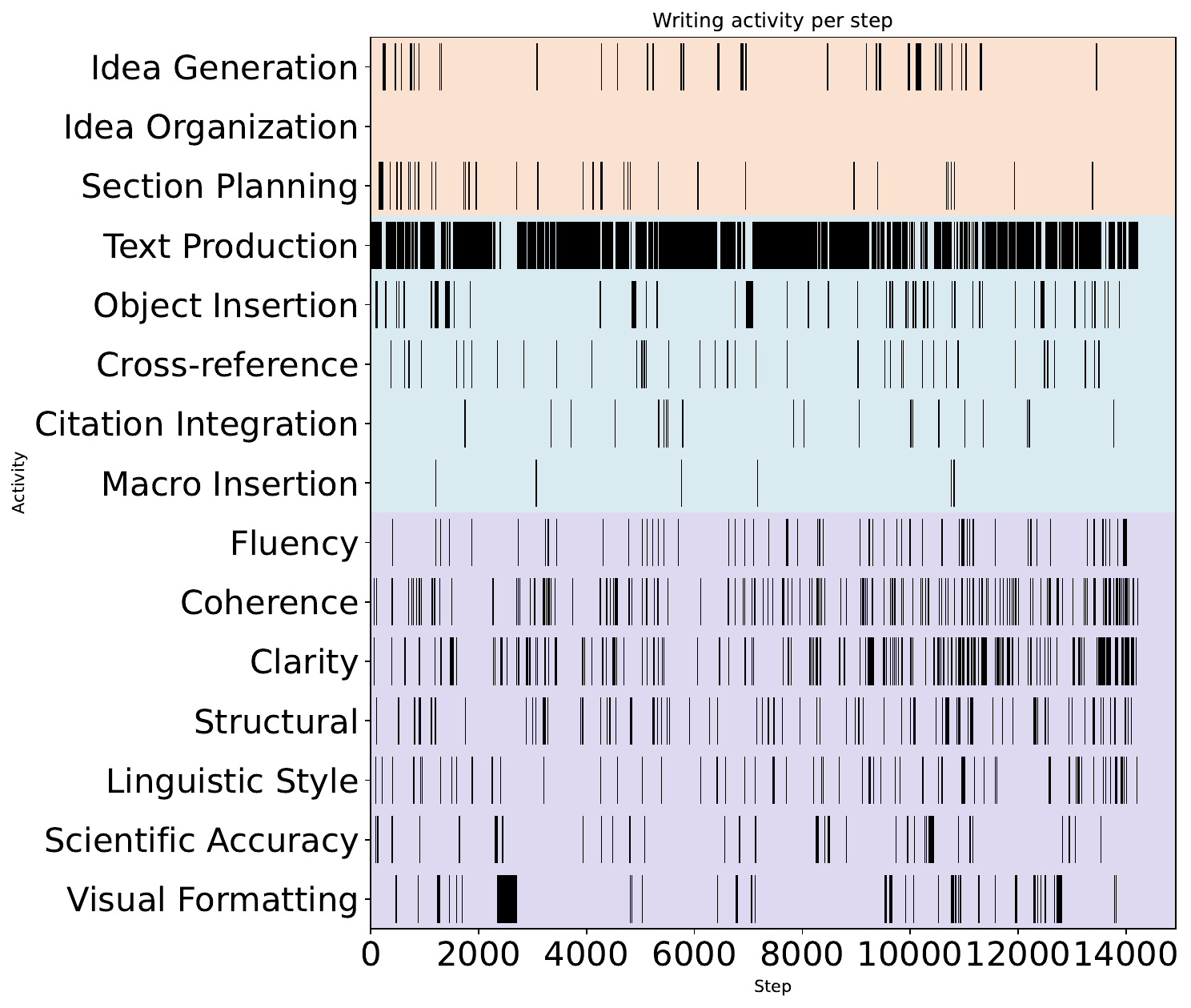}
        \caption{Project 1}
    \end{subfigure}
    \begin{subfigure}{0.32\textwidth}
        \includegraphics[width=\textwidth]{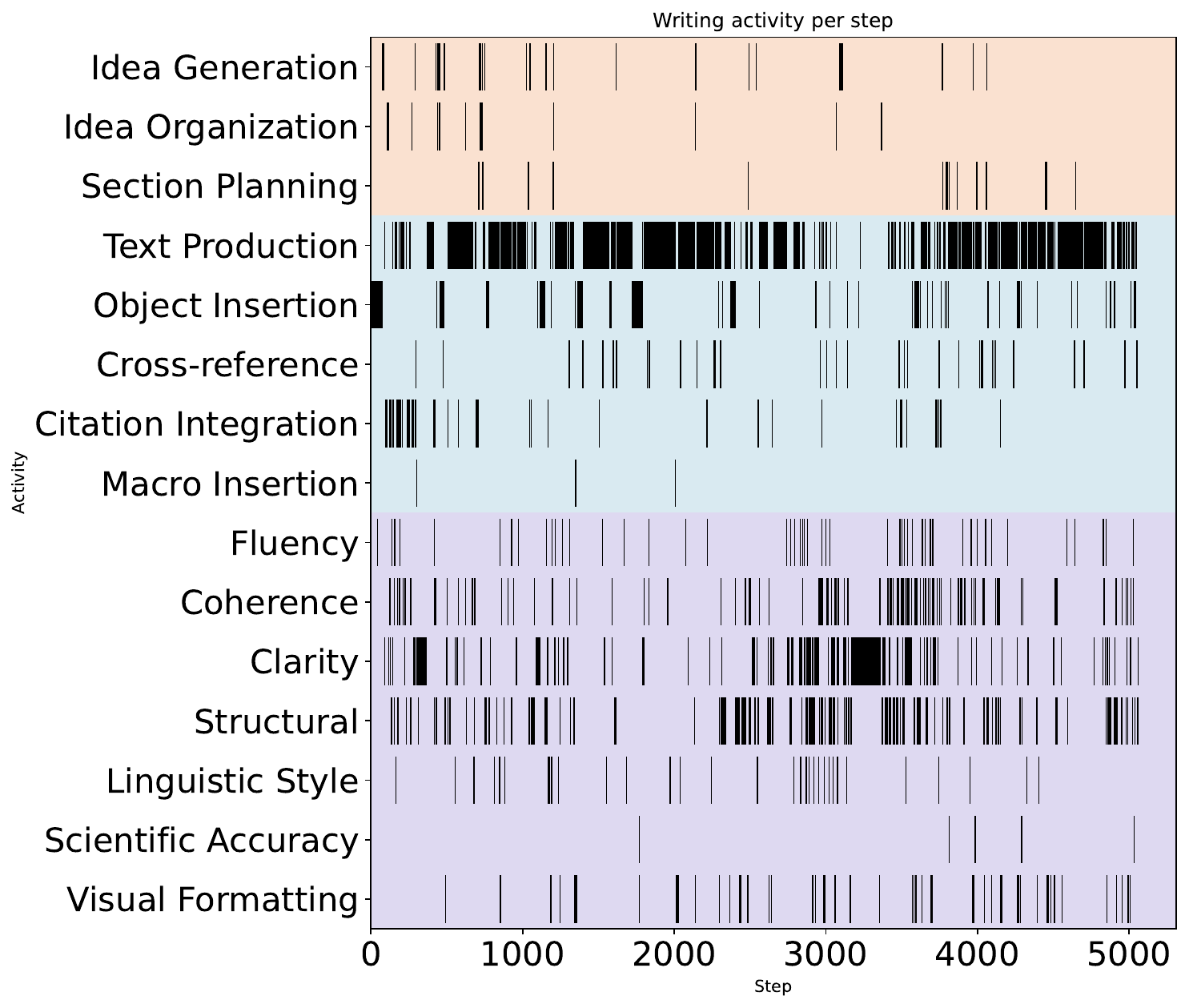}
        \caption{Project 2}
    \end{subfigure}
    \begin{subfigure}{0.32\textwidth}
        \includegraphics[width=\textwidth]{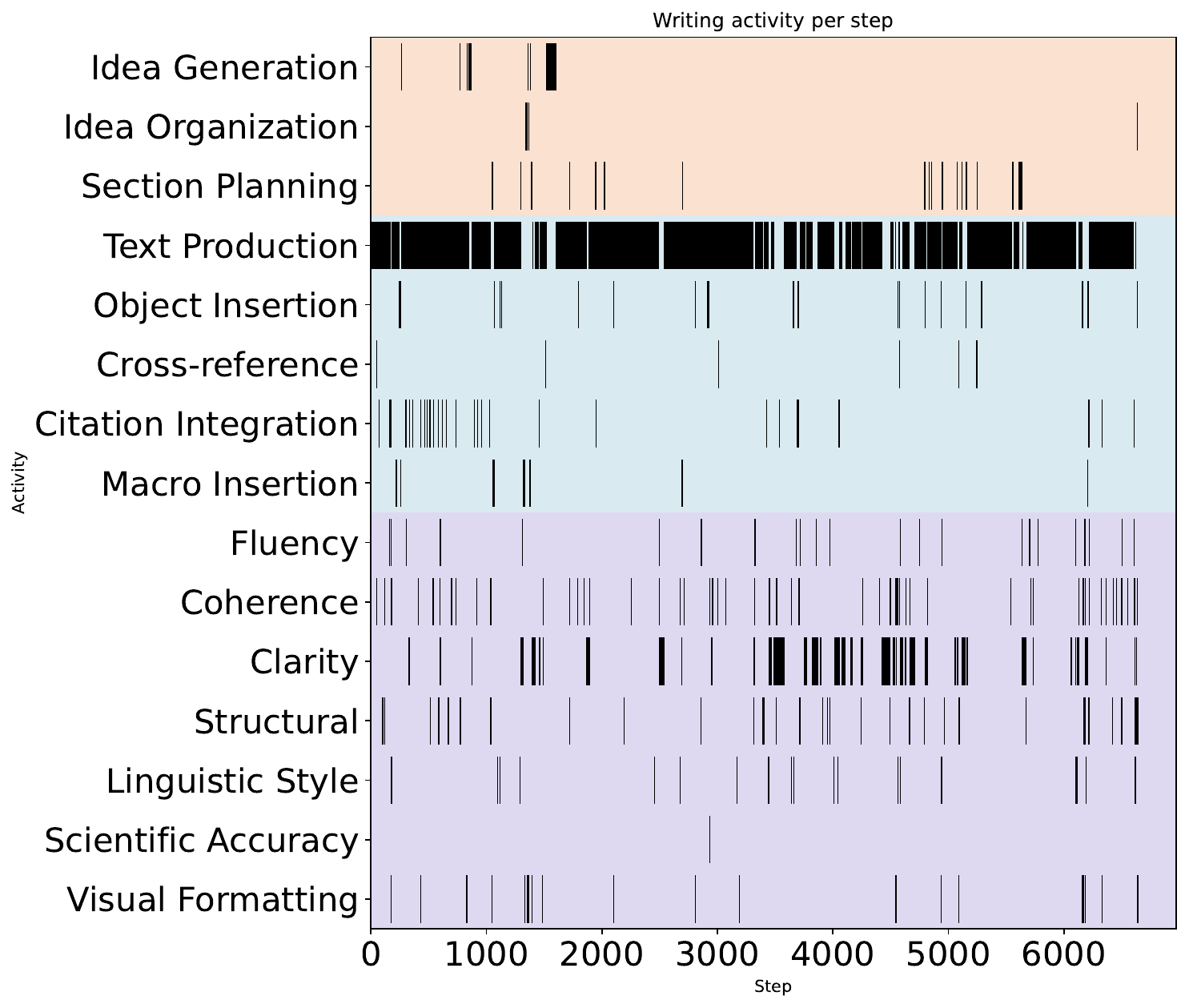}
        \caption{Project 3}
    \end{subfigure}
    \begin{subfigure}{0.32\textwidth}
        \includegraphics[width=\textwidth]{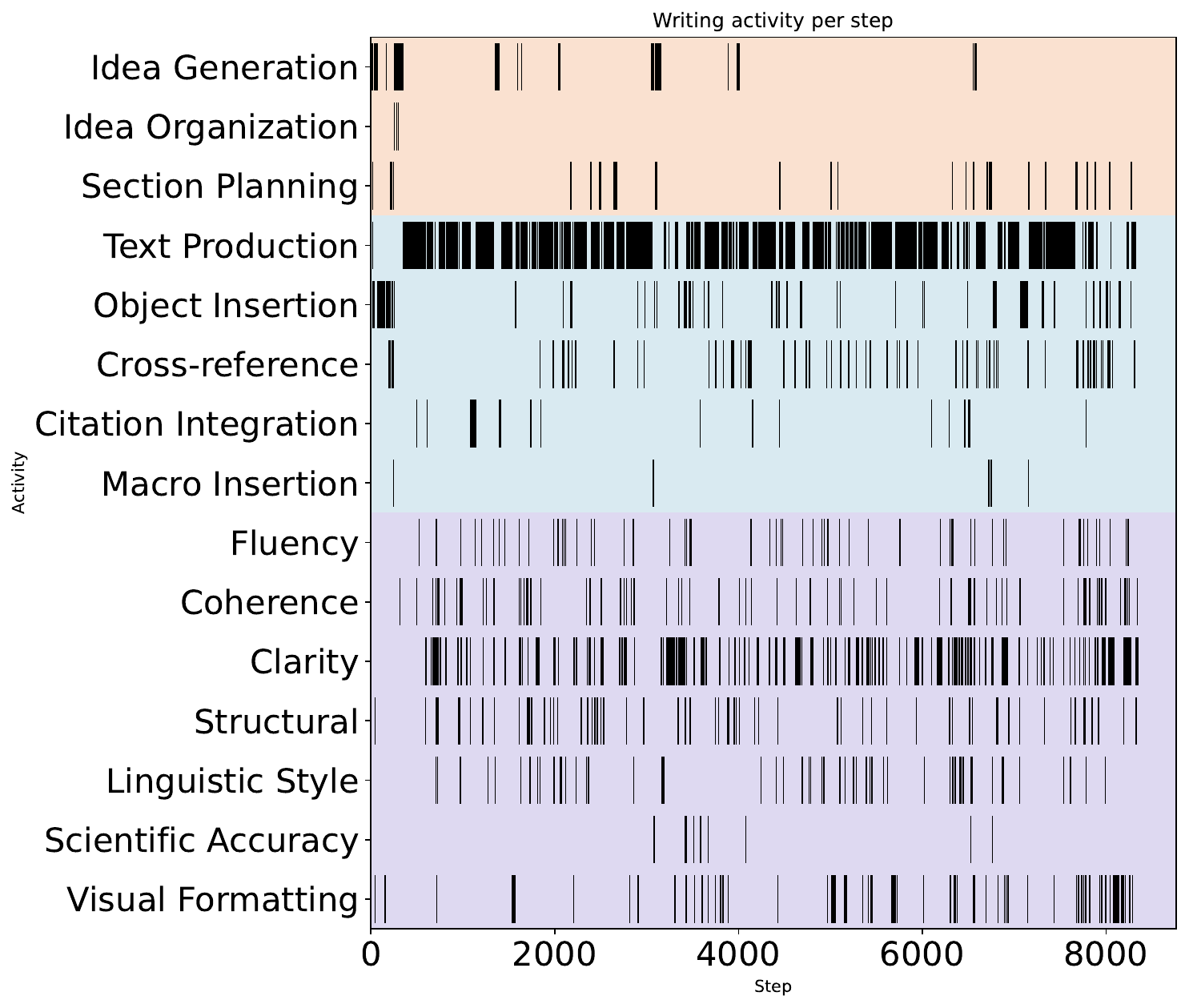}
        \caption{Project 4}
    \end{subfigure}
    \begin{subfigure}{0.32\textwidth}
        \includegraphics[width=\textwidth]{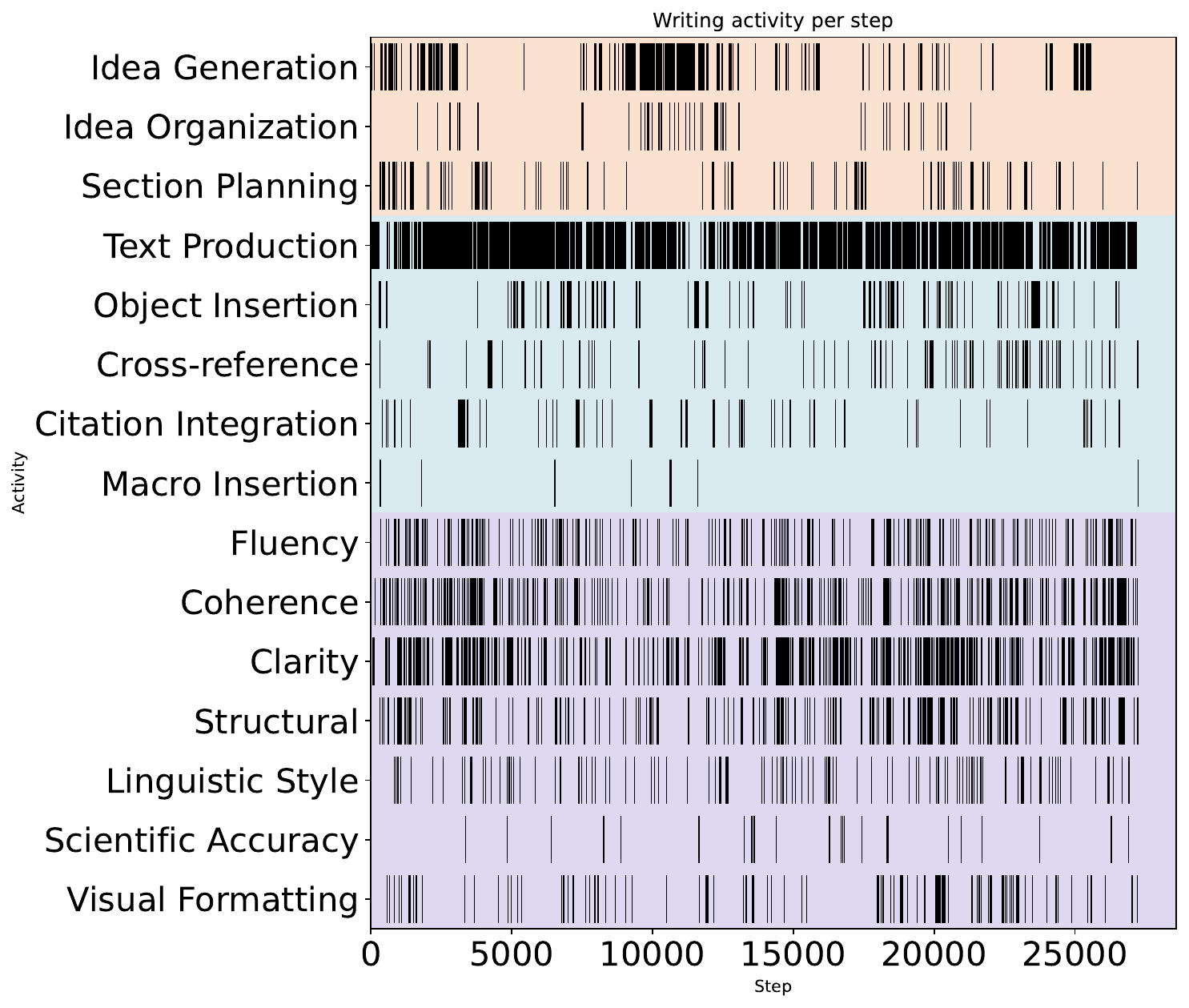}
        \caption{Project 5}
    \end{subfigure}
    \caption{Distribution of Per-intention writing activities over time. Orange, Blue and Purple represent Planning, Implementation, and Revision writing actions respectively.}
    \label{fig:writing-step-detailed-all}
\end{figure*}

Figures \ref{fig:dist-to-uni-all} to \ref{fig:writing-step-detailed-all} show several characteristics of human writing process, analyzed from \textsc{ScholaWrite dataset}: (1) Figure \ref{fig:dist-to-uni-all} -  the average Wasserstein distance between each intention distribution and uniform distribution; (2) Figure \ref{fig:label-dist-all} - distribution of labels over time; (3) Figure \ref{fig:writing-step-broad-all} for high-level intention distribution over time; and (4) Figure \ref{fig:writing-step-detailed-all} for intention-wise writing activity distribution over time. 

\begin{figure*}
    \centering
    \begin{subfigure}{0.32\textwidth}
        \includegraphics[width=\textwidth]{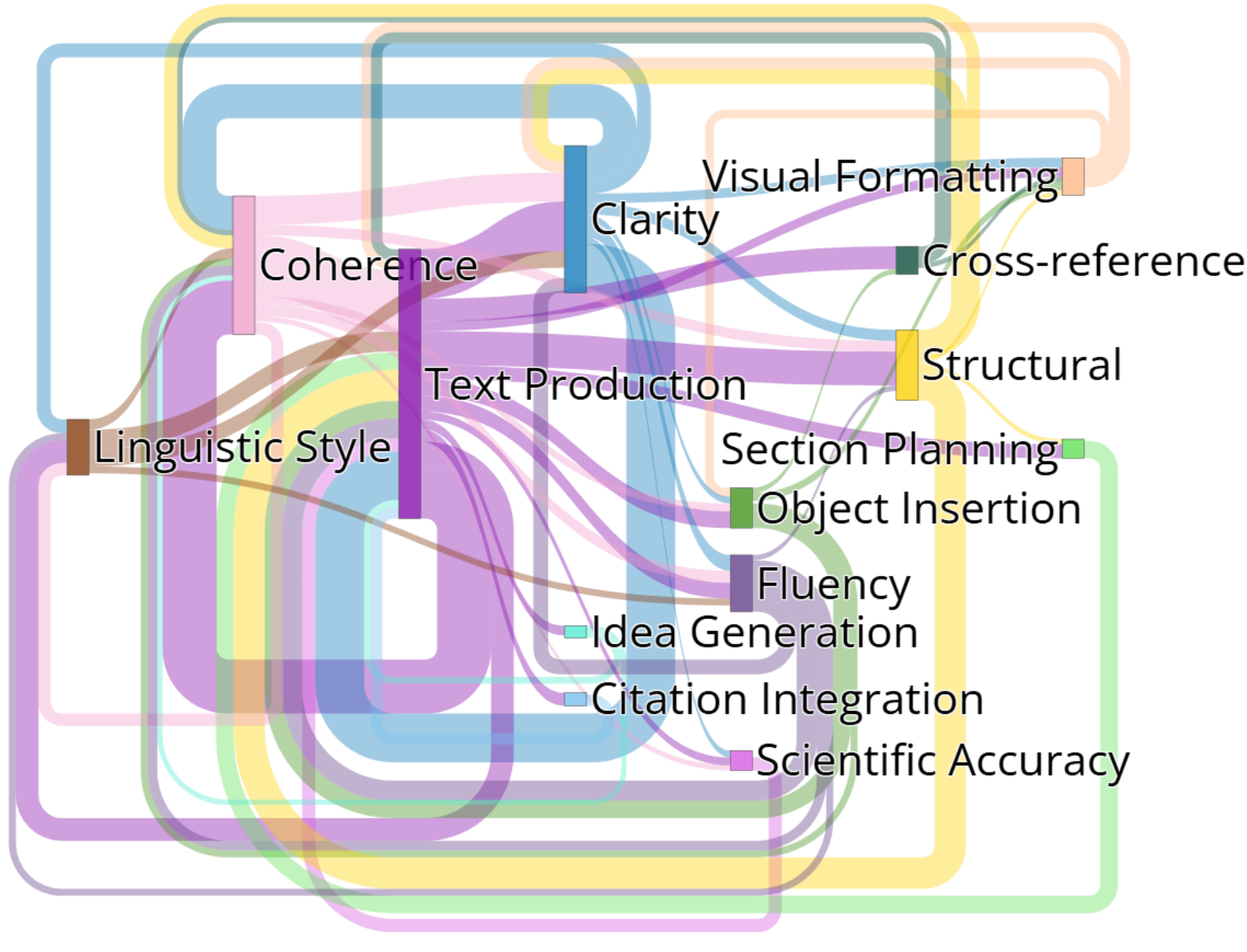}
        \caption{Project 1}
    \end{subfigure}
    \begin{subfigure}{0.32\textwidth}
        \includegraphics[width=\textwidth]{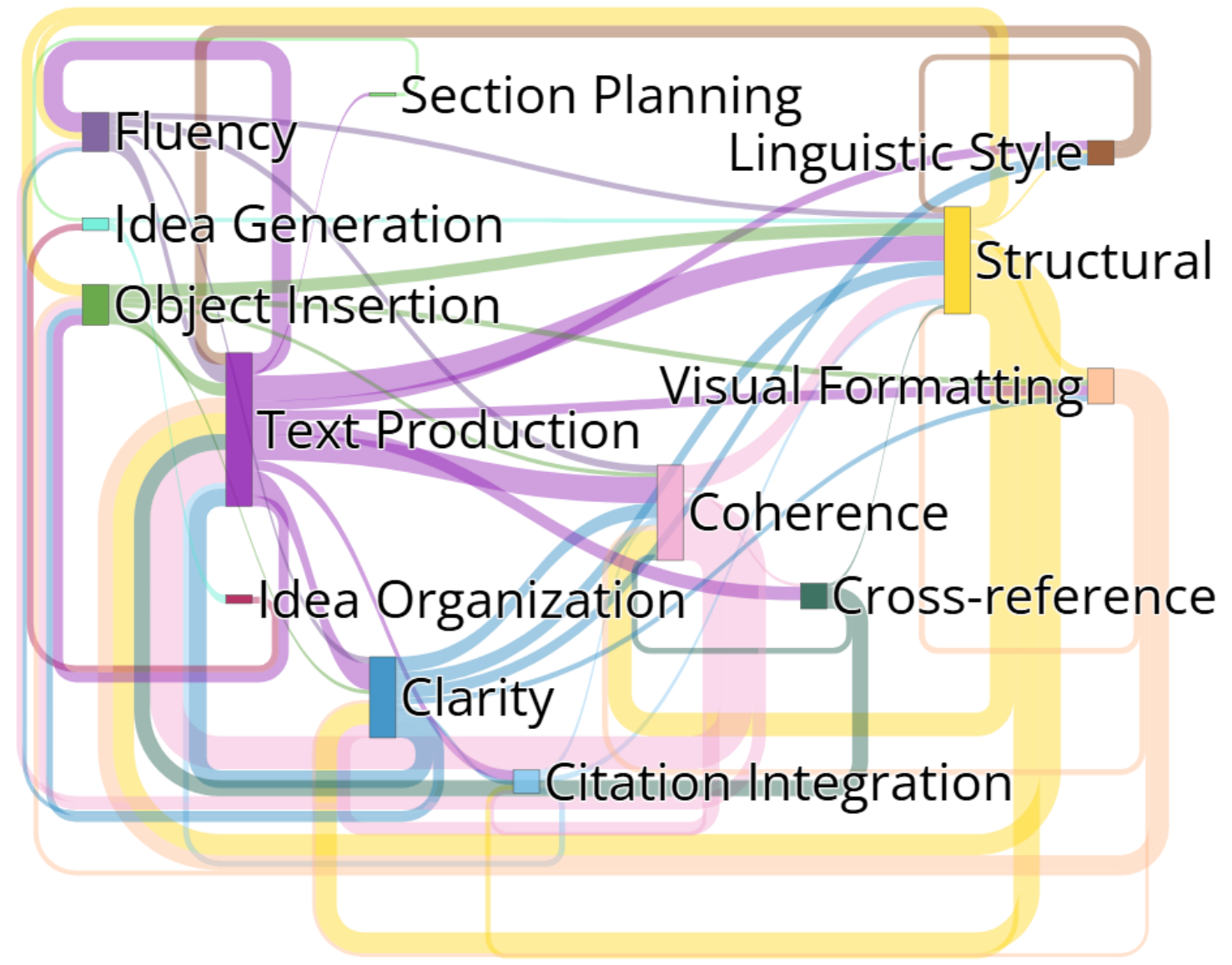}
        \caption{Project 2}
    \end{subfigure}
    \begin{subfigure}{0.32\textwidth}
        \includegraphics[width=\textwidth]{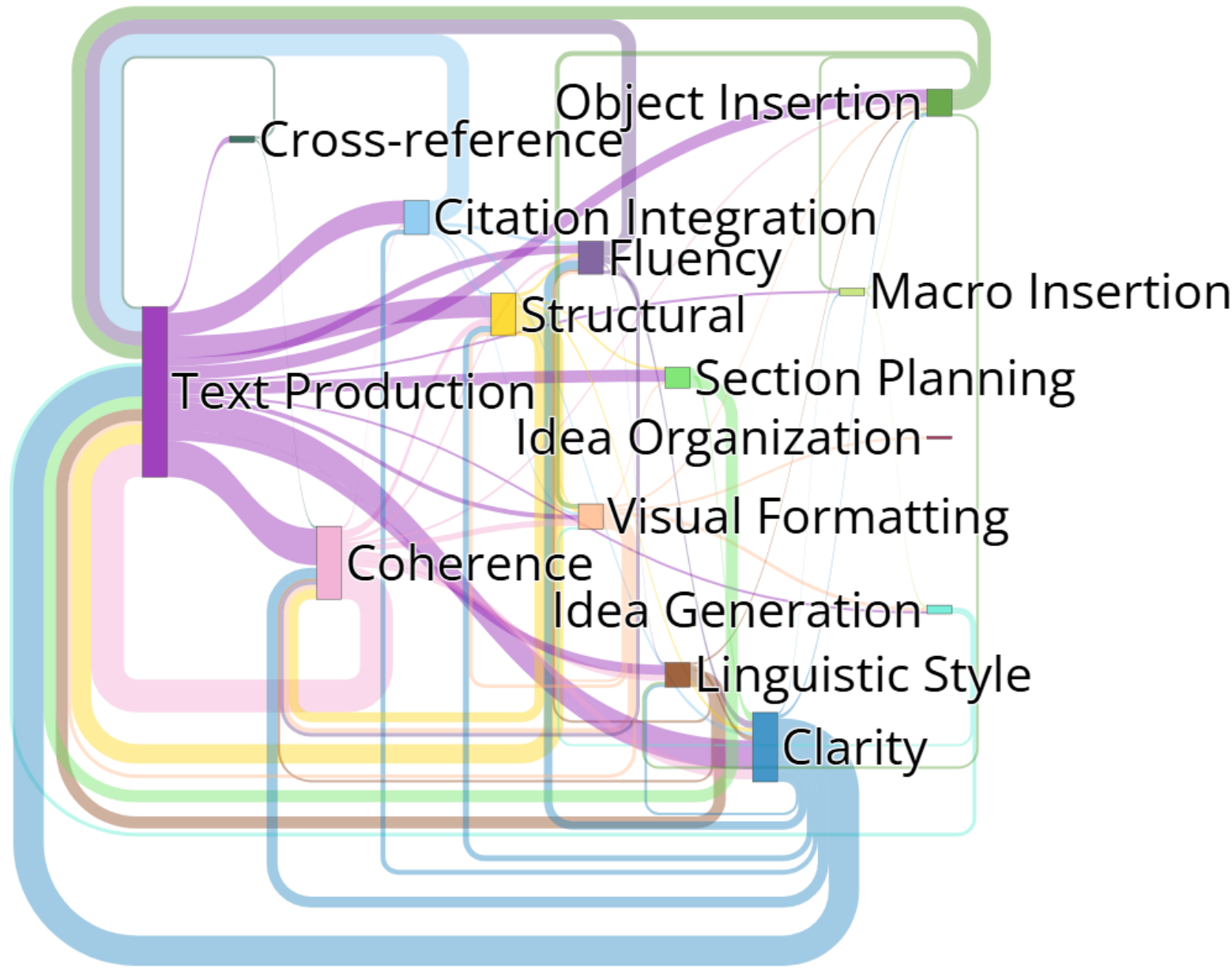}
        \caption{Project 3}
    \end{subfigure}
    \begin{subfigure}{0.32\textwidth}
        \includegraphics[width=\textwidth]{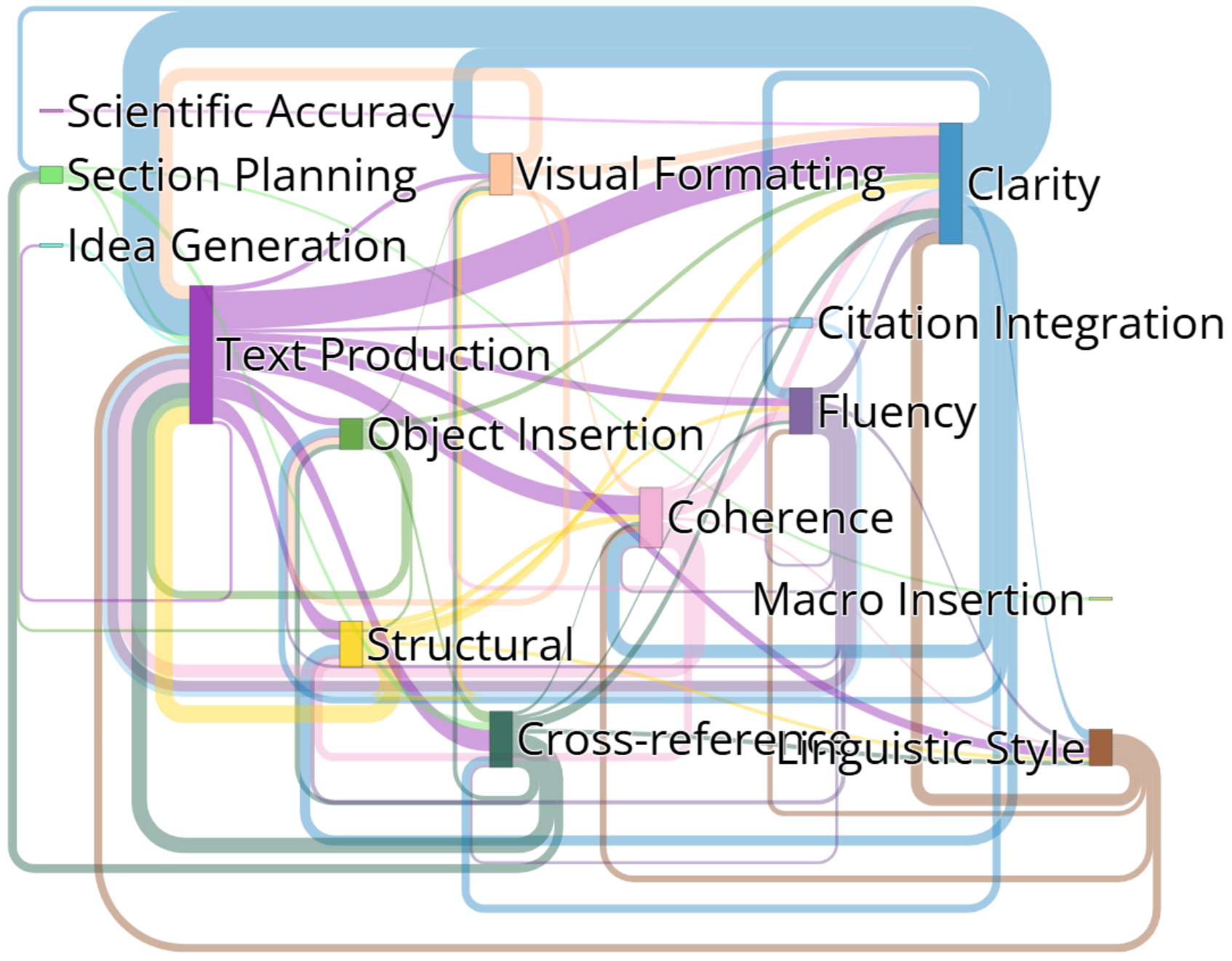}
        \caption{Project 4}
    \end{subfigure}
    \begin{subfigure}{0.32\textwidth}
        \includegraphics[width=\textwidth]{figures/sankey/project5.pdf}
        \caption{Project 5}
    \end{subfigure}
    \caption{Sankey diagrams representing the intention flow of each project. Figure (a) to (d) generated from all intentions. Figure (e) generated from first 10K intentions due to computational constraint.}
    \label{fig:writing-sankey-all}
\end{figure*}

\section{More about Analyses}
\begin{table}[ht!]
\centering
\footnotesize
\resizebox{\columnwidth}{!}{
\begin{tabular}{@{}llr llr@{}}
\toprule
\textbf{Intention} & & \textbf{\% Time} & \textbf{Intention} & & \textbf{\% Time} \\
\midrule
\colorbox{planningcolor}{Idea Generation}     & & 4.8  &
\colorbox{revisioncolor}{Fluency}             & & 3.3 \\
\colorbox{planningcolor}{Idea Organization}   & & 0.6  &
\colorbox{revisioncolor}{Coherence}           & & \textbf{6.9} \\
\colorbox{planningcolor}{Section Planning}    & & 2.6  &
\colorbox{revisioncolor}{Clarity}             & & \textbf{13.4} \\
\colorbox{implementationcolor}{Text Production} & & \textbf{40.3} &
\colorbox{revisioncolor}{Structural}          & & 4.0 \\
\colorbox{implementationcolor}{Object Insertion} & & \textbf{8.4} &
\colorbox{revisioncolor}{Linguistic Style}    & & 2.6 \\
\colorbox{implementationcolor}{Citation Integration} & & 2.4 &
\colorbox{revisioncolor}{Scientific Accuracy} & & 1.2 \\
\colorbox{implementationcolor}{Cross-reference} & & 2.4 &
\colorbox{revisioncolor}{Visual Formatting}   & & \textbf{6.1} \\
\colorbox{implementationcolor}{Macro Insertion} & & 0.8 & & & \\
\bottomrule
\end{tabular}}
\caption{Average percentage of total writing time spent. 
\textcolor{implementationcolor}{Implementation} intentions dominate overall effort, while 
\textcolor{revisioncolor}{Revision} tasks show sustained cognitive load.}\vspace{-3mm}
\label{table:intention-time-avg}
\end{table}

\begin{table}[ht!]
\centering
\footnotesize
\begin{tabular}{@{}lrrrr@{}}
\toprule
\textbf{Intention} & \textbf{Mean} & \textbf{SD} & \textbf{Max} & \textbf{Count} \\
\midrule
\colorbox{planningcolor}{Idea Generation}     & 1.87 & 1.94 & 11.40 & 76 \\
\colorbox{planningcolor}{Idea Organization}   & \textbf{3.71} & 4.58 & 13.79 & 7 \\
\colorbox{planningcolor}{Section Planning}    & 1.37 & 1.57 & 9.36  & 38 \\
\colorbox{implementationcolor}{Text Production}     & 2.20 & 2.53 & \textbf{28.14} & \textbf{724} \\
\colorbox{implementationcolor}{Object Insertion}    & \textbf{2.88} & 3.05 & 21.61 & 108 \\
\colorbox{implementationcolor}{Citation Integration} & 1.70 & 1.53 & 7.53  & 34 \\
\colorbox{implementationcolor}{Cross-reference}     & 1.08 & 0.77 & 3.98  & 38 \\
\colorbox{implementationcolor}{Macro Insertion}     & 1.43 & 1.05 & 3.71  & 7 \\
\colorbox{revisioncolor}{Fluency}             & 0.66 & 0.18 & 1.06  & 10 \\
\colorbox{revisioncolor}{Coherence}           & 1.46 & 1.71 & 9.62  & 55 \\
\colorbox{revisioncolor}{Clarity}             & 1.64 & 1.61 & 9.53  & 243 \\
\colorbox{revisioncolor}{Structural}          & 1.50 & 1.43 & 8.42  & 53 \\
\colorbox{revisioncolor}{Linguistic Style}    & 2.05 & 2.55 & 10.81 & 24 \\
\colorbox{revisioncolor}{Scientific Accuracy} & 1.97 & 1.60 & 6.44  & 20 \\
\colorbox{revisioncolor}{Visual Formatting}   & \textbf{2.22} & 2.66 & 14.25 & 87 \\
\bottomrule
\end{tabular}
\caption{Session-level time statistics in minutes. 
Longer sessions for \textcolor{planningcolor}{Idea Organization}, 
\textcolor{implementationcolor}{Object Insertion}, and 
\textcolor{revisioncolor}{Visual Formatting} reflect deeper cognitive engagement.}
\vspace{-4mm}
\label{table:intention-session-stats}
\end{table}

\begin{figure}[ht!]
    \centering
    \includegraphics[width=0.5\textwidth,trim=0 0 0 0,clip]{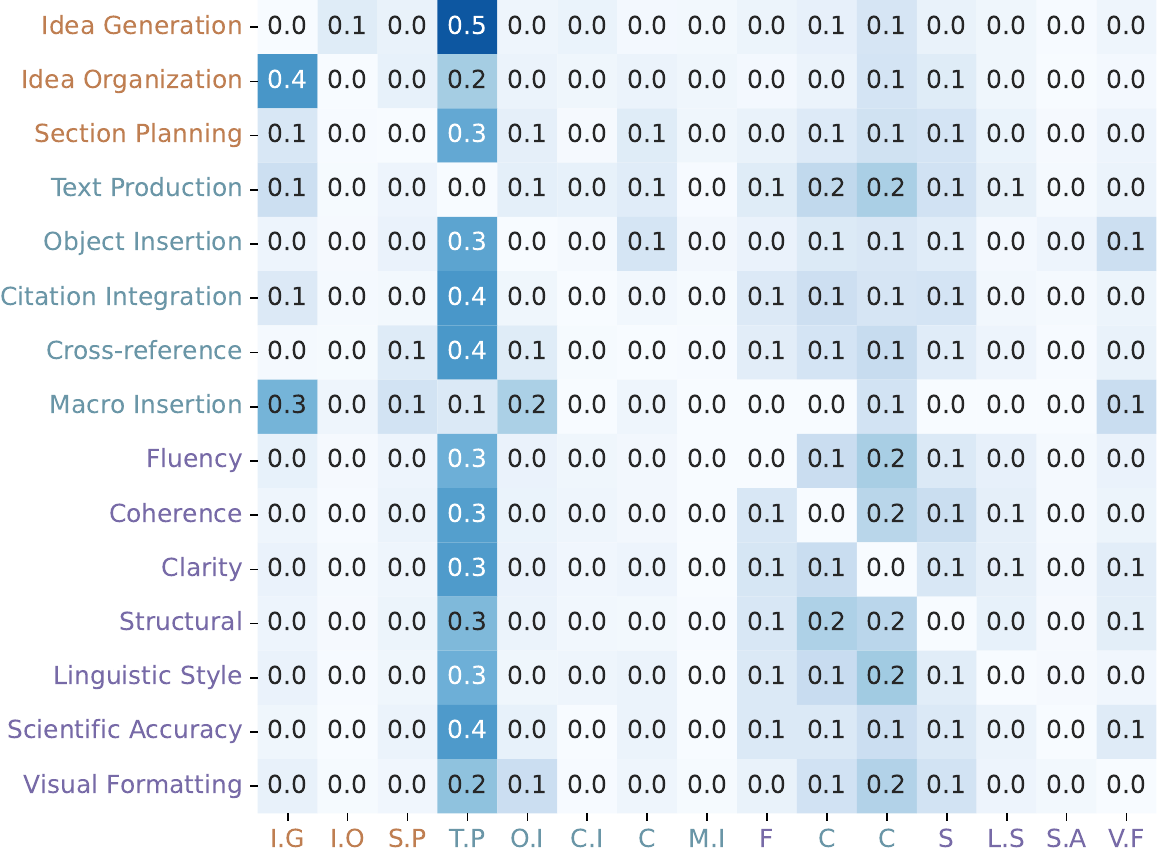}
    \caption{Transition probability matrix between writing intentions. Each cell shows the likelihood that a session with the current intention (y-axis) is followed by a session with the next intention (x-axis).
    }
    \label{fig:trans-prob-all}\vspace{-3mm}
\end{figure}
\begin{figure*}[ht!]
    \centering
    \begin{subfigure}[t]{0.48\textwidth}
        \centering
        \includegraphics[width=\linewidth]{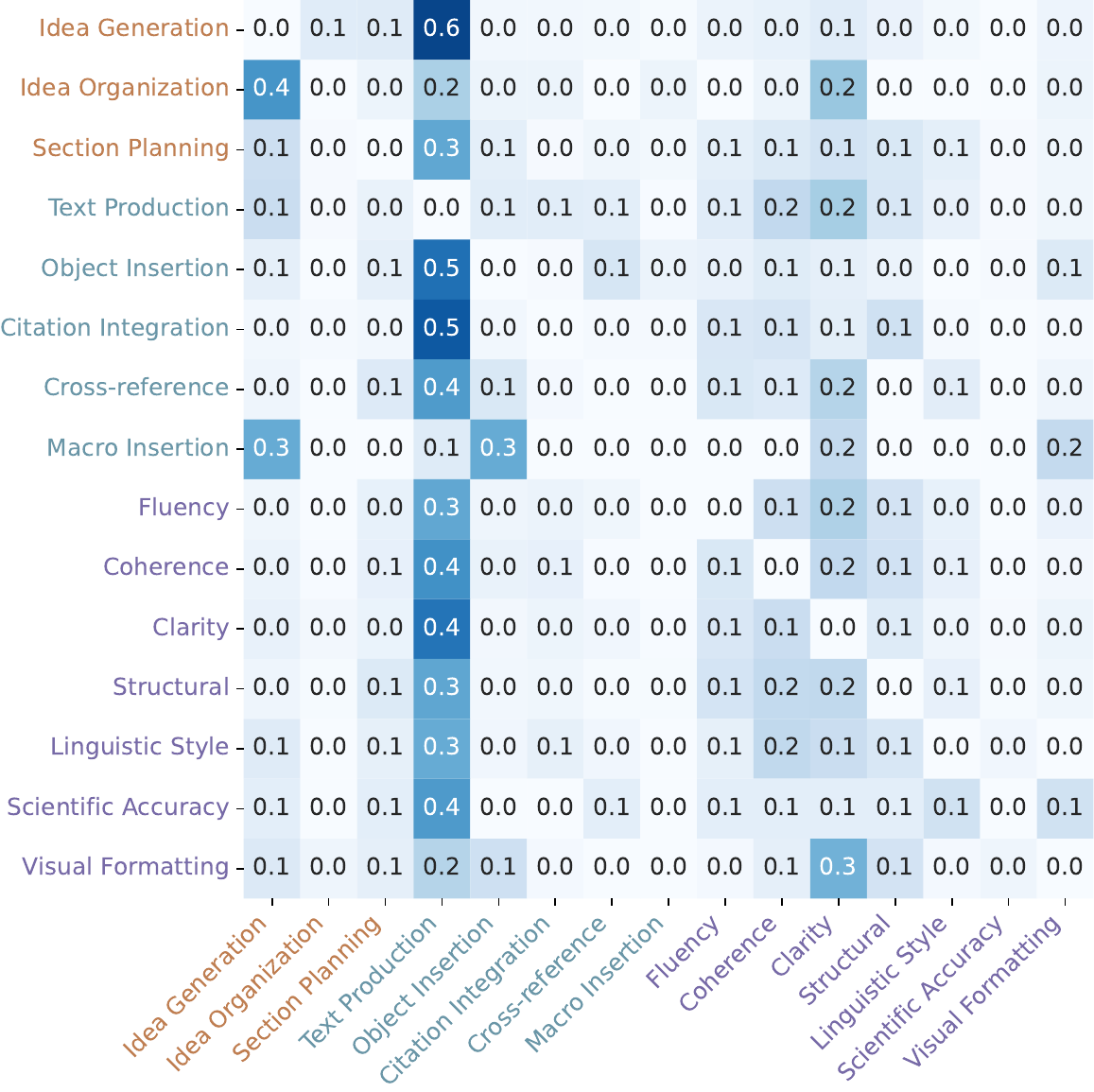}
        \caption{Early stage}
        \label{fig:transition-early}
    \end{subfigure}
    \hfill
    \begin{subfigure}[t]{0.48\textwidth}
        \centering
        \includegraphics[width=\linewidth]{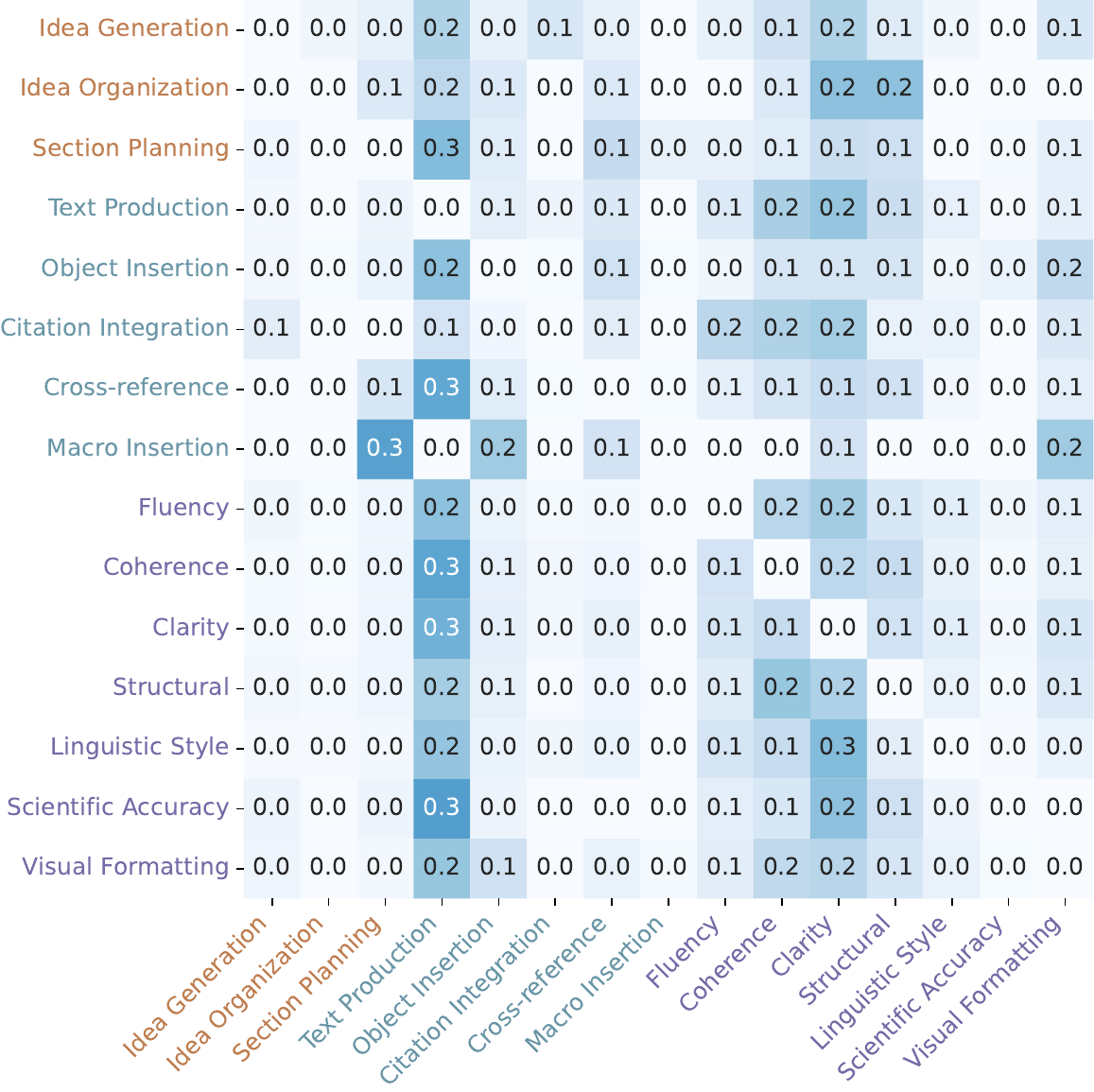}
        \caption{Late stage}
        \label{fig:transition-late}
    \end{subfigure}
    
    \caption{Transition probabilities across time. (a) Early sessions vs. (b) Late sessions.}
    \label{fig:transition-prob-accross-time}
\end{figure*}

Figure \ref{fig:transition-prob-accross-time} shows transition probability between pair of intentions in early stage and later stage. 

Table \ref{table:ngram-seqs} shows top 6-gram writing intention sequences by session coverage (\%)

\begin{table*}[ht!]
\centering
\footnotesize
\begin{tabular}{@{}p{15cm}r@{}}
\toprule
\textbf{6-gram Sequence} & \textbf{Percentage} \\
\midrule
Text Production $\rightarrow$ Clarity $\rightarrow$ Text Production $\rightarrow$ Clarity $\rightarrow$ Text Production $\rightarrow$ Clarity & 3.22 \\
Clarity $\rightarrow$ Text Production $\rightarrow$ Clarity $\rightarrow$ Text Production $\rightarrow$ Clarity $\rightarrow$ Text Production & 3.22 \\
Text Production $\rightarrow$ Object Insertion $\rightarrow$ Text Production $\rightarrow$ Object Insertion $\rightarrow$ Text Production $\rightarrow$ Object Insertion & 2.34 \\
Object Insertion $\rightarrow$ Text Production $\rightarrow$ Object Insertion $\rightarrow$ Text Production $\rightarrow$ Object Insertion $\rightarrow$ Text Production & 2.34 \\
Idea Generation $\rightarrow$ Idea Organization $\rightarrow$ Idea Generation $\rightarrow$ Idea Organization $\rightarrow$ Idea Generation $\rightarrow$ Idea Organization & 2.05 \\
\bottomrule
\end{tabular}
\caption{Top 6-gram writing intention sequences by session coverage (\%)}
\label{table:ngram-seqs}
\end{table*}

\section{More about Predicting next writing intention}\label{sec:appendix:pred-intent}
\subsection{Training environments}
\paragraph{BERT \& RoBERTa} We fine-tuned BERT and RoBERTa with the following hyperparameter setups: (1) a learning rate of $2e^{-5}$; (2) training batch size per device of 8; (3) evaluation batch size per device of 8; (4) the number of training epochs of 10; and (5) a weight decay of $0.01$. For each model, it took approximately 3.5 hours on one NVIDIA RTX A6000. 

\paragraph{Llama} For all experiments, we used baseline models of 4-bit quantized  Llama-8B-Instruct\footnote{\url{https://huggingface.co/unsloth/Meta-Llama-3.1-8B-Instruct-bnb-4bit}}, using unsloth library\footnote{\url{https://github.com/unslothai/unsloth}}. 

For \textbf{the intention prediction task} (Sec. \ref{sec:next_predict}), here are the hyperparameter setups for the Llama models: (1) only one epoch of training; (2) a weight decay of 0.01; (3) warm-up steps of 5; (4) learning rate of $2e^{-4}$; and (5) AdamW 8-bit optimizer. Due to computational constraints, we were able to run only one epoch for fine-tuning Llama models on our \textsc{ScholaWrite} dataset. 
For Llama-8B, it took approximately 8 hours on one RTX A5000. 

\subsection{Details About Finetuning Process}
The fine-tuning prompt included all possible labels with definitions, task instructions, the ``before-text'' chunk, and the corresponding human-annotated intention label, asking the model to predict the intention label based on the ``before-text''. Differences in prompts were limited to only task instructions.

To achieve optimal performance while minimizing memory usage, we employed QLoRA \citep{dettmers2024qlora} to fine-tune all linear modules of a 4-bit quantized Llama. During fine-tuning, we utilized the `\texttt{train\_on\_response\_only}' function provided by the \texttt{unsloth} library, which masks the task instructions, intention label definitions, and ``before'' text with -100s. This ensures the model is trained exclusively on the response portion of the fine-tuning prompt (i.e., the predicted intention label), without being influenced by the instructional components of the input. The model was fine-tuned for one epoch with a batch size of 2, 4 gradient accumulation steps, and the AdamW 8-bit optimizer.

\subsection{Prompt Templates}
We present the prompt templates used for the next writing intention prediction experiments in Listings~\ref{listing:prediction-prompt-zero}--\ref{listing:prediction-prompt-gpt}.

\begin{table}[ht!]
\centering
\begin{minipage}{\linewidth}
\lstset{
    basicstyle=\ttfamily\footnotesize, 
    breaklines=true, 
    frame=single, 
    columns=fullflexible, 
    captionpos=b 
}
\begin{lstlisting}
Here are all the possible writing intention labels:

    - Idea Generation: Formulate and record initial thoughts and concepts.
    - Idea Organization: Select the most useful materials and demarcate those generated ideas in a visually formatted way.
    - Section Planning: Initially create sections and sub-level structures.
    - Text Production: Translate their ideas into full languages, either from the writers' language or borrowed sentences from an external source. 
    - Object Insertion: Insert visual claims of their arguments (e.g., figures, tables, equations, footnotes, itemized lists, etc.). 
    - Cross-reference: Link different sections, figures, tables, or other elements within a document through referencing commands. 
    - Citation Integration: Incorporate bibliographic references into a document and systematically link these references using citation commands. 
    - Macro Insertion: Incorporate predefined commands or packages into a LaTeX document to alter its formatting. 
    - Fluency: Fix grammatical or syntactic errors in the text or LaTeX commands.
    - Coherence: Logically link (1) any of the two or multiple sentences within the same paragraph; (2) any two subsequent paragraphs; or (3) objects to be consistent as a whole. 
    - Structural: Improve the flow of information by modifying the location of texts and objects. 
    - Clarity: Improve the semantic relationships between texts to be more straightforward and concise. 
    - Linguistic Style: Modify texts with the writer's writing preferences regarding styles and word choices, etc. 
    - Scientific Accuracy: Update or correct scientific evidence (e.g., numbers, equations) for more accurate claims. 
    - Visual Formatting: Modify the stylistic formatting of texts, objects, and citations.

Identify the most likely next writing intention of a graduate researcher when editing the following LaTex paper draft. Your output should only be a label from the list above.

## Input: {before_text}
## Output: 

\end{lstlisting}
\vspace{-3mm}
\captionof{lstlisting}{Prediction prompt for \textcolor{teal}{Llama-8B-Zero}}
\label{listing:prediction-prompt-zero}
\end{minipage}
\end{table}

\begin{table}[ht!]
\centering
\begin{minipage}{\linewidth}
\lstset{
    basicstyle=\ttfamily\footnotesize, 
    breaklines=true, 
    frame=single, 
    columns=fullflexible, 
    captionpos=b 
}
\begin{lstlisting}
Here are all the possible writing intention labels:
    - Idea Generation: Formulate and record initial thoughts and concepts.
    - Idea Organization: Select the most useful materials and demarcate those generated ideas in a visually formatted way.
    - Section Planning: Initially create sections and sub-level structures.
    - Text Production: Translate their ideas into full languages, either from the writers' language or borrowed sentences from an external source. 
    - Object Insertion: Insert visual claims of their arguments (e.g., figures, tables, equations, footnotes, itemized lists, etc.). 
    - Cross-reference: Link different sections, figures, tables, or other elements within a document through referencing commands. 
    - Citation Integration: Incorporate bibliographic references into a document and systematically link these references using citation commands. 
    - Macro Insertion: Incorporate predefined commands or packages into a LaTeX document to alter its formatting. 
    - Fluency: Fix grammatical or syntactic errors in the text or LaTeX commands.
    - Coherence: Logically link (1) any of the two or multiple sentences within the same paragraph; (2) any two subsequent paragraphs; or (3) objects to be consistent as a whole. 
    - Structural: Improve the flow of information by modifying the location of texts and objects. 
    - Clarity: Improve the semantic relationships between texts to be more straightforward and concise. 
    - Linguistic Style: Modify texts with the writer's writing preferences regarding styles and word choices, etc. 
    - Scientific Accuracy: Update or correct scientific evidence (e.g., numbers, equations) for more accurate claims. 
    - Visual Formatting: Modify the stylistic formatting of texts, objects, and citations.

Identify the most likely next writing intention of a graduate researcher when writing the following LaTex paper draft. Your output should only be a label from the list above.

## Input: {before_text}
## Output:
\end{lstlisting}
\vspace{-3mm}
\captionof{lstlisting}{Prediction prompt for \textcolor{magenta}{Llama-8B-SW}}
\label{listing:prediction-prompt-sw}
\end{minipage}
\end{table}

\begin{table}[ht!]
\centering
\begin{minipage}{\linewidth}
\lstset{
    basicstyle=\ttfamily\footnotesize, 
    breaklines=true, 
    frame=single, 
    columns=fullflexible, 
    captionpos=b 
}
\begin{lstlisting}
You are a classifier that identify the most likely next writing intention. You will be given a list of all possible writing intention labels with definitions, and an in-progress LaTex paper draft written by a graduate student. Please strictly follow user's instruction to identify the most likely next writing intention. 

Here are the verbalizers of all the possible writing intention labels:

    - Idea Generation: Formulate and record initial thoughts and concepts.
    - Idea Organization: Select the most useful materials and demarcate those generated ideas in a visually formatted way.
    - Section Planning: Initially create sections and sub-level structures.
    - Text Production: Translate their ideas into full languages, either from the writers' language or borrowed sentences from an external source. 
    - Object Insertion: Insert visual claims of their arguments (e.g., figures, tables, equations, footnotes, itemized lists, etc.). 
    - Cross-reference: Link different sections, figures, tables, or other elements within a document through referencing commands. 
    - Citation Integration: Incorporate bibliographic references into a document and systematically link these references using citation commands. 
    - Macro Insertion: Incorporate predefined commands or packages into a LaTeX document to alter its formatting. 
    - Fluency: Fix grammatical or syntactic errors in the text or LaTeX commands.
    - Coherence: Logically link (1) any of the two or multiple sentences within the same paragraph; (2) any two subsequent paragraphs; or (3) objects to be consistent as a whole. 
    - Structural: Improve the flow of information by modifying the location of texts and objects. 
    - Clarity: Improve the semantic relationships between texts to be more straightforward and concise. 
    - Linguistic Style: Modify texts with the writer's writing preferences regarding styles and word choices, etc. 
    - Scientific Accuracy: Update or correct scientific evidence (e.g., numbers, equations) for more accurate claims. 
    - Visual Formatting: Modify the stylistic formatting of texts, objects, and citations.

Identify the most likely next writing intention of a graduate researcher when editing the following LaTex paper draft. Your output should only be a label from the list above.

## Input: {before_text}
## Output:
\end{lstlisting}
\vspace{-3mm}
\captionof{lstlisting}{Prediction prompt for \textcolor{blue}{GPT-5}}
\label{listing:prediction-prompt-gpt}
\end{minipage}
\end{table}

\subsection{Results}\label{sec:intention:f1}
Table~\ref{tab:per-class-f1-bert} reports the per-class F1 scores for the fine-tuned BERT model.
\begin{table}[ht!]
\centering
\small
\begin{tabular}{@{}lc@{}}
\toprule
\textbf{Writing Intention} & \textbf{F1-score} \\
\midrule
Idea Generation & 0.72 \\
Idea Organization & 0.47 \\
Section Planning & 0.83 \\
Text Production & 0.63 \\
Object Insertion & 0.79 \\
Citation Integration & 0.52 \\
Cross-reference & 0.10 \\
Macro Insertion & 0.67 \\
Fluency & 0.13 \\
Coherence & 0.32 \\
Clarity & 0.59 \\
Structural & 0.39 \\
Linguistic Style & 0.27 \\
Visual Formatting & 0.44 \\
\midrule
Accuracy & 0.56 \\
Macro Avg & 0.49 \\
\bottomrule
\end{tabular}
\caption{Per-class F1 scores for BERT-finetuned.}
\label{tab:per-class-f1-bert}
\vspace{-3mm}
\end{table}

According to Table \ref{table:intention-prediction}, none of them is reaching 0.7 in F1. This is likely due to the intricate nature of the task. The model is asked to predict the next intention by only giving the before text. In the annotation task, the annotator labels the data by looking through multiple consecutive before and after text pairs to determine the current intention rather than looking at the current text to predict the next intention. Another reason is that the next intention chosen by the author does not necessarily mean that it is the only correct intention.
We also noticed that BERT and RoBERTa perform much better than the vanilla Llama-8b-Instruct. This is likely due to Llama 8b being under-trained as they have a larger size of parameters, while we fine-tuned it on our data for only one epoch.

\section{More about Coarse-grained output alignment}\label{sec:appendix:coarse-alignment}

\subsection{Prompt Templates}
We present the prompt templates used for the session output alignment experiments in Listings~\ref{listing:ouput-align-prompt-single}--\ref{listing:ouput-align-prompt-multi}.

\begin{table}[ht!]
\centering
\begin{minipage}{\linewidth}
\lstset{
    basicstyle=\ttfamily\footnotesize, 
    breaklines=true, 
    frame=single, 
    columns=fullflexible, 
    captionpos=b 
}
\begin{lstlisting}
You are an expert writing assistant.
Your role is to improve user-provided text according to specific criteria. 
The input will be LaTeX text. 
Always output only the improved LaTeX text, with no explanations or notes.
Improve the following LaTeX text according to this criteria:
- {intention\}: {intention definition}
Text to improve:
{before_text}
\end{lstlisting}
\vspace{-3mm}
\captionof{lstlisting}{Single-intention Session Output Generation prompt}
\label{listing:ouput-align-prompt-single}
\end{minipage}
\end{table}

\begin{table}[ht!]
\centering
\begin{minipage}{\linewidth}
\lstset{
    basicstyle=\ttfamily\footnotesize, 
    breaklines=true, 
    frame=single, 
    columns=fullflexible, 
    captionpos=b 
}
\begin{lstlisting}
You are an expert writing assistant.
Your role is to improve user-provided text according to specific criteria.
The input will be LaTeX text.
Always output only the improved LaTeX text, with no explanations or notes.
Improve the following LaTeX text according to these criteria:
{list of intentions and their definitions}
Text to improve: {before_text}
\end{lstlisting}
\vspace{-3mm}
\captionof{lstlisting}{Multi-intention Session Output Generation prompt}
\label{listing:ouput-align-prompt-multi}
\end{minipage}
\end{table}
\subsection{Results}\label{sec:appendix:coarse-fig}

\begin{figure*}[ht!]
    \centering

    \begin{subfigure}[t]{0.24\textwidth}
        \centering
        \includegraphics[width=\linewidth, trim={0.9cm 3.5cm 5.5cm 0cm}, clip]{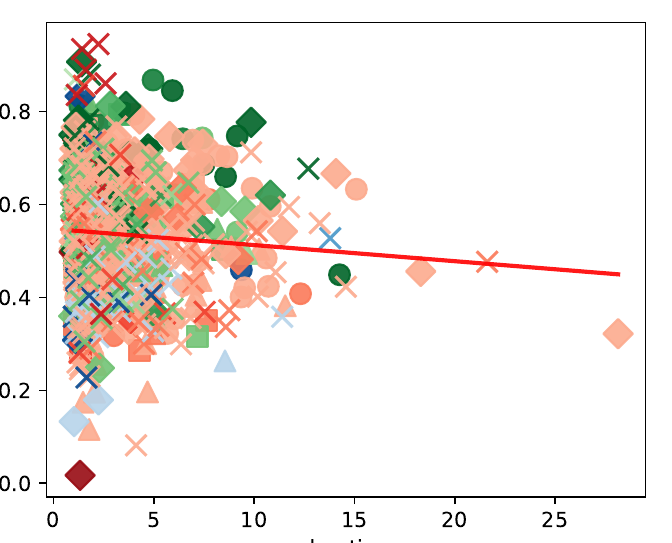}
        \caption{Levenshtein ratio (GPT-5)}
    \end{subfigure}
    \hfill
    \begin{subfigure}[t]{0.24\textwidth}
        \centering
        \includegraphics[width=\linewidth, trim={0.9cm 3.5cm  5.5cm 0cm}, clip]{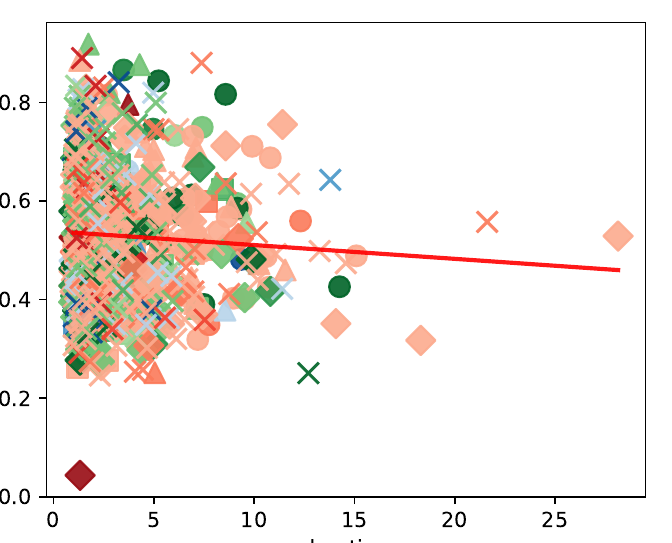}
        \caption{Levenshtein ratio (Qwen)}
    \end{subfigure}
    \hfill
    \begin{subfigure}[t]{0.24\textwidth}
        \centering
        \includegraphics[width=\linewidth, trim={0.9cm 3.5cm 5.5cm 0cm}, clip]{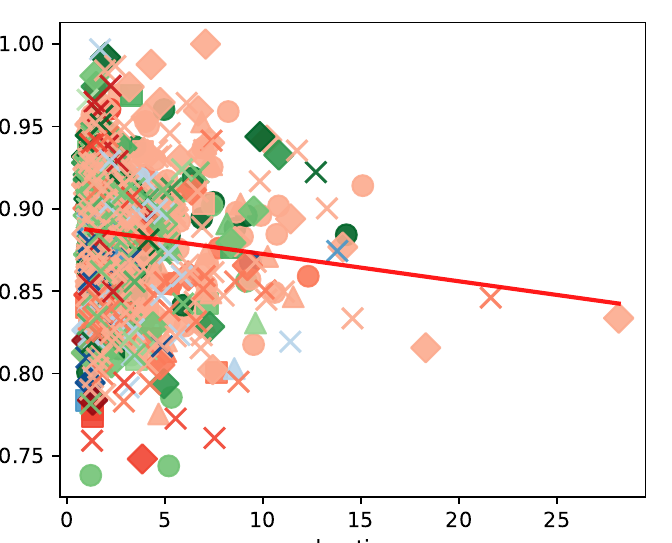}
        \caption{BERTScore-F1 (GPT-5)}
    \end{subfigure}
    \hfill
    \begin{subfigure}[t]{0.24\textwidth}
        \centering
        \includegraphics[width=\linewidth, trim={0.9cm 3.5cm 5.5cm 0cm}, clip]{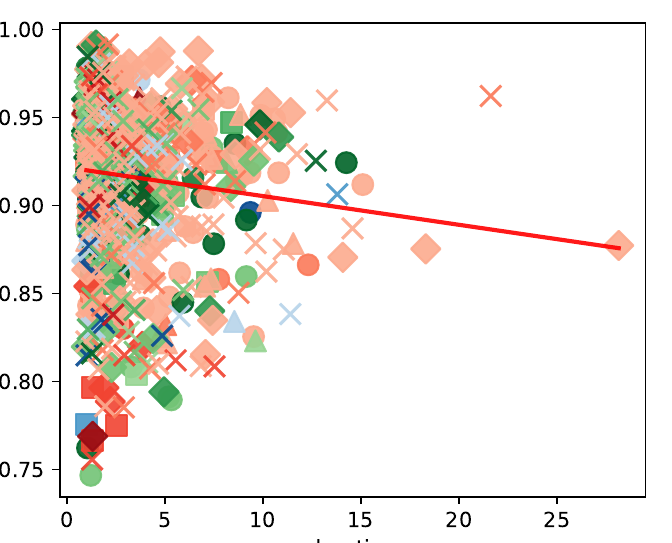}
        \caption{BERTScore-F1 (Qwen)}
    \end{subfigure}

    \caption{Writing duration (minutes) vs. model alignment scores (Levenshtein ratio and BERTScore-F1)}
    \label{fig:intention-session-all}
\end{figure*}
\begin{figure*}[ht!]
    \centering

    \begin{subfigure}[t]{0.48\textwidth}
        \centering
        \includegraphics[width=\linewidth, trim={0.9cm 0.9cm 0cm 0cm}, clip]{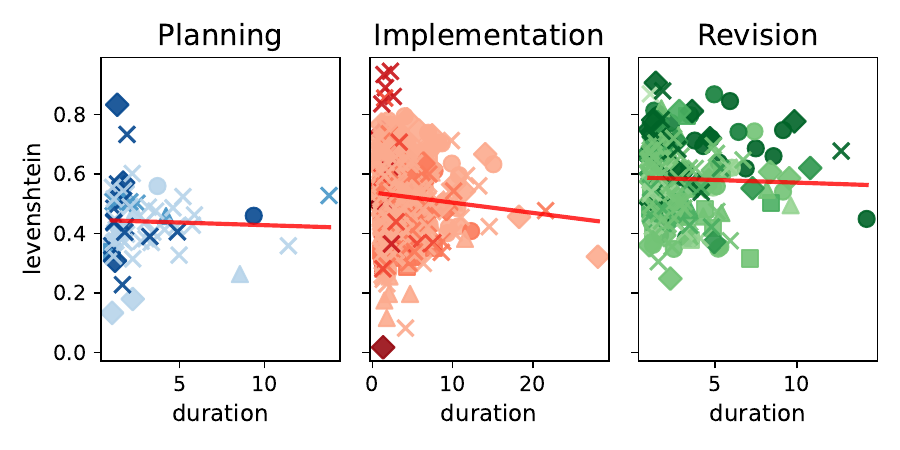}
        \caption{Levenshtein ratio (GPT-5)}
    \end{subfigure}
    \hfill
    \begin{subfigure}[t]{0.48\textwidth}
        \centering
        \includegraphics[width=\linewidth, trim={0.9cm 0.9cm 0cm 0cm}, clip]{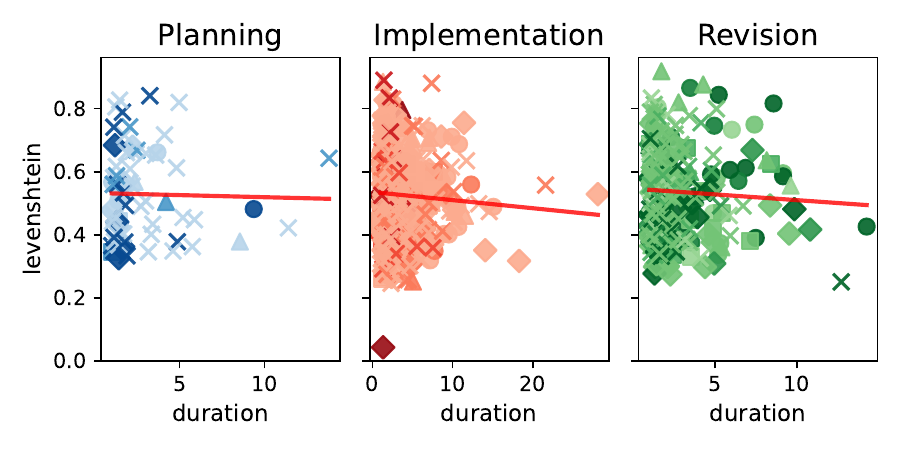}
        \caption{Levenshtein ratio (Qwen)}
    \end{subfigure}

    \vspace{1em}
    \begin{subfigure}[t]{0.48\textwidth}
        \centering
        \includegraphics[width=\linewidth, trim={0.9cm 0.9cm 0cm 0cm}, clip]{figures/experiments/intention_session_duration_vs_bertscore_by_category_gpt5.pdf}
        \caption{BERTScore-F1 (GPT-5)}
    \end{subfigure}
    \hfill
    \begin{subfigure}[t]{0.48\textwidth}
        \centering
        \includegraphics[width=\linewidth, trim={0.9cm 0.9cm 0cm 0cm}, clip]{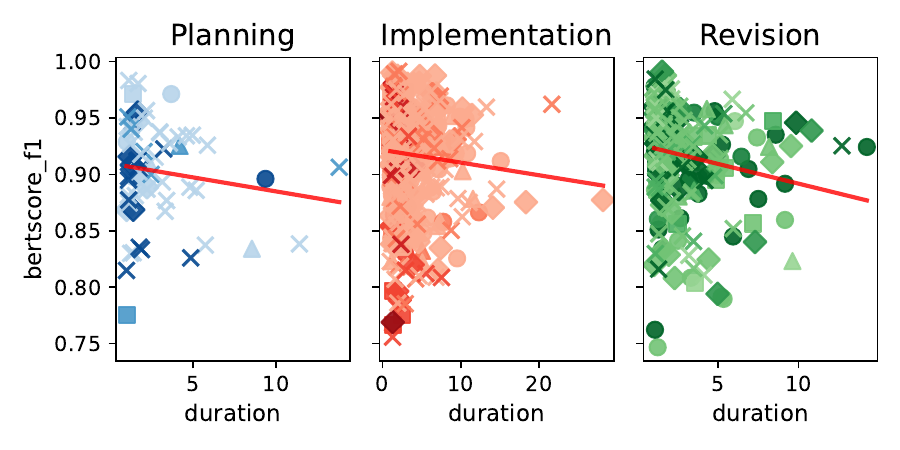}
        \caption{BERTScore-F1 (Qwen)}
    \end{subfigure}

    \caption{Writing duration (minutes) vs. model alignment scores (Levenshtein ratio and BERTScore-F1) across different intention categories }
    \label{fig:intention-session-category}
\end{figure*}
\begin{figure*}[ht!]
    \centering

    \begin{subfigure}[t]{0.24\textwidth}
        \centering
        \includegraphics[width=\linewidth, trim={0.9cm 0.9cm 3.5cm 0cm}, clip]{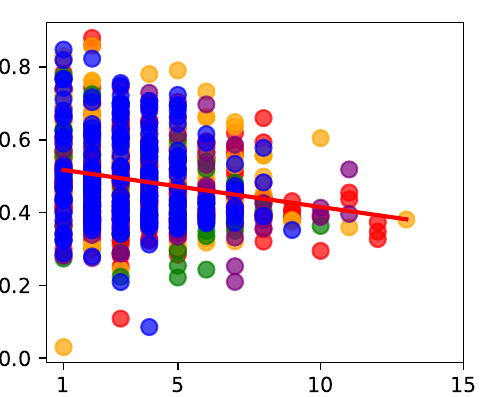}
        \caption{Levenshtein ratio (GPT-5)}
    \end{subfigure}
    \hfill
    \begin{subfigure}[t]{0.24\textwidth}
        \centering
        \includegraphics[width=\linewidth, trim={0.9cm 0.9cm 3.5cm 0cm}, clip]{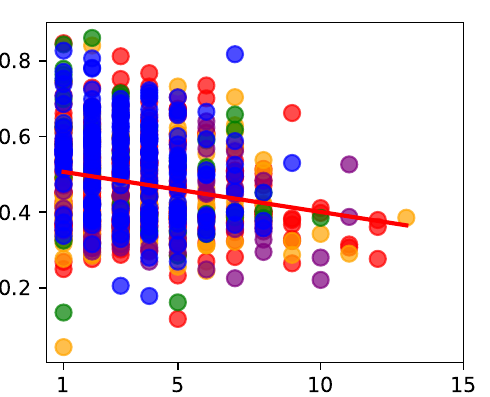}
        \caption{Levenshtein ratio (Qwen)}
    \end{subfigure}
    \hfill
    \begin{subfigure}[t]{0.24\textwidth}
        \centering
        \includegraphics[width=\linewidth, trim={0.9cm 0.9cm 3.5cm 0cm}, clip]{figures/experiments/session_len_label_bertscore_gpt5.pdf}
        \caption{BERTScore-F1 (GPT-5)}
    \end{subfigure}
    \hfill
    \begin{subfigure}[t]{0.24\textwidth}
        \centering
        \includegraphics[width=\linewidth, trim={0.9cm 0.9cm 3.5cm 0cm}, clip]{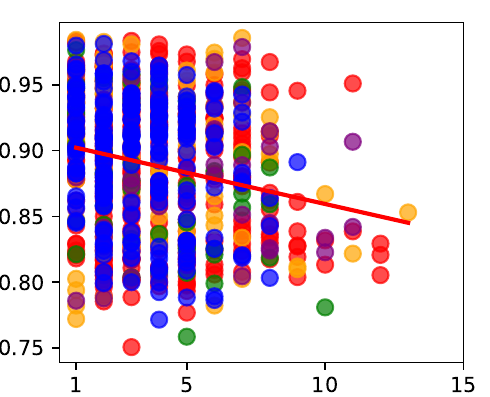}
        \caption{BERTScore-F1 (Qwen)}
    \end{subfigure}

    \caption{Number of writing intentions vs. alignment scores (Levenshtein ratio and BERTScore-F1) for GPT-5 and Qwen-2.5-14B}
    \label{fig:session-all}
\end{figure*}

We provide figures supporting the findings discussed in the main text. 
Figure~\ref{fig:intention-session-all} shows the relationship between single-intention session duration and alignment score for GPT-5 and Qwen-2.5-14B. Figure~\ref{fig:intention-session-category} breaks this down across different intention categories. Figure~\ref{fig:session-all} shows the relationship between the number of co-occurring intentions in a multi-intention session and alignment score for both models. Alignment is measured using both Levenshtein ratio and BERTScore-F1.


\section{More about Fine-grained output alignment}\label{sec:appendix:fine-alignment}
\subsection{Training environments}

\paragraph{BERT \& RoBERTa} We fine-tuned BERT and RoBERTa with the following hyperparameter setups: (1) a learning rate of $2e^{-5}$; (2) training batch size per device of 8; (3) evaluation batch size per device of 8; (4) the number of training epochs of 10; and (5) a weight decay of $0.01$. For each model, it took approximately 3.5 hours on one NVIDIA RTX A6000. 

\paragraph{Llama} For all experiments, we used baseline models of 4-bit quantized  Llama-8B-Instruct\footnote{\url{https://huggingface.co/unsloth/Meta-Llama-3.1-8B-Instruct-bnb-4bit}}, using unsloth library\footnote{\url{https://github.com/unslothai/unsloth}}.

For \textbf{the `after' text generation subtask} experiment, we used the following hyperparameter setups for the fine-tuned Llama-8B-Instruct: (1) only one epoch of training; (2) a weight decay of 0.01; (3) warm-up steps of 10; (4) learning rate of $3e^{-4}$; and (5) AdamW 8-bit optimizer. Due to computational constraints, we were able to run only one epoch for fine-tuning Llama models on our \textsc{ScholaWrite} dataset. Also, it took approximately 12 hours on one NVIDIA L40s.

\subsection{Details About Finetuning Process}\label{sec:appendix:finetuning}

The structure of the ``after'' text differs slightly to help the model learn where and what edits to make. We used the \texttt{diff\_match\_patch} library to generate a word-level difference array between the  ``before'' and ``after'' texts. Special tokens (\texttt{<same>}, \texttt{</same>}, \texttt{<del>}, \texttt{</del>}, \texttt{<add>}, \texttt{</add>}) were added to the tokenizer, and the difference array was converted into text wrapped with these tokens. For example, given a ``before'' text of \texttt{``Bad dog''} and an ``after'' text of \texttt{``Good dog''}, the difference array would be \texttt{[(-1, `Bad'), (1, `Good '), (0, `dog')]}. This is converted into: \texttt{<del>Bad </del><add>Good </add><same>dog</same>}. This transformation was applied only to the ``after'' text, while the ``before'' text remained as plain LaTeX text.

For finetuning, we randomly split the \textsc{ScholaWrite} dataset into training (80\%) and testing (20\%) sets. From each intention label in the test set, we sample up to 300 keystroke entries due to budget constraints. 

The fine-tuning prompt included task instructions, a verbalizer derived from human-annotated labels, and the ``before'' text. For fine-tuning, we used QLoRA \citep{dettmers2024qlora} to optimize all linear modules of a 4-bit quantized model while maintaining a small memory footprint. Additionally, the \texttt{embed\_tokens} and \texttt{lm\_head} modules were set as trainable and saved in the final checkpoint. To focus training on the response portion (the ``after'' text), we used the \texttt{train\_on\_response\_only} function, masking the task instructions, verbalizer, and ``before'' text with \texttt{-100s}. This ensures the model learns to generate the "after" text without being influenced by instructional input. The model was trained for one epoch with a batch size of 1, 4 gradient accumulation steps, and the AdamW 8-bit optimizer.

\section{Iterative Writing}\label{sec:appendix:iterative-writing}
\subsection{Setups}

For model setups, we created two pairs of models for each prediction and generation subtasks as follows: 

\begin{itemize}
    \item \textcolor{magenta}{LLama-8B-SW}: LLama-8B-Instruct fine-tuned on \textsc{ScholaWrite} dataset (Prediction) \& LLama-8B-Instruct fine-tuned on \textsc{ScholaWrite} dataset (Generation), independently
    \item \textcolor{teal}{LLama-8B-Instruct}: Vanilla LLama-8B-Instruct (Prediction) \& Vanilla LLama-8B-Instruct (Generation), independently
\end{itemize}

During iterative writing, we performed 100 iterations, treating the model’s output under one intention as a single iteration. If the intention predicted by the classification model (fine-tuned Llama3.1-8B-Instruct, as described in Sec \ref{sec:next_predict}) matched the current predicted intention, the model was prompted to edit the text again. In this case, the newly generated output was not treated as final output in the iteration, and the iteration did not proceed. We moved to the next iteration only when the intention prediction model generated a different intention label than the previous one.

Also, due to budget constraints, models had different revision strategies. \textsc{Llama-8B-SW-*} and \textsc{Llama-8B-Instruct} continued revision until the next predicted intention changed.

\subsection{Prompt Templates}

We present the generation prompt templates used for the iterative writing experiments in Listings \ref{prompt:iter-sw}--\ref{prompt:iter-zero}.

\begin{table}[ht!]
\centering
\begin{minipage}{\linewidth}
\lstset{
    basicstyle=\ttfamily\footnotesize, 
    breaklines=true, 
    frame=single, 
    columns=fullflexible, 
    captionpos=b 
}
\begin{lstlisting}
You are a computer science researcher with extensive experience of scholarly writing. Here, you are writing a research paper in natural language processing using LaTeX languages.

You currently want to \texttt{{``put the verbalizer of the predicted intention label''} (e.g., ``initially create sections and sub-level structures'' if the predicted label was section planning)}. 

Below is the paper you have written so far. Please strictly follow the writing intention given above and insert, delete, or revise at appropriate places in the paper given below.

Your writing should relate to the paper given below. Do not generate text other than paper content. Do not describe the changes you are making or your reasoning.

## Input: {before_text}
\end{lstlisting}
\vspace{-3mm}
\captionof{lstlisting}{Iterative writing generation prompt for \textcolor{magenta}{LLama-8B-SW}}
\label{prompt:iter-sw}
\end{minipage}
\end{table}

\begin{table}[ht!]
\centering
\begin{minipage}{\linewidth}
\lstset{
    basicstyle=\ttfamily\footnotesize, 
    breaklines=true, 
    frame=single, 
    columns=fullflexible, 
    captionpos=b 
}
\begin{lstlisting}
You are a computer science researcher with extensive experience in scholarly writing. Here, you are writing a research paper in natural language processing using LaTeX.

You currently want to \texttt{{``put the verbalizer of the predicted intention label''} (e.g., ``initially create sections and sub-level structures'' if the predicted label was section planning)}. 

Below is the paper you have written so far. Given the paper information below and the corresponding scholarly writing intention, please revise or add to the text to fulfill this writing intention.

You may insert, delete, or revise text at appropriate places in the given paper.

Please provide a complete output. Do not generate text that is nonsensical or unrelated to the given paper information.

## Input: {before_text}
\end{lstlisting}
\vspace{-3mm}
\captionof{lstlisting}{Iterative writing generation prompt for \textcolor{teal}{LLama-8B-Zero}}
\label{prompt:iter-zero}
\end{minipage}
\end{table}

\subsection{Seed Documents for Iterative Self-Writing}

Seed documents were derived from LaTeX-formatted abstracts of four award-winning NLP papers on diverse topics \citep{zeng-etal-2024-johnny, lu-etal-2024-semisupervised, du-etal-2022-read, etxaniz-etal-2024-latxa}, as shown in Listings \ref{table:seed-entry-johnny}-\ref{table:seed-entry-latxa}.

\begin{table}[ht!]
\centering
\begin{minipage}{\linewidth}
\lstset{
    basicstyle=\ttfamily\footnotesize, 
    breaklines=true, 
    frame=single, 
    columns=fullflexible, 
    captionpos=b 
}
\begin{lstlisting}

\begin{document}
\maketitle

\title{How Johnny Can Persuade LLMs to Jailbreak Them: Rethinking Persuasion to Challenge AI Safety by Humanizing LLMs}
\author{}
\date{}

\begin{abstract}
Most traditional AI safety research has approached AI models as machines and centered on algorithm-focused attacks developed by security experts. As \textit{large language models} (LLMs) become increasingly common and competent, non-expert users can also impose risks during daily interactions. This paper introduces a new perspective on jailbreaking LLMs as human-like communicators to explore this overlooked intersection between everyday language interaction and AI safety. Specifically, we study how to persuade LLMs to jailbreak them. First, we propose a persuasion taxonomy derived from decades of social science research. Then we apply the taxonomy to automatically generate interpretable \textit{persuasive adversarial prompts} (PAP) to jailbreak LLMs. Results show that persuasion significantly increases the jailbreak performance across all risk categories: PAP consistently achieves an attack success rate of over $92\%$ on Llama 2-7b Chat, GPT-3.5, and GPT-4 in $10$ trials, surpassing recent algorithm-focused attacks. On the defense side, we explore various mechanisms against PAP, find a significant gap in existing defenses, and advocate for more fundamental mitigation for highly interactive LLMs.
\end{abstract}

\end{document}

\end{lstlisting}
\vspace{-3mm}
\captionof{lstlisting}{An example seed document \citep{zeng-etal-2024-johnny} as shown in LaTeX codes to begin iterative self-writing.}
\label{table:seed-entry-johnny}
\end{minipage}
\end{table}

\begin{table}[ht!]
\centering
\begin{minipage}{\linewidth}
\lstset{
    basicstyle=\ttfamily\footnotesize, 
    breaklines=true, 
    frame=single, 
    columns=fullflexible, 
    captionpos=b 
}
\begin{lstlisting}
\begin{document}
\maketitle

\title{Read, Revise, Repeat: A System Demonstration for Human-in-the-loop Iterative Text Revision}
\author{}
\date{}

\begin{abstract}
Revision is an essential part of the human writing process. It tends to be strategic, adaptive, and, more importantly, \textit{iterative} in nature. Despite the success of large language models on text revision tasks, they are limited to non-iterative, one-shot revisions. Examining and evaluating the capability of large language models for making continuous revisions and collaborating with human writers is a critical step towards building effective writing assistants. In this work, we present a human-in-the-loop iterative text revision system, $\mathcal{R}$ead, $\mathcal{R}$evise, $\mathcal{R}$epeat (\textsc{$\mathcal{R}3$}), which aims at achieving high quality text revisions with minimal human efforts by reading model-generated revisions and user feedbacks, revising documents, and repeating human-machine interactions. In \method, a text revision model provides text editing suggestions for human writers, who can accept or reject the suggested edits. The accepted edits are then incorporated into the model for the next iteration of document revision. Writers can therefore revise documents iteratively by interacting with the system and simply accepting/rejecting its suggested edits until the text revision model stops making further revisions or reaches a predefined maximum number of revisions. Empirical experiments show that \method can generate revisions with comparable acceptance rate to human writers at early revision depths, and the human-machine interaction can get higher quality revisions with fewer iterations and edits. 
\end{abstract}
\end{document}

\end{lstlisting}
\vspace{-3mm}
\captionof{lstlisting}{An example seed document \cite{du-etal-2022-read} as shown in LaTeX codes to begin iterative self-writing.}
\label{table:seed-entry-read}
\end{minipage}
\end{table}

\begin{table}[ht!]
\centering
\begin{minipage}{\linewidth}
\lstset{
    basicstyle=\ttfamily\footnotesize, 
    breaklines=true, 
    frame=single, 
    columns=fullflexible, 
    captionpos=b 
}
\begin{lstlisting}
\begin{document}
\maketitle

\title{Semisupervised Neural Proto-Language Reconstruction}
\author{}
\date{}

\begin{abstract}
Existing work implementing comparative reconstruction of ancestral languages (proto-languages) has usually required full supervision. However, historical reconstruction models are only of practical value if they can be trained with a limited amount of labeled data. We propose a semisupervised historical reconstruction task in which the model is trained on only a small amount of labeled data (cognate sets with proto-forms) and a large amount of unlabeled data (cognate sets without proto-forms). We propose a neural architecture for comparative reconstruction (DPD-BiReconstructor) incorporating an essential insight from linguists' comparative method: that reconstructed words should not only be reconstructable from their daughter words, but also deterministically transformable back into their daughter words. We show that this architecture is able to leverage unlabeled cognate sets to outperform strong semisupervised baselines on this novel task.
\end{abstract}

\end{document}

\end{lstlisting}
\vspace{-3mm}
\captionof{lstlisting}{An example seed document \citep{lu-etal-2024-semisupervised} as shown in LaTeX codes to begin iterative self-writing.}
\label{table:seed-entry-semi}
\end{minipage}
\end{table}

\begin{table}[ht!]
\centering
\begin{minipage}{\linewidth}
\lstset{
    basicstyle=\ttfamily\footnotesize, 
    breaklines=true, 
    frame=single, 
    columns=fullflexible, 
    captionpos=b 
}
\begin{lstlisting}
\begin{document}
\maketitle

\title{Latxa: An Open Language Model and Evaluation Suite for Basque}
\author{}
\date{}

\begin{abstract}
We introduce Latxa, a family of large language models for Basque ranging from 7 to 70 billion parameters.
Latxa is based on Llama 2, which we continue pretraining on a new Basque corpus comprising 4.3M documents and 4.2B tokens. Addressing the scarcity of high-quality benchmarks for Basque, we further introduce 4 multiple choice evaluation datasets: EusProficiency, comprising 5,169 questions from official language proficiency exams; EusReading, comprising 352 reading comprehension questions; EusTrivia, comprising 1,715 trivia questions from 5 knowledge areas; and EusExams, comprising 16,774 questions from public examinations. In our extensive evaluation, Latxa outperforms all previous open models we compare to by a large margin. In addition, it is competitive with GPT-4 Turbo in language proficiency and understanding, despite lagging behind in reading comprehension and knowledge-intensive tasks. Both the Latxa family of models, as well as our new pretraining corpora and evaluation datasets, are publicly available under open licenses. Our suite enables reproducible research on methods to build LLMs for low-resource languages.
\end{abstract}

\end{document}

\end{lstlisting}
\vspace{-3mm}
\captionof{lstlisting}{An example seed document \citep{etxaniz-etal-2024-latxa} as shown in LaTeX codes to begin iterative self-writing.}
\label{table:seed-entry-latxa}
\end{minipage}
\end{table}

\begin{figure}
    \centering
    \begin{subfigure}{0.4\textwidth}
        \includegraphics[width=\textwidth]{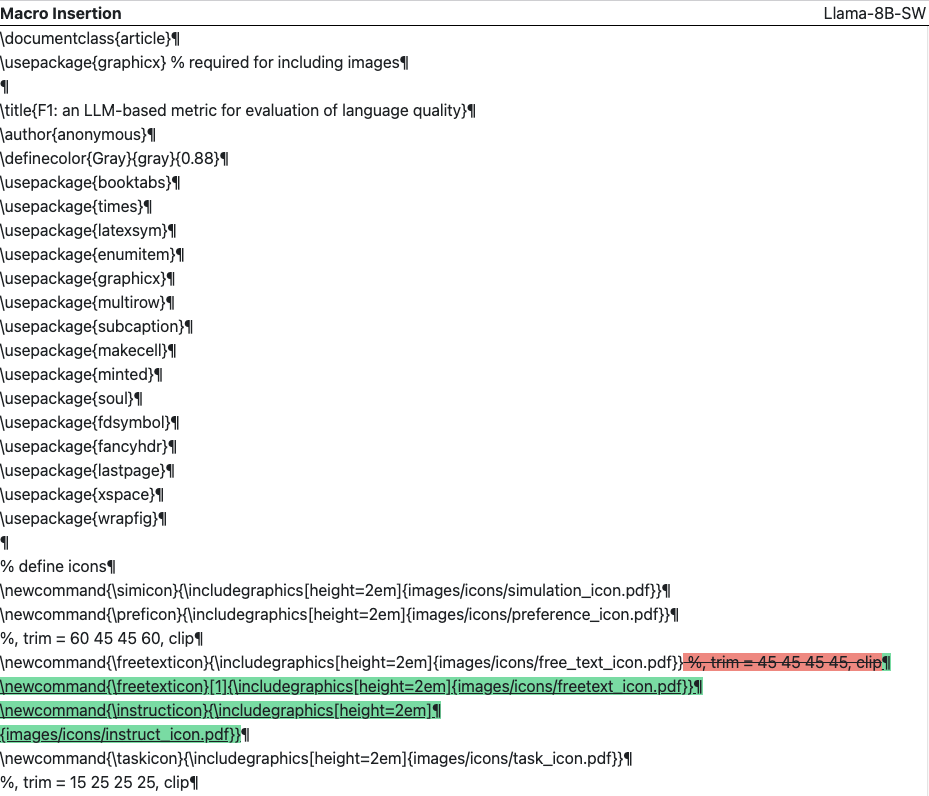}
        \caption{Llama writing inference for Macro Insertion activity. The model successfully added several LaTeX commands for custom actions.}
        \label{fig:llama_iter_sub_a}
    \end{subfigure}
    \hspace{0.05\textwidth}
    \begin{subfigure}{0.4\textwidth}
        \includegraphics[width=\textwidth]{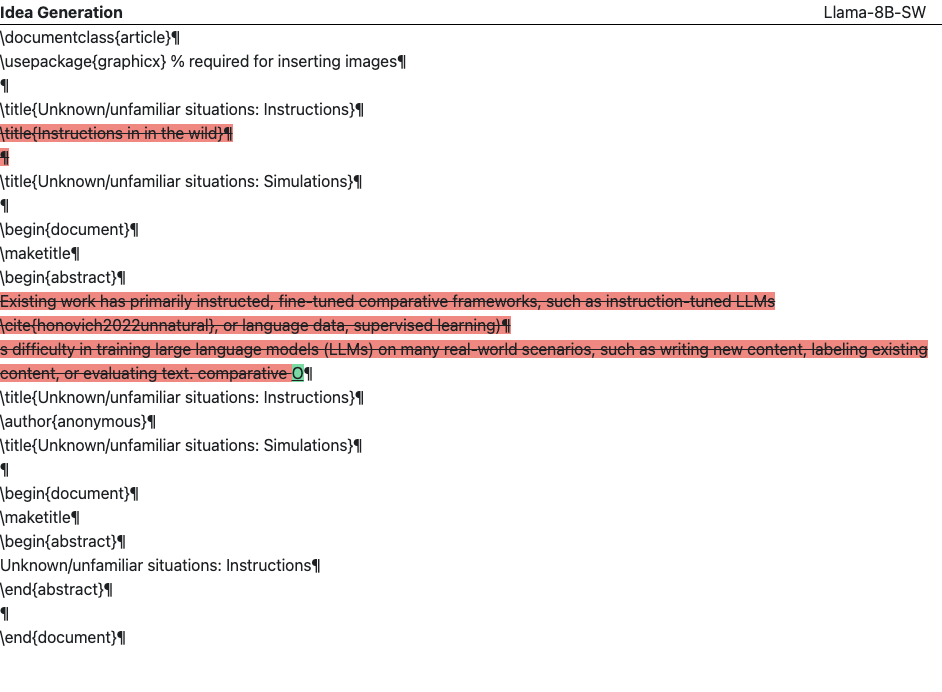}
        \caption{Llama writing inference for Idea Generation. The model failed to provide generated ideas and instead deleted abstract.}
        \label{fig:llama_iter_sub_d}
    \end{subfigure}
    \begin{subfigure}{0.43\textwidth}
        \includegraphics[width=\textwidth]{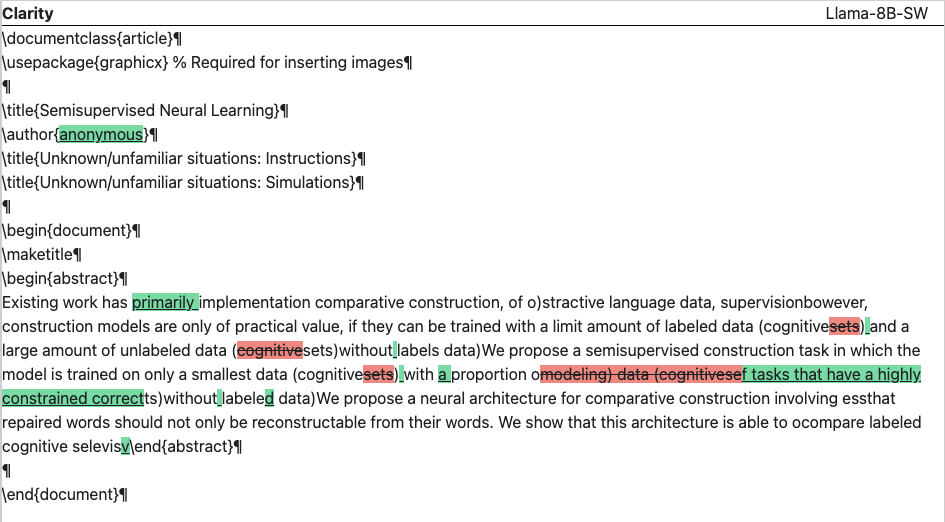}
        \caption{Llama writing inference for Clarity. The model successfully revised words and phrases for clearer delivery.}
        \label{fig:llama_iter_sub_b}
    \end{subfigure}
    \caption{Sample outputs from \textcolor{magenta}{Llama-8B-SW} during the self-writing experiment.}
    \label{fig:llama_iterative_output}
\end{figure}

\subsection{Iterative Training Sample Outputs}

Figure \ref{fig:llama_iterative_output} presents the sample outputs from the fine-tuned Llama-8B model during the iterative self-writing experiment. The model successfully was able to add several LaTeX commands to put some custom icon images (``Macro Insertion''). Also, it successfully revised several words and phrases in a paragraph for better clarity (``Clarity''). However, it struggled with understanding the definition of ``Idea Generation,'' and the model just deleted all paragraphs instead.

\subsection{Evaluation}
\subsubsection{Automatic Evaluation}
\paragraph{Definition of Quantitative Metrics for Iterative Self-Writing}

\begin{itemize}
    \item \textit{Lexical diversity}: Assess the unique tokens model generated in the final iteration of writing, measured by the number of unique tokens divided by the total tokens generated. 
    \item \textit{Topic consistency}: Cosine similarity between the seed document and output from the final iteration of writing. 
    \item \textit{Intention coverage}: Assess the diversity of the model's writing intention, measured by the number of unique labels predicted through the entire 100 iterations divided by all 15 intended labels available in our taxonomy.
\end{itemize}
\subsubsection{Human Evaluation}\label{sec:appendix:human-eval}

\paragraph{Study Procedures}

Human evaluation was conducted on the outputs of two models: the LLama-8B-Instruct\footnote{`\texttt{Unsloth/Meta-Llama-3.1-8B-Instruct-bnb-4bit'}}(\textcolor{teal}{Llama-8B-Zero}) and its finetuned counterpart (\textcolor{magenta}{Llama-8B-SW}). Each evaluation session lasted approximately two hours and was conducted via a Zoom call. Participants were three graduate students of an R1 University in the United States, with extensive experience in Overleaf-based writing, and they were compensated with a US 40 dollar gift card for their effort.

Before the evaluation began, the author shared their screen to present the following information:

\begin{itemize}
    \item A brief explanation of the research and an overview of the task (e.g., evaluating outputs from two models) using Google Slides.
    \item An explanation of all intention labels and their corresponding definitions in our taxonomy.
    \item A walkthrough of the evaluation web application, including (1) information displayed in the user interface (UI); (2) the locations on the UI where the participants should focus, such as the intention label; and (3) how to move between different seed documents.
    \item A tutorial on how to complete the evaluation of the five metrics (accuracy, alignment, fluency, coherence, relevance) using the provided Google Sheet.
\end{itemize}

During the evaluation, participants were required to share their screen while using the evaluation web application in their browser. The author remained muted but monitored the participants' shared screens and the Google Sheet to ensure the process was proceeding smoothly. The author only interacted with participants to address questions, resolve technical issues, or clarify instructions. No unsolicited interaction was allowed.

\begin{figure*}
    \centering
    \includegraphics[width=\linewidth]{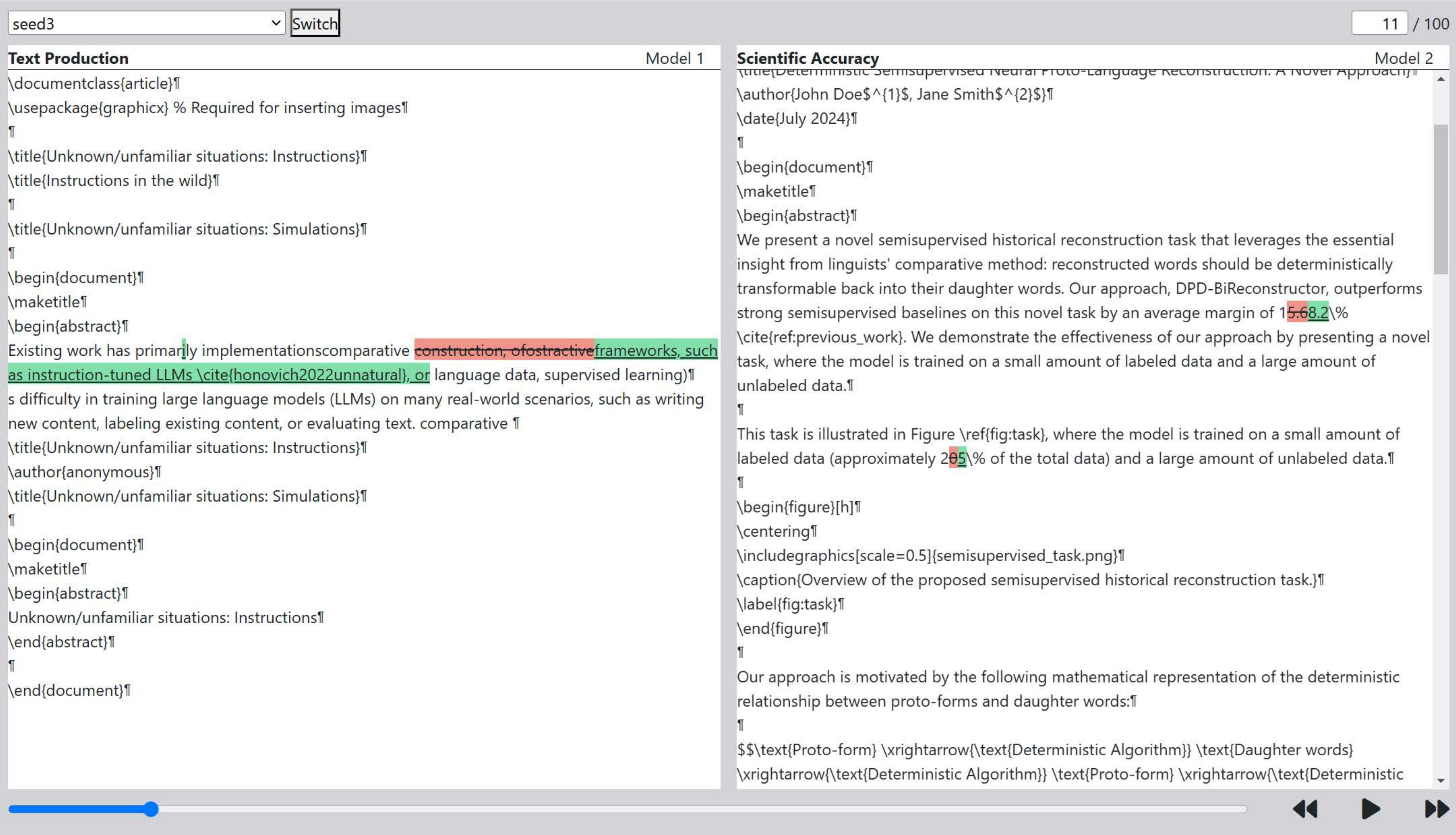}
    \caption{The user interface for the human evaluation process. }
    \label{fig:human-eval-interface}
\end{figure*}

The model identities were hidden from the participants. Instead, the models were labeled as ``Model 1'' and ``Model 2.'' Specifically, ``Model 1'' corresponded to the \textcolor{magenta}{Llama-8B-SW}, and ``Model 2'' corresponded to the \textcolor{teal}{Llama-8B-Zero}.
The evaluation web app displayed two text boxes side by side, with an intention label predicted by the respective model shown in the top-left corner of each text box. Please refer to Figure \ref{fig:human-eval-interface} as a screenshot of the user interface that we developed for the human evaluation process. 

After completing the evaluation, the authors thanked the participants for their time and effort, marking the conclusion of the Zoom call.

\paragraph{Definitions of Evaluation Metrics}

Here are the full description of our human evaluation metrics for the iterative self-writing experiment:

\begin{itemize}
    \item \textit{Accuracy}: Out of 100,  how many is the number of generated outputs that align with the provided intention? 
    \item \textit{Alignment}: Which model's whole writing process throughout the entire 100 iterations looks more human-like behaviors?
    \item \textit{Fluency}: Which model’s final writing sounds more grammatically correct?
    \item \textit{Coherence}: Which model’s final writing sounds more logical? 
    \item \textit{Relevance}: Does the final writing from each model contain related contents to the original seed document? 
\end{itemize}

\paragraph{Results \& Discussion}

Tables \ref{table:human-eval-seed1} to \ref{table:human-eval-seed4} present the human evaluation of results for each seed document. We observe that the fine-tuned \textcolor{magenta}{Llama-8B-SW} did not outperform the baseline vanilla counterpart (\textcolor{teal}{Llama-8B-Zero}) across all metrics for all four seed settings. 

This discrepancy may be attributed to the way the prompt used for training isolates individual writing actions from the continuous, interconnected process typical of human writing. A single writing action involves referencing the text before making an edit, the text after the edit, and the intention behind the edit. In human writing, these actions are cognitively and logically linked as part of a cohesive sequence. However, our model struggles to capture these connections due to the prompt structure, potentially causing it to become stuck in a local minimum. 

\begin{table*}[ht!]
\small 
\centering
\begin{tabular}{@{}p{2cm}c|c|c|c@{}}
\toprule
\textbf{Metrics} & \textbf{Model} & \textbf{Evaluator 1} & \textbf{Evaluator 2} & \textbf{Evaluator 3} \\ \midrule
 \multirow{2}{*}{Accuracy} &  \textcolor{magenta}{SW} & 43 & 3 & 17\\
\cmidrule(r){2-5}
& {\textcolor{teal}{Zero}} & 47 & 22 & 38\\
\midrule
\multirow{2}{*}{Alignment} &  \textcolor{magenta}{SW} &  &  \\
\cmidrule(r){2-5}
& {\textcolor{teal}{Zero}} & X & X & X\\
\midrule
\multirow{2}{*}{Fluency} &  \textcolor{magenta}{SW} &  &  &\\
\cmidrule(r){2-5}
& {\textcolor{teal}{Zero}} & X & X & X\\
\midrule
\multirow{2}{*}{Coherence} &  \textcolor{magenta}{SW} &  &  &\\
\cmidrule(r){2-5}
& {\textcolor{teal}{Zero}} & X & X & X\\
\midrule
\multirow{2}{*}{Relevance} &  \textcolor{magenta}{SW} & Yes & No & No\\
\cmidrule(r){2-5}
& {\textcolor{teal}{Zero}} & Yes & Yes & Yes\\
\bottomrule
\end{tabular}
\caption{Human evaluation results for the seed document 1.}
\label{table:human-eval-seed1}
\end{table*}

\begin{table*}[ht!]
\small 
\centering
\begin{tabular}{@{}p{2cm}c|c|c|c@{}}
\toprule
\textbf{Metrics} & \textbf{Model} & \textbf{Evaluator 1} & \textbf{Evaluator 2} & \textbf{Evaluator 3} \\ \midrule
 \multirow{2}{*}{Accuracy} &  \textcolor{magenta}{SW} & 26 & 0 & 5\\
\cmidrule(r){2-5}
& {\textcolor{teal}{Zero}} & 48 & 12 & 29\\
\midrule
\multirow{2}{*}{Alignment} &  \textcolor{magenta}{SW} &  &  \\
\cmidrule(r){2-5}
& {\textcolor{teal}{Zero}} & X & X & X\\
\midrule
\multirow{2}{*}{Fluency} &  \textcolor{magenta}{SW} &  &  &\\
\cmidrule(r){2-5}
& {\textcolor{teal}{Zero}} & X & X & X\\
\midrule
\multirow{2}{*}{Coherence} &  \textcolor{magenta}{SW} &  &  &\\
\cmidrule(r){2-5}
& {\textcolor{teal}{Zero}} & X & X & X\\
\midrule
\multirow{2}{*}{Relevance} &   \textcolor{magenta}{SW} & Yes & Yes & Yes\\
\cmidrule(r){2-5}
& {\textcolor{teal}{Zero}} & Yes & Yes & Yes\\
\bottomrule
\end{tabular}
\caption{Human evaluation results for the seed document 2.}
\label{table:human-eval-seed2}
\end{table*}

\begin{table*}[ht!]
\small 
\centering
\begin{tabular}{@{}p{2cm}c|c|c|c@{}}
\toprule
\textbf{Metrics} & \textbf{Model} & \textbf{Evaluator 1} & \textbf{Evaluator 2} & \textbf{Evaluator 3} \\ \midrule
 \multirow{2}{*}{Accuracy} &   \textcolor{magenta}{SW} & 52 & 0 & 3\\
\cmidrule(r){2-5}
& {\textcolor{teal}{Zero}} & 70 & 23 & 43\\
\midrule
\multirow{2}{*}{Alignment} &   \textcolor{magenta}{SW} &  &  \\
\cmidrule(r){2-5}
& {\textcolor{teal}{Zero}} & X & X & X\\
\midrule
\multirow{2}{*}{Fluency} &   \textcolor{magenta}{SW} &  &  &\\
\cmidrule(r){2-5}
& {\textcolor{teal}{Zero}} & X & X & X\\
\midrule
\multirow{2}{*}{Coherence} &   \textcolor{magenta}{SW} &  &  &\\
\cmidrule(r){2-5}
& {\textcolor{teal}{Zero}} & X & X & X\\
\midrule
\multirow{2}{*}{Relevance} &   \textcolor{magenta}{SW} & Yes & Yes & No\\
\cmidrule(r){2-5}
& {\textcolor{teal}{Zero}} & Yes & Yes & Yes\\
\bottomrule
\end{tabular}
\caption{Human evaluation results for the seed document 3.}
\label{table:human-eval-seed3}
\end{table*}

\begin{table*}[ht!]
\small 
\centering
\begin{tabular}{@{}p{2cm}c|c|c|c@{}}
\toprule
\textbf{Metrics} & \textbf{Model} & \textbf{Evaluator 1} & \textbf{Evaluator 2} & \textbf{Evaluator 3} \\ \midrule
 \multirow{2}{*}{Accuracy} &  \textcolor{magenta}{SW} & 37 & 3 & 6\\
\cmidrule(r){2-5}
& {\textcolor{teal}{Zero}} & 60 & 22 & 48\\
\midrule
\multirow{2}{*}{Alignment} &  \textcolor{magenta}{SW} &  &  \\
\cmidrule(r){2-5}
& {\textcolor{teal}{Zero}} & X & X & X\\
\midrule
\multirow{2}{*}{Fluency} &  \textcolor{magenta}{SW} &  &  &\\
\cmidrule(r){2-5}
& {\textcolor{teal}{Zero}} & X & X & X\\
\midrule
\multirow{2}{*}{Coherence} &  \textcolor{magenta}{SW} &  &  &\\
\cmidrule(r){2-5}
& {\textcolor{teal}{Zero}} & X & X & X\\
\midrule
\multirow{2}{*}{Relevance} &  \textcolor{magenta}{SW} & Yes & No & No\\
\cmidrule(r){2-5}
& {\textcolor{teal}{Zero}} & Yes & Yes & Yes\\
\bottomrule
\end{tabular}
\caption{Human evaluation results for the seed document 4.}
\label{table:human-eval-seed4}
\end{table*}

\begin{figure*}
    \centering
    \makebox[0.33\textwidth]{\small \textcolor{magenta}{Llama-8B-ScholaWrite}}
    \makebox[0.32\textwidth]{\small \textcolor{teal}{Llama-8B-Zero}}
    \makebox[0.32\textwidth]{\small \textcolor{blue}{GPT-4o}} 
    \\[1em]
        \includegraphics[width=0.32\textwidth]{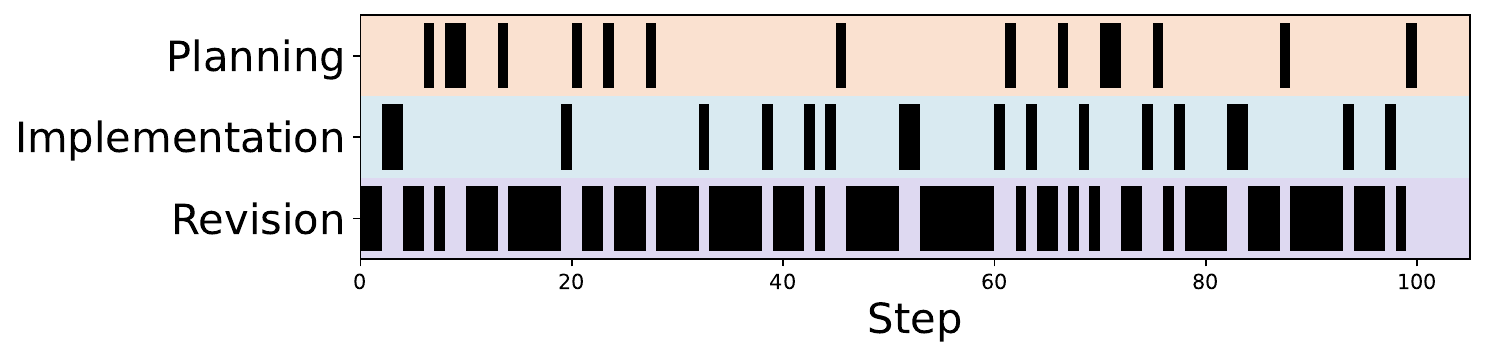}
        \includegraphics[width=0.32\textwidth]{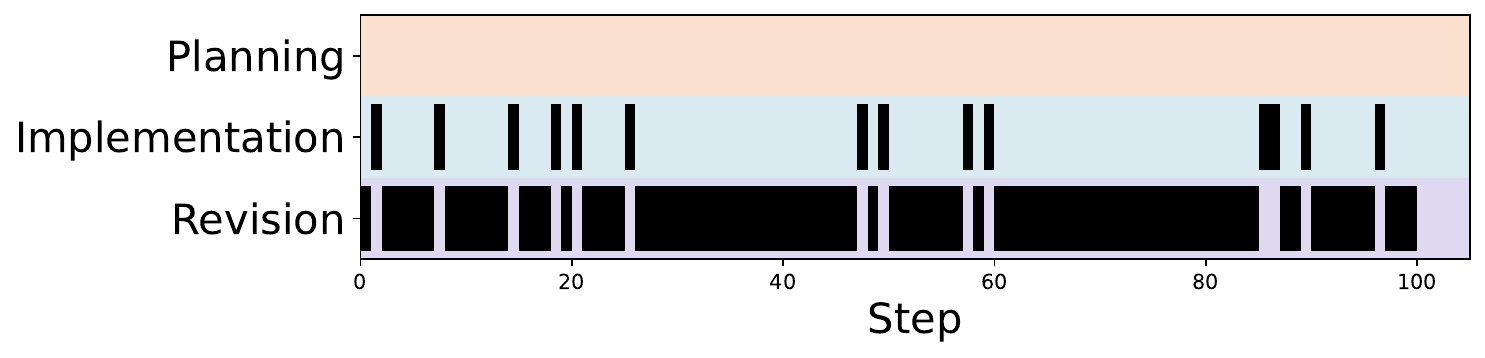}
        \includegraphics[width=0.32\textwidth]{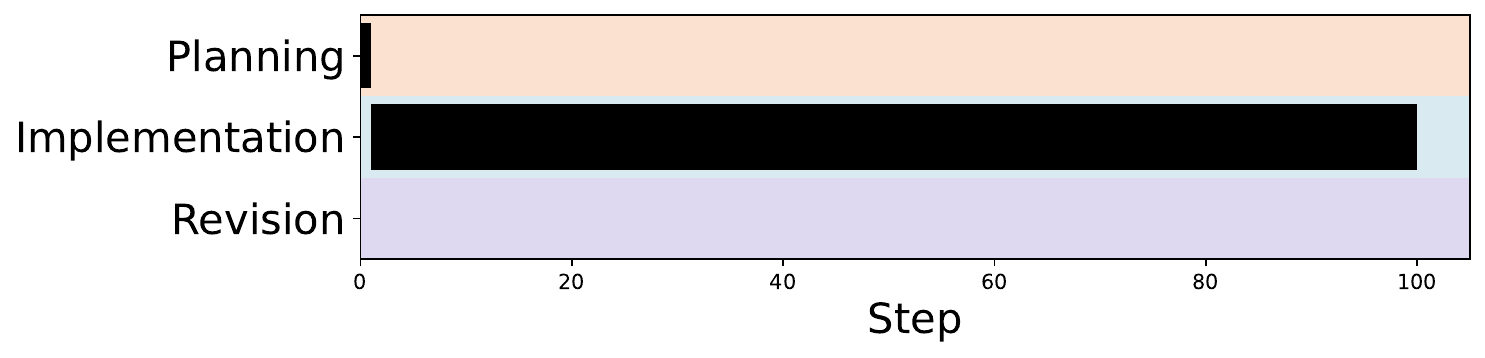}
        \\[2pt]
        \makebox[\textwidth]{\small (a) Seed 1 } 
    \\[0.5em]
        \includegraphics[width=0.32\textwidth]{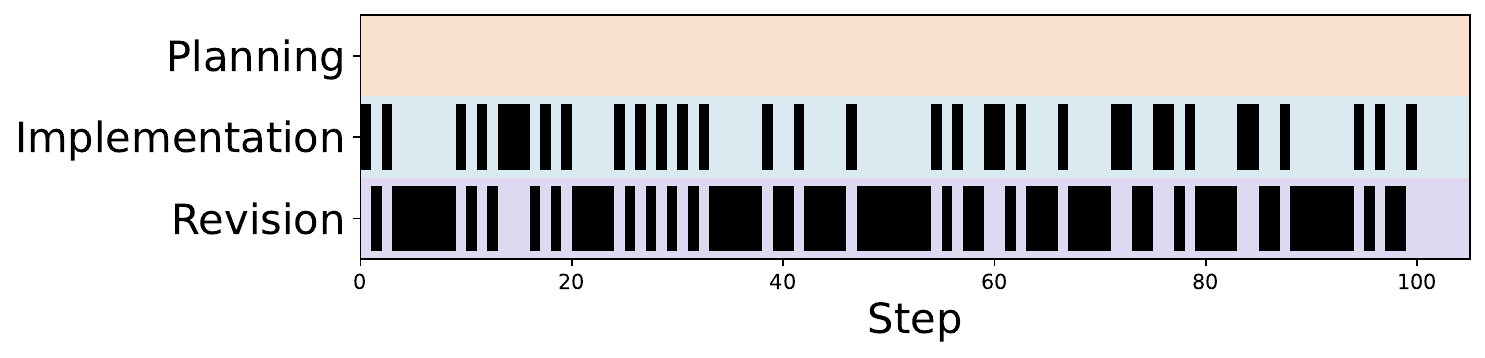}
        \includegraphics[width=0.32\textwidth]{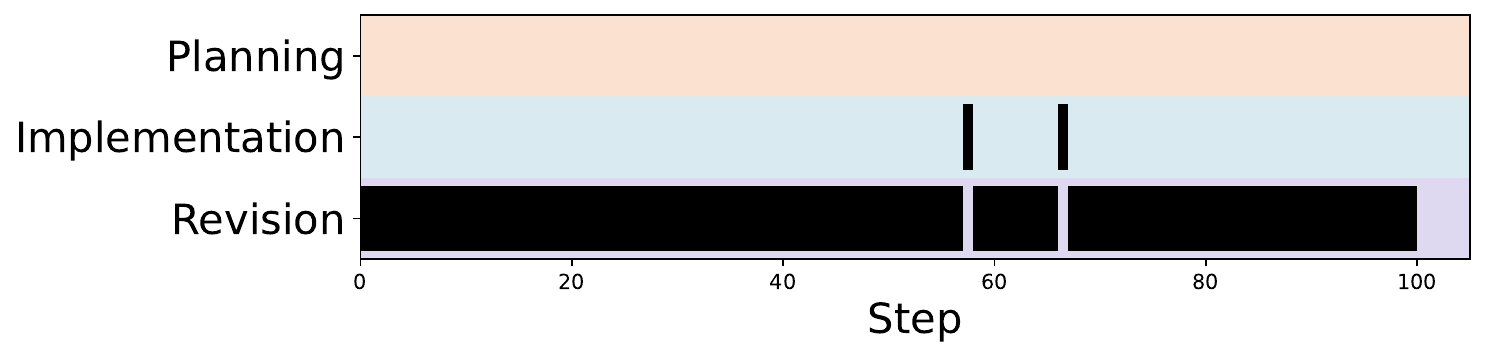}
        \includegraphics[width=0.32\textwidth]{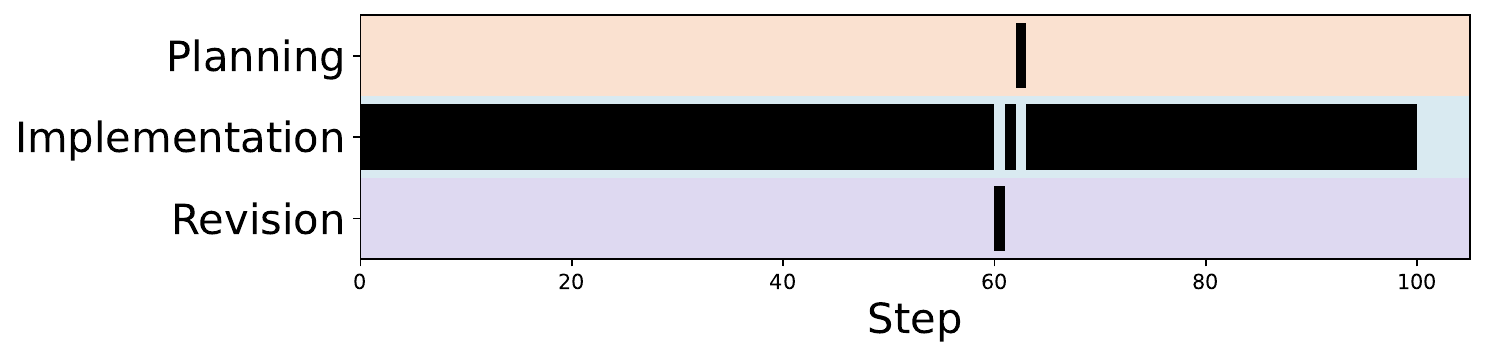}
        \\[2pt]
        \makebox[\textwidth]{\small (b) Seed 2 } 
    \\[0.5em]
        \includegraphics[width=0.32\textwidth]{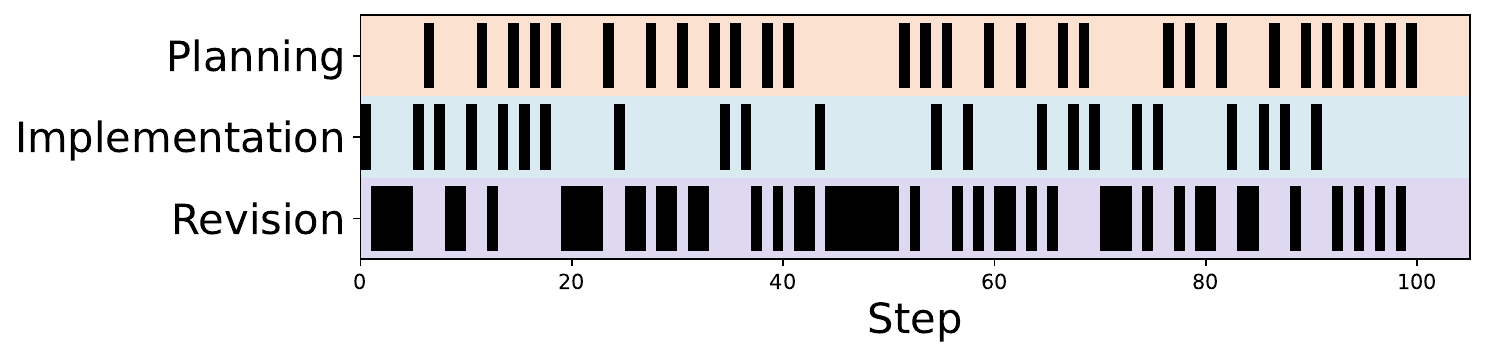}
        \includegraphics[width=0.32\textwidth]{figures/llama8_meta_output_broad_seed3.pdf}
        \includegraphics[width=0.32\textwidth]{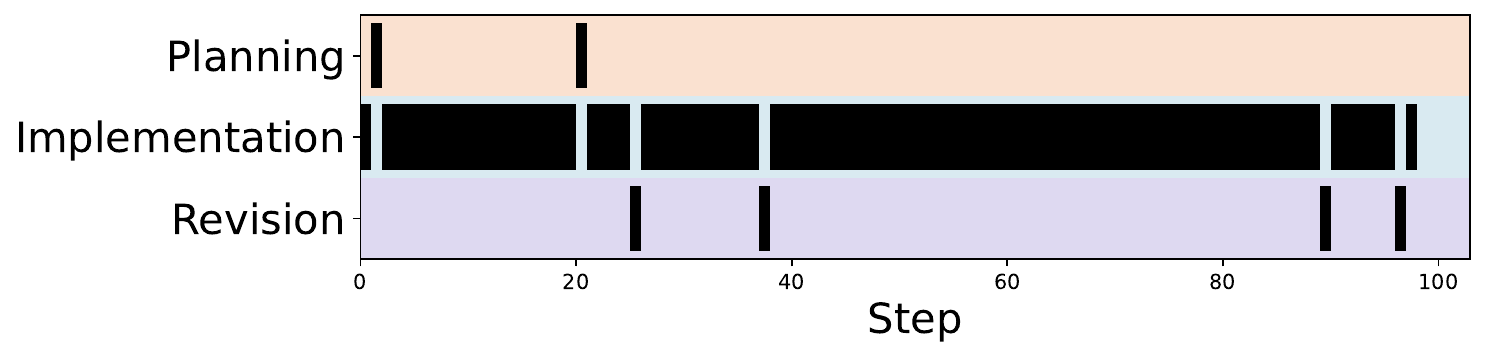}
        \\[2pt]
        \makebox[\textwidth]{\small (c) Seed 3 } 
    \\[0.5em]
        \includegraphics[width=0.32\textwidth]{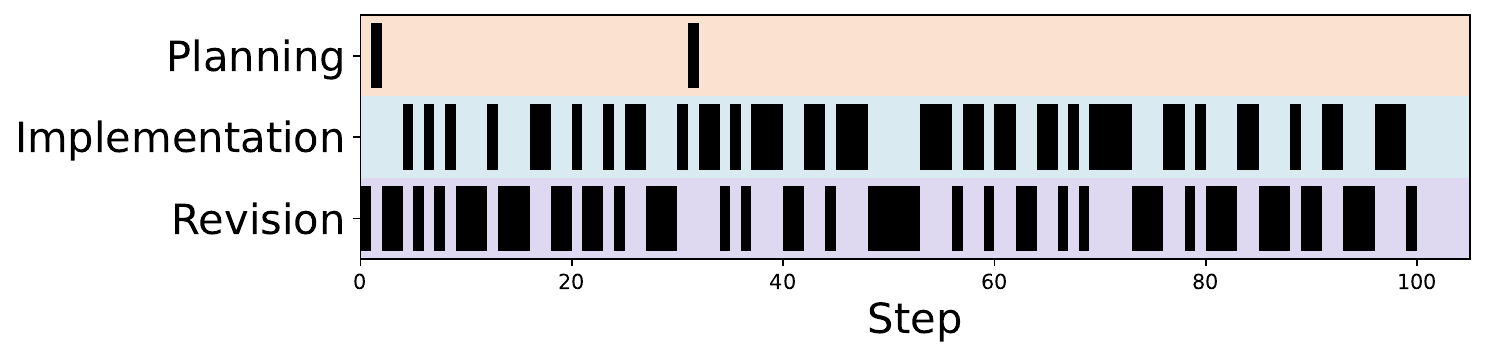}
        \includegraphics[width=0.32\textwidth]{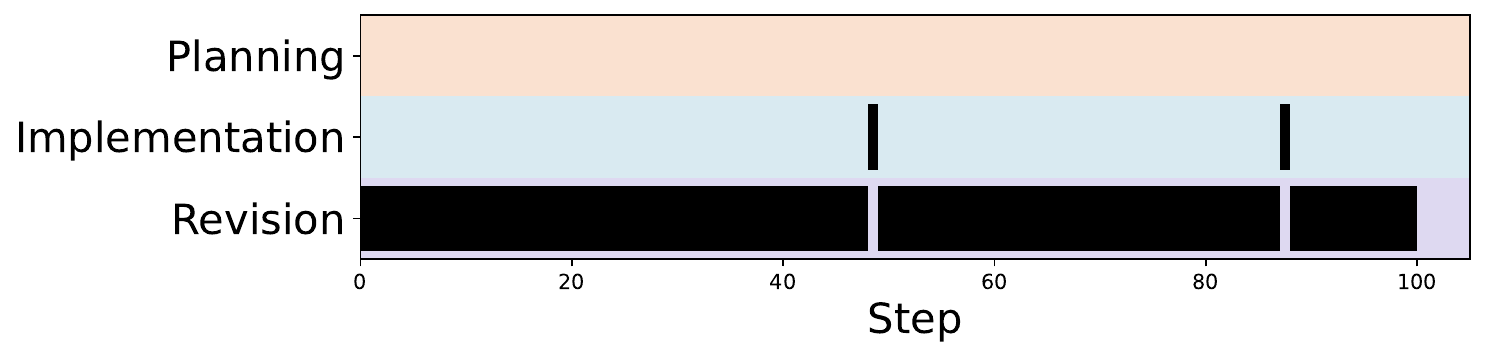}
        \includegraphics[width=0.32\textwidth]{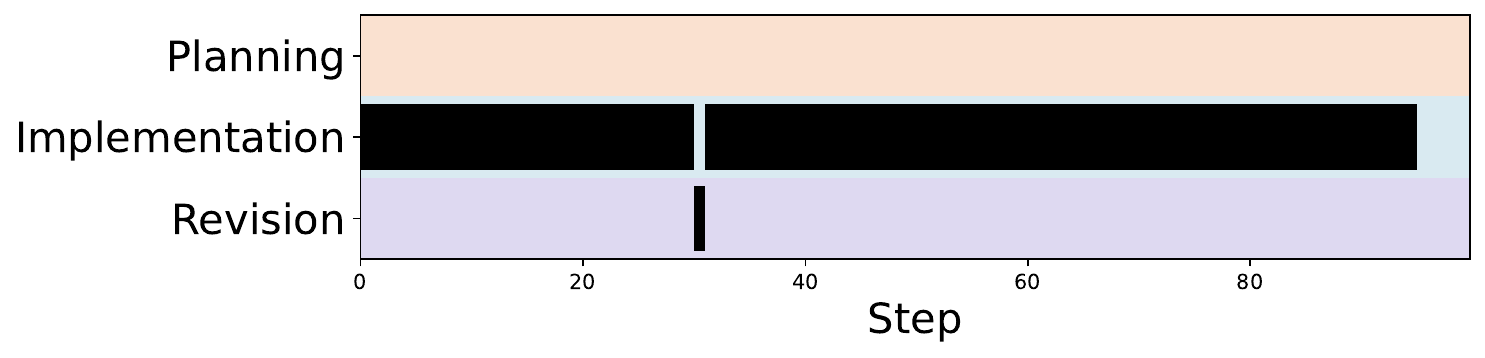}
        \\[2pt]
        \makebox[\textwidth]{\small (d) Seed 4 } 
    \\[0.5em]
    \caption{Distribution of high-level writing activities over time by models - \textcolor{magenta}{Llama-8B-ScholaWrite} (left); \textcolor{teal}{Llama-8B-Zero} (middle); \textcolor{blue}{GPT-4o} (right). Orange, Blue, and Purple represent Planning, Implementation, and Revision writing actions respectively. }
    \label{fig:writing-step-broad-all-model}
\end{figure*}

\begin{figure*}
    \centering
    \makebox[0.33\textwidth]{\small \textcolor{magenta}{Llama-8B-ScholaWrite}}
    \makebox[0.32\textwidth]{\small \textcolor{teal}{Llama-8B-Zero}}
    \makebox[0.32\textwidth]{\small \textcolor{blue}{GPT-4o}} 
    \\[1em]
        \includegraphics[width=0.32\textwidth]{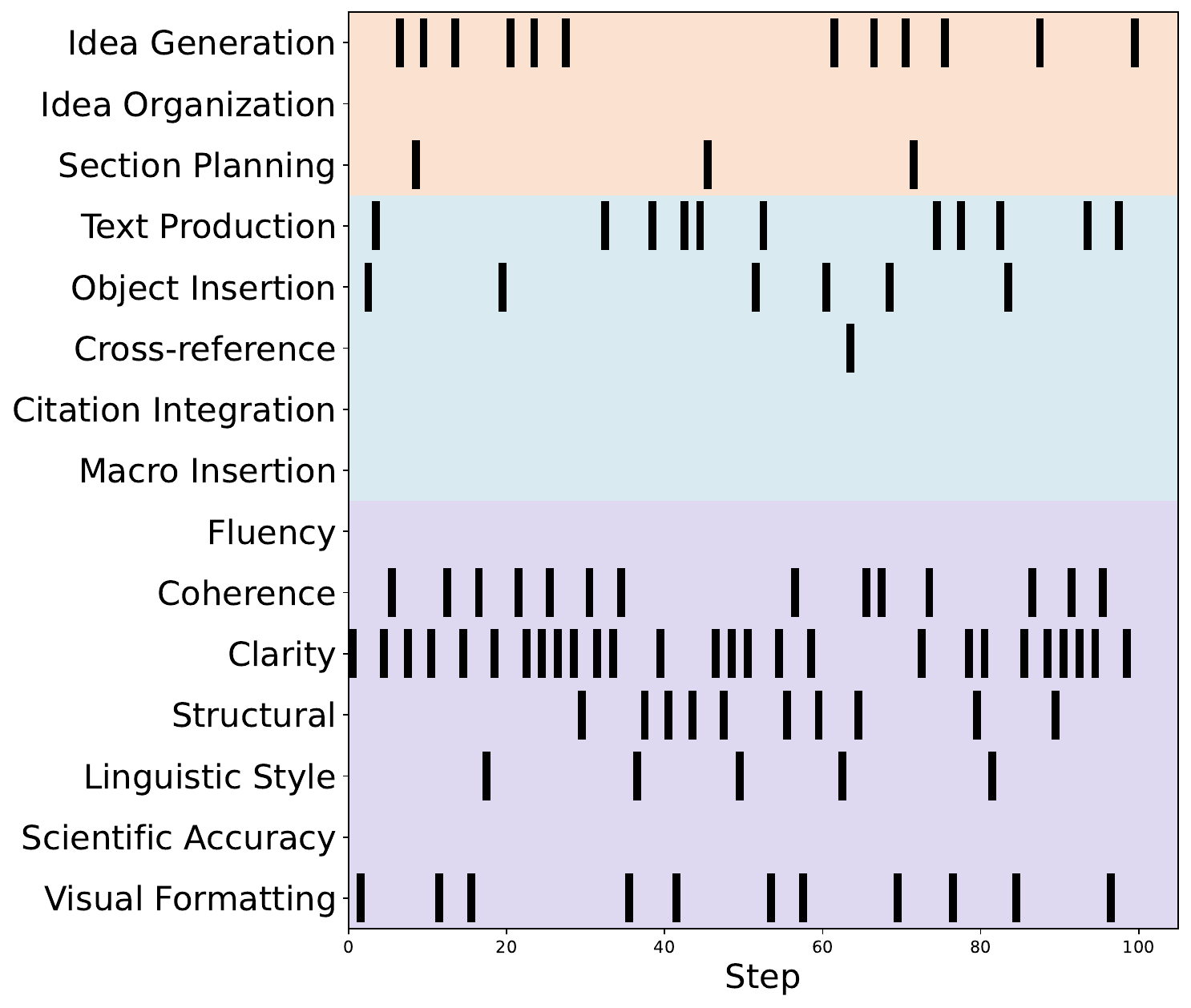}
        \includegraphics[width=0.32\textwidth]{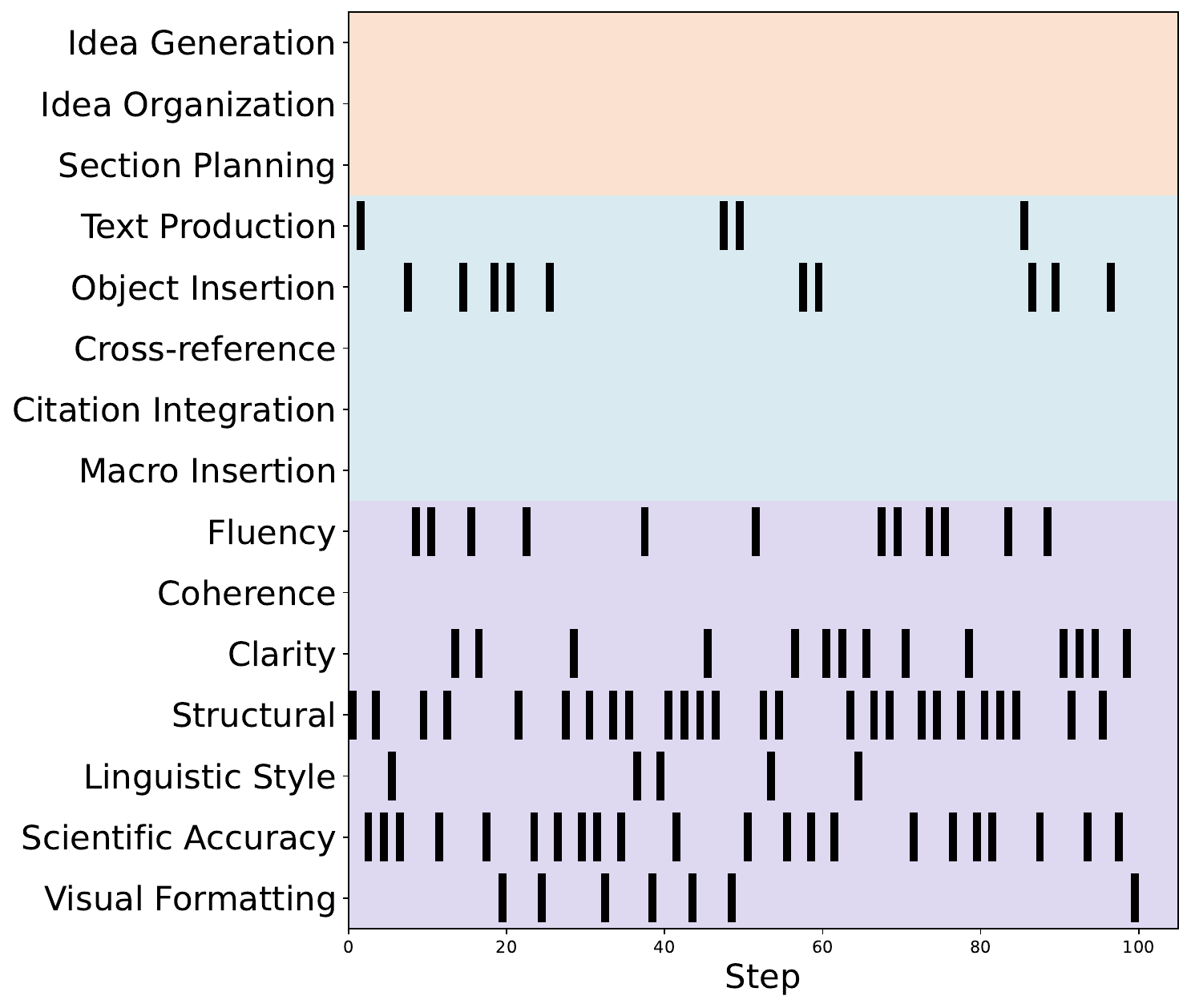}
        \includegraphics[width=0.32\textwidth]{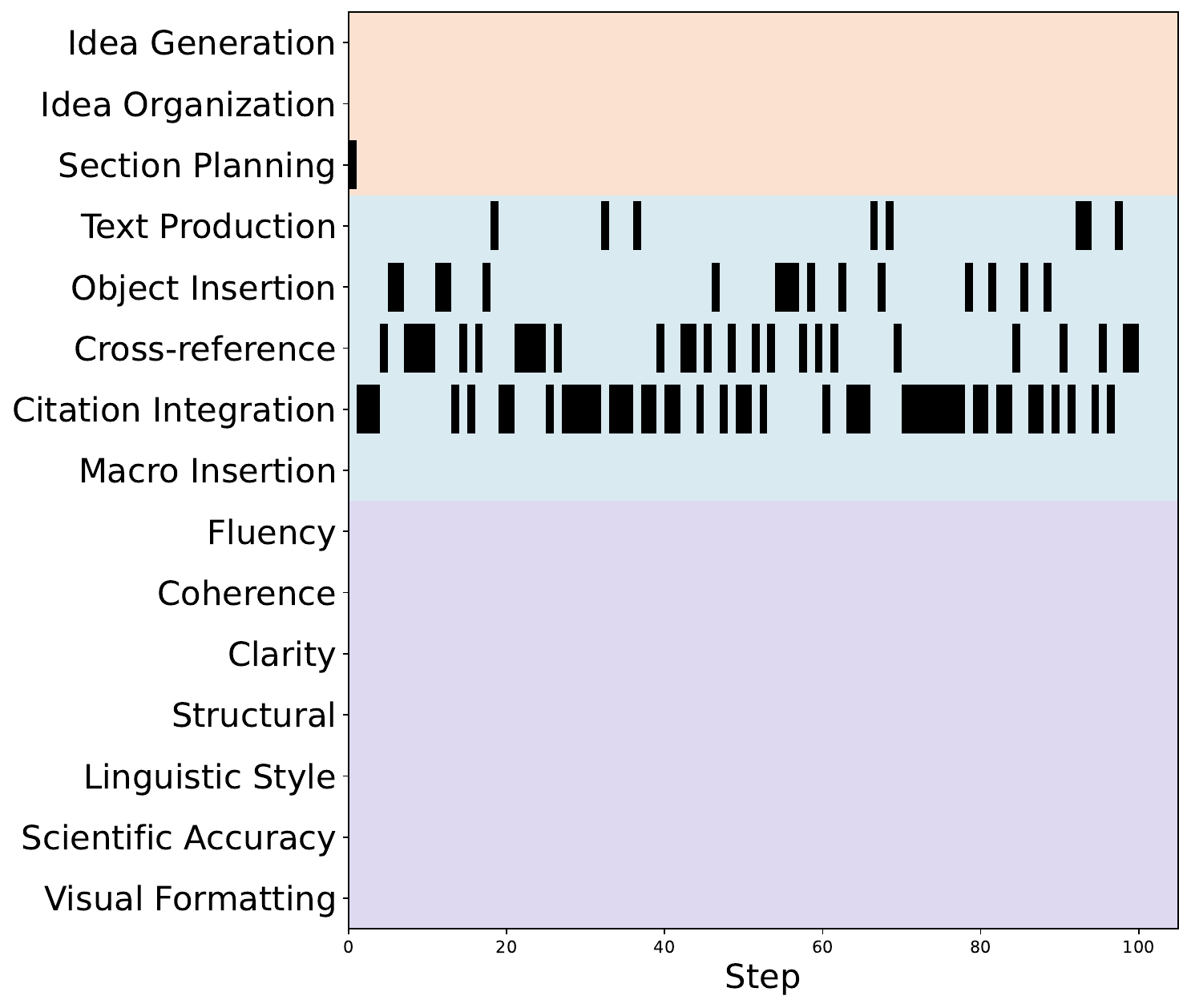}
        \\[2pt]
        \makebox[\textwidth]{\small (a) Seed 1 } 
    \\[0.5em]
        \includegraphics[width=0.32\textwidth]{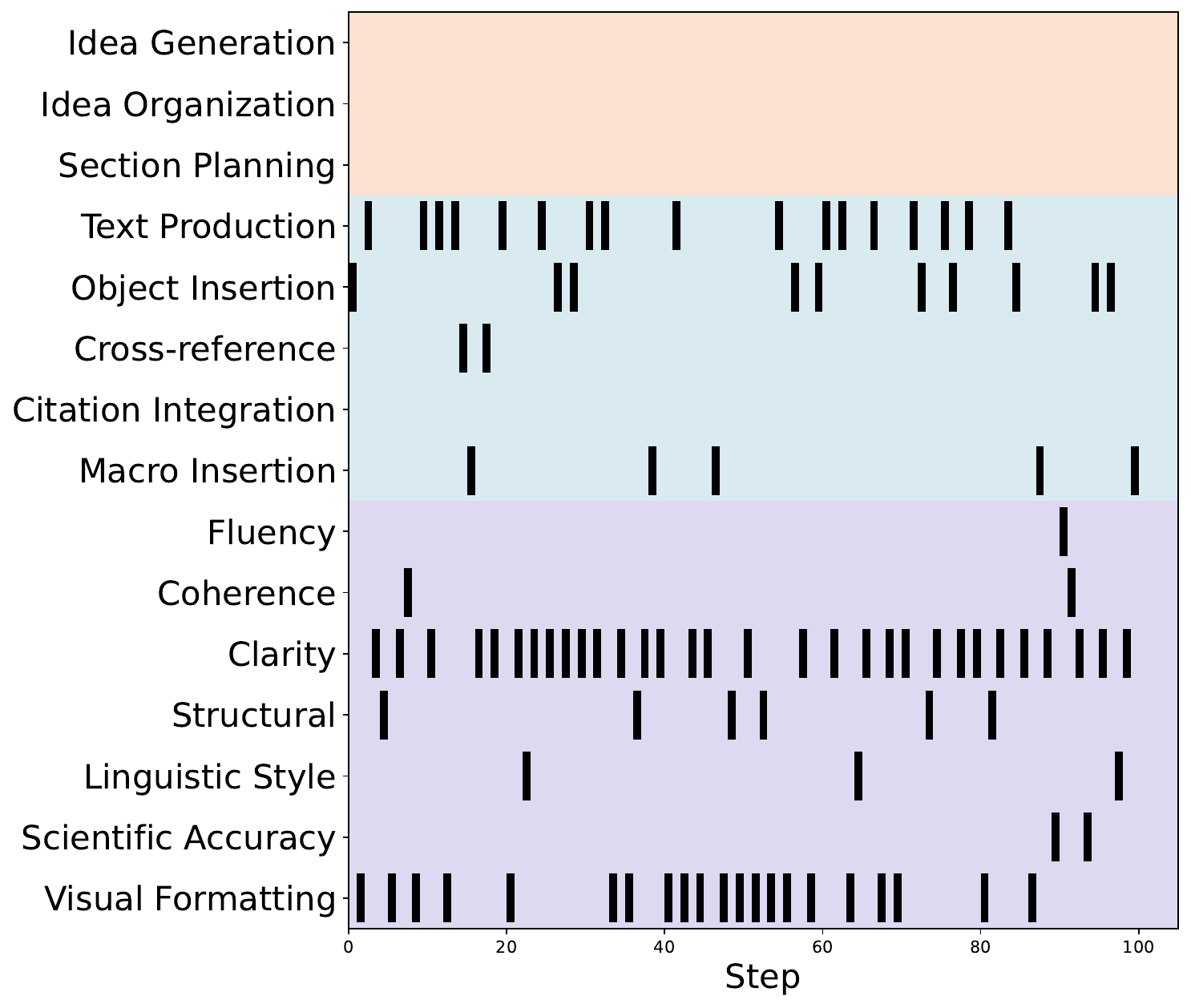}
        \includegraphics[width=0.32\textwidth]{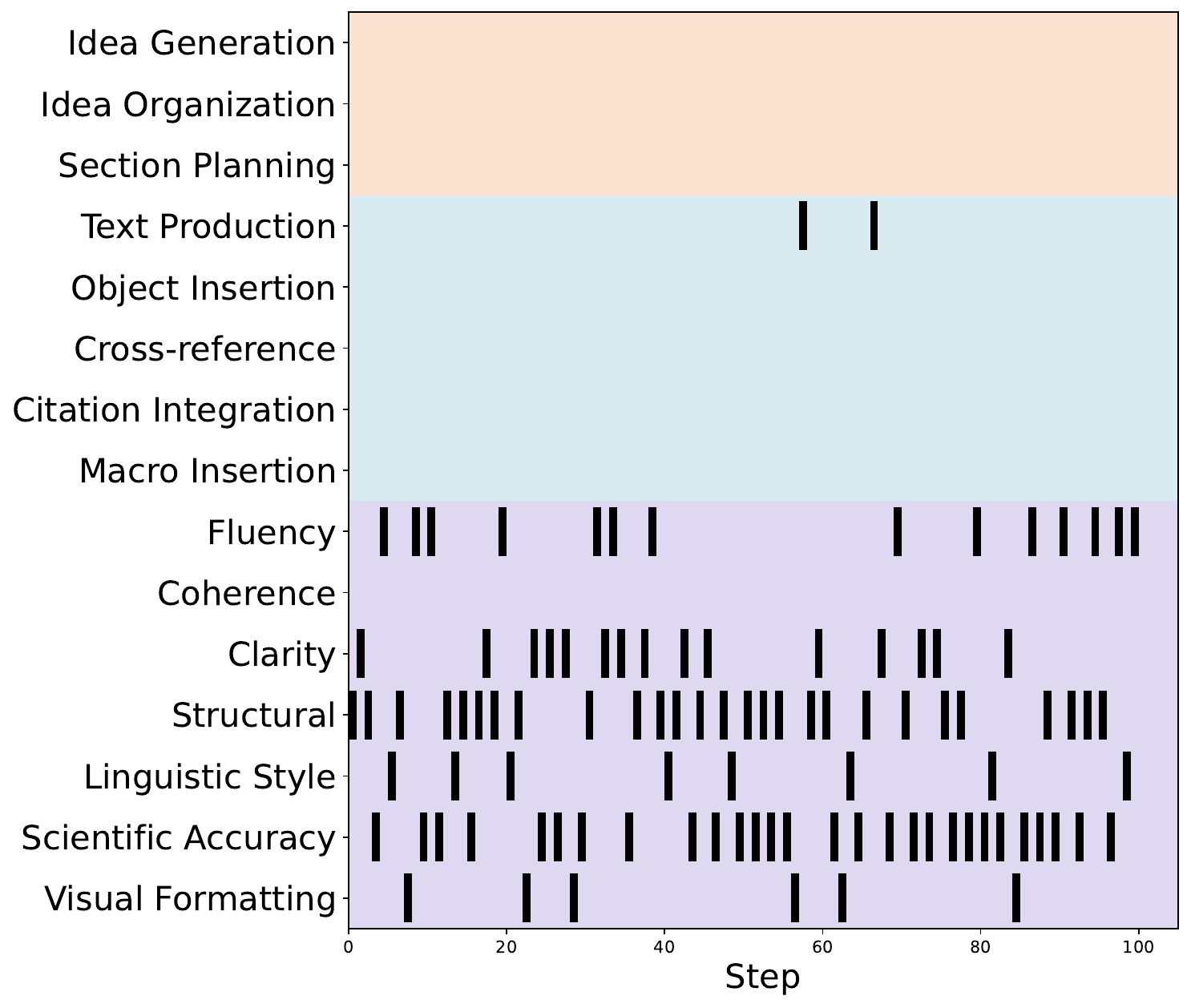}
        \includegraphics[width=0.32\textwidth]{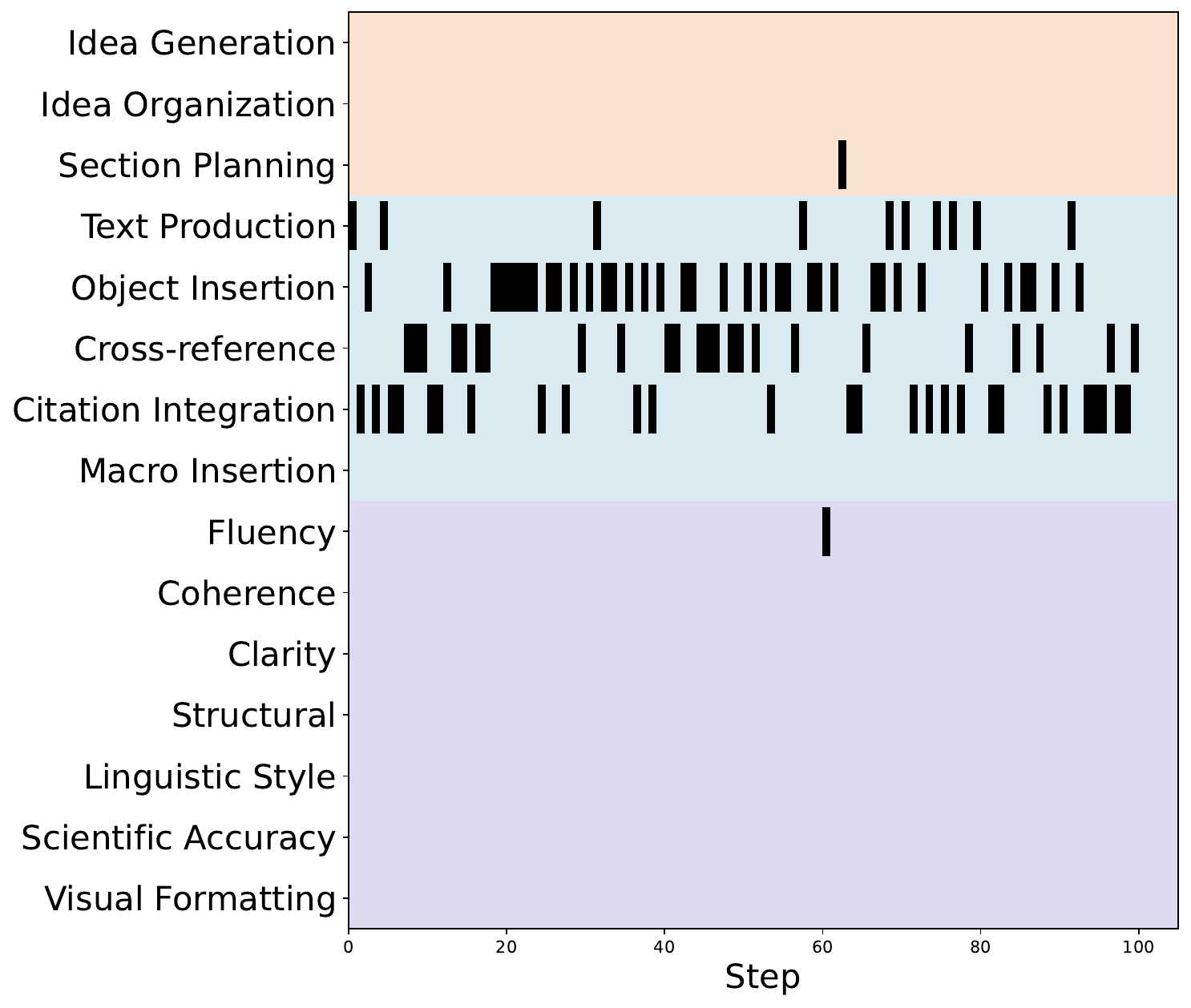}
        \\[2pt]
        \makebox[\textwidth]{\small (b) Seed 2 } 
    \\[0.5em]
        \includegraphics[width=0.32\textwidth]{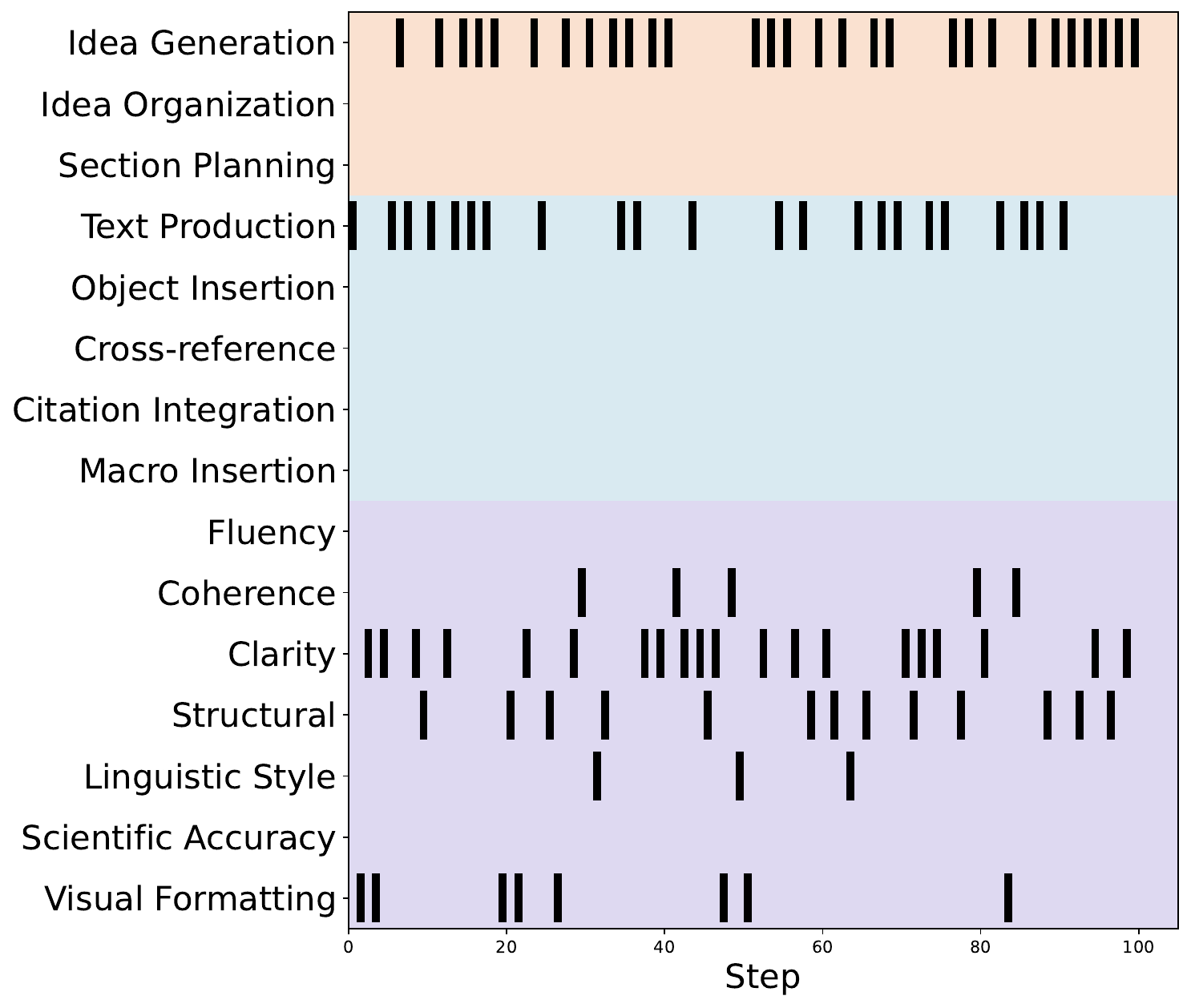}
        \includegraphics[width=0.32\textwidth]{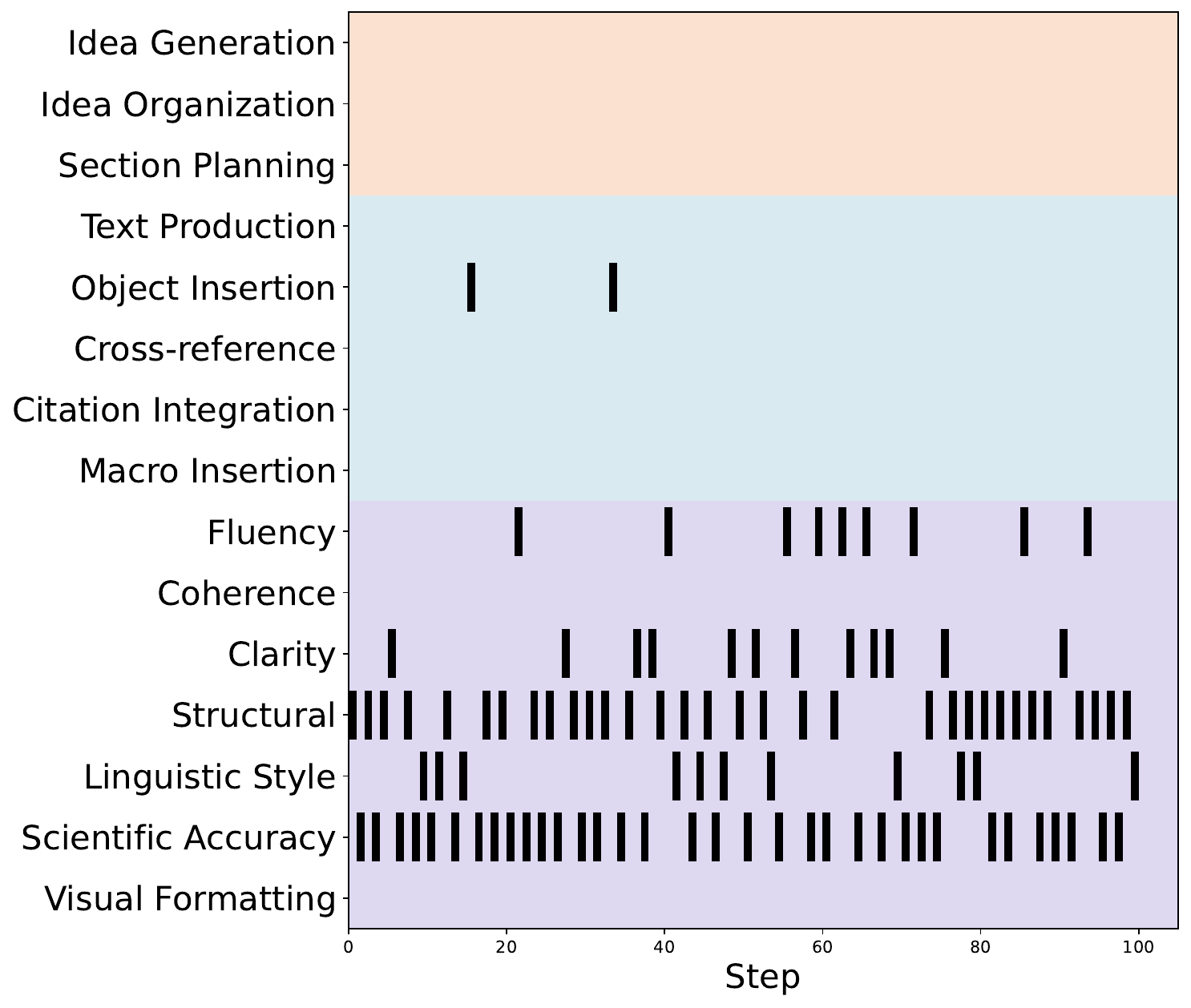}
        \includegraphics[width=0.32\textwidth]{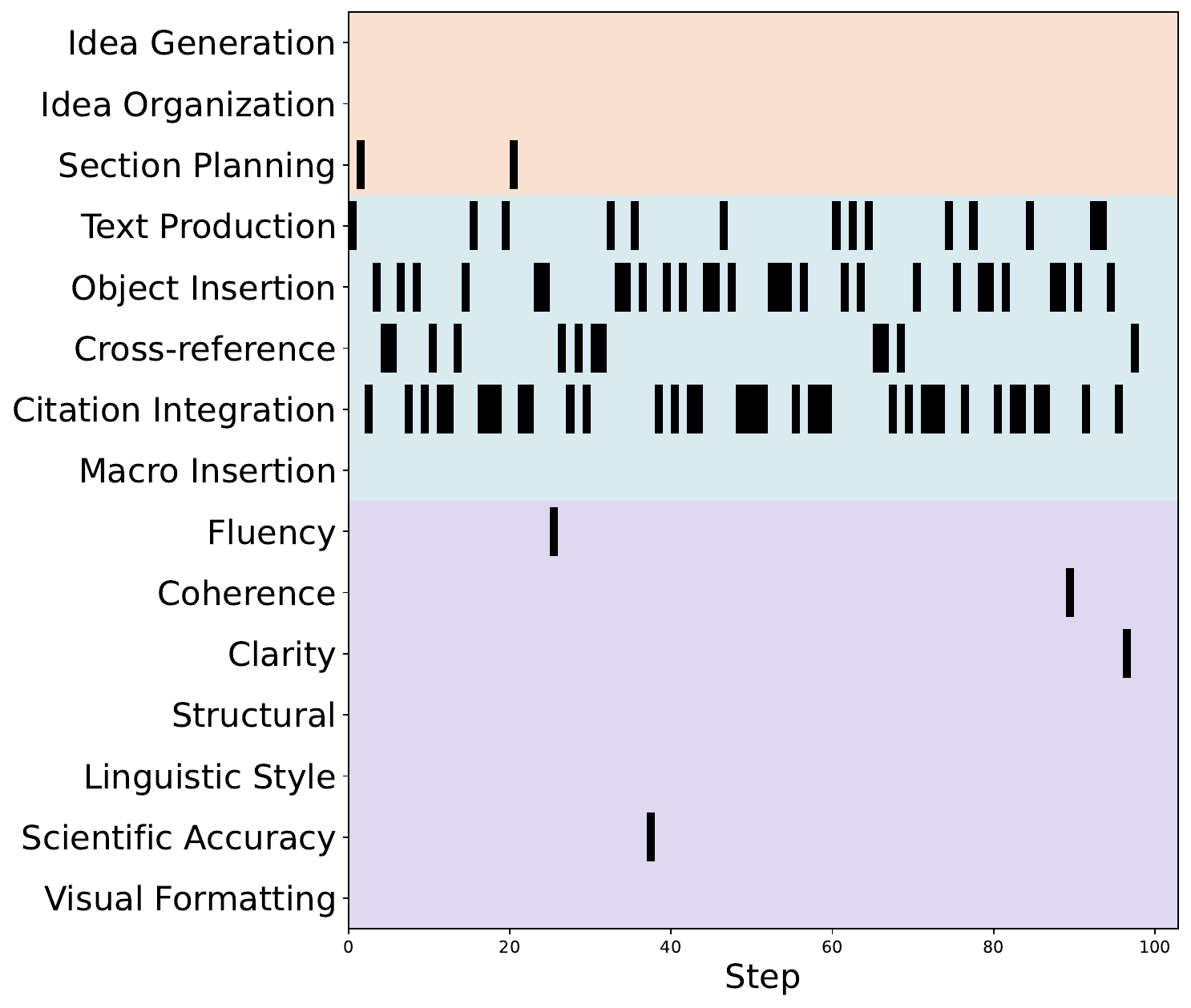}
        \\[2pt]
        \makebox[\textwidth]{\small (c) Seed 3 } 
    \\[0.5em]
        \includegraphics[width=0.32\textwidth]{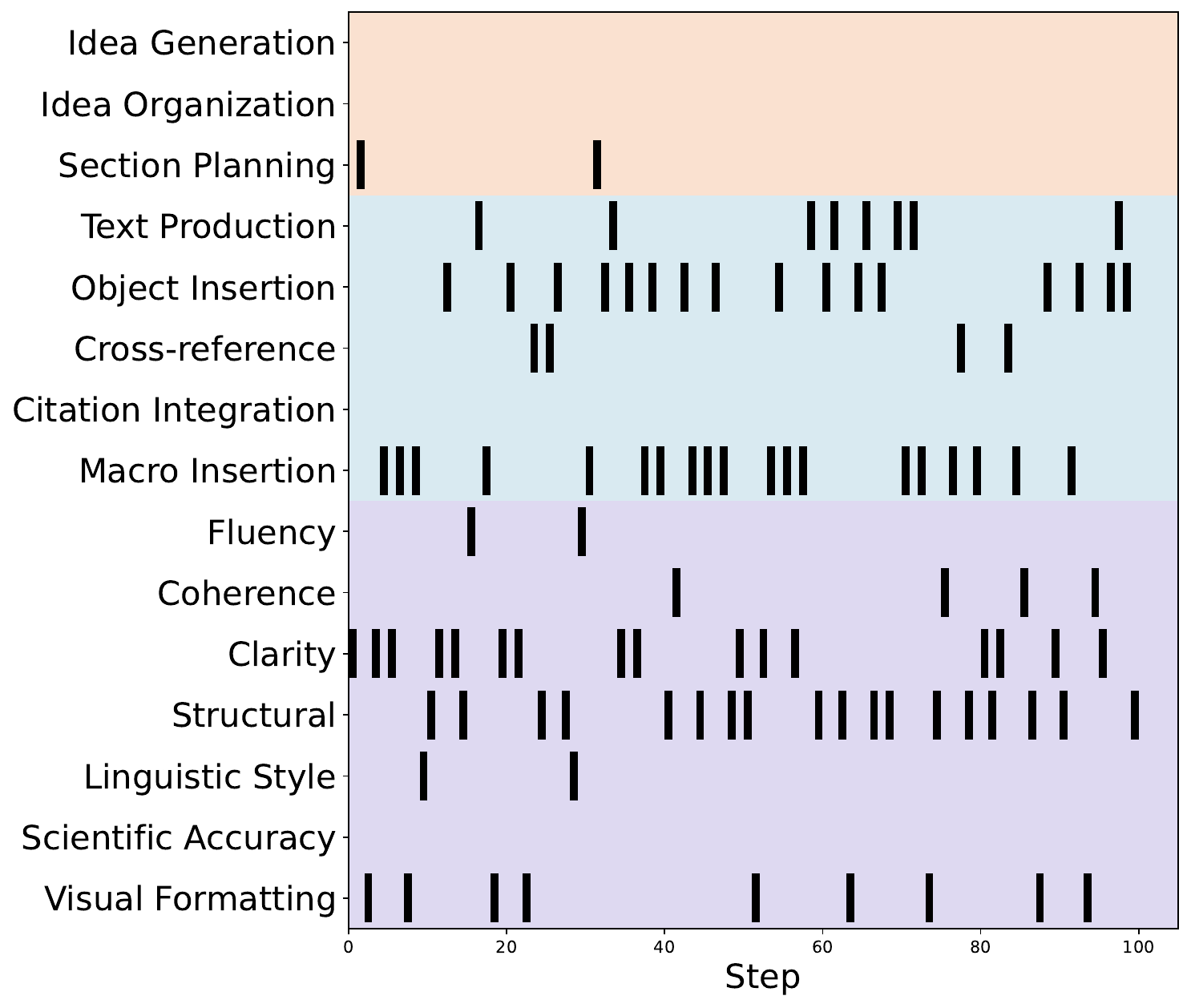}
        \includegraphics[width=0.32\textwidth]{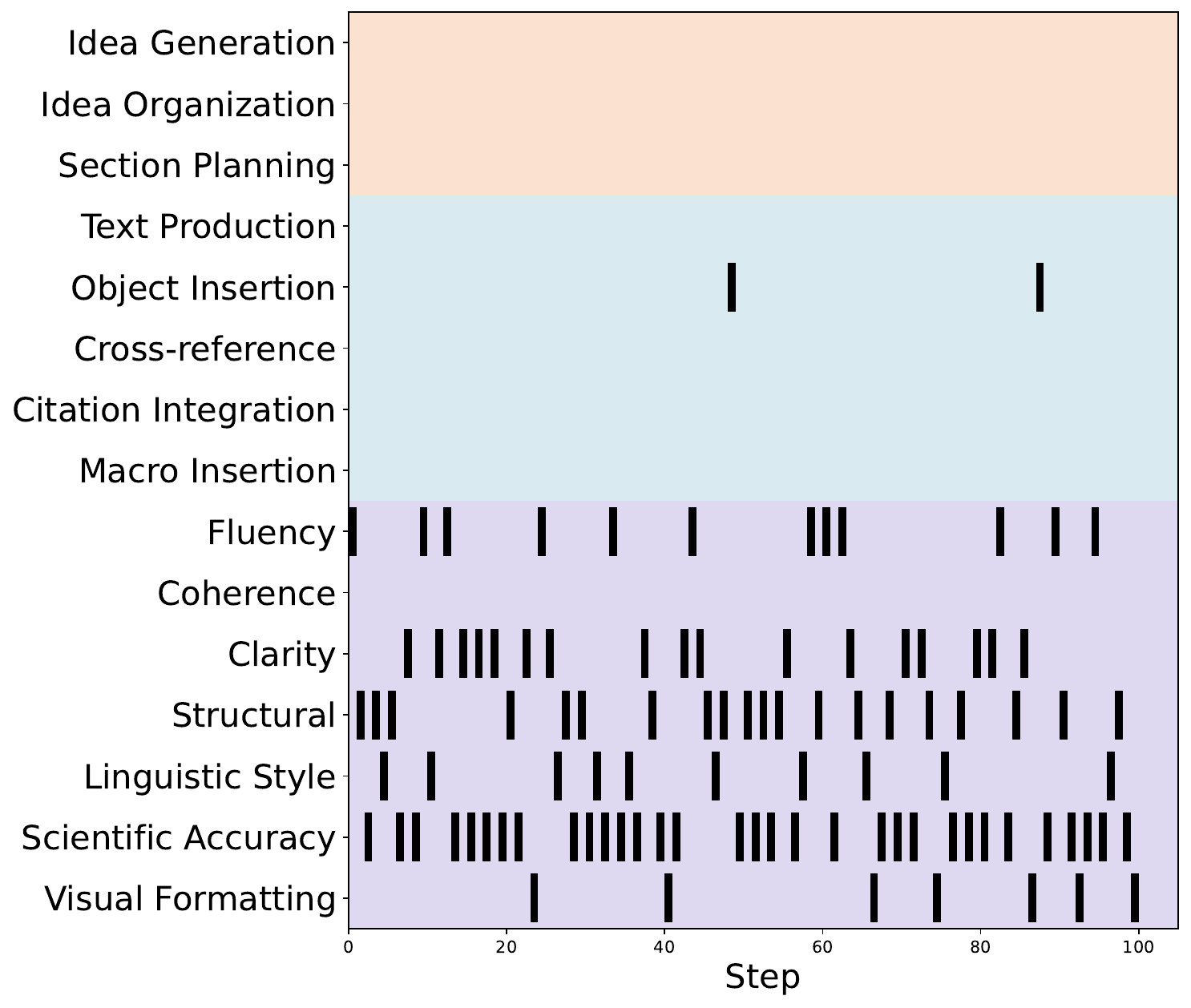}
        \includegraphics[width=0.32\textwidth]{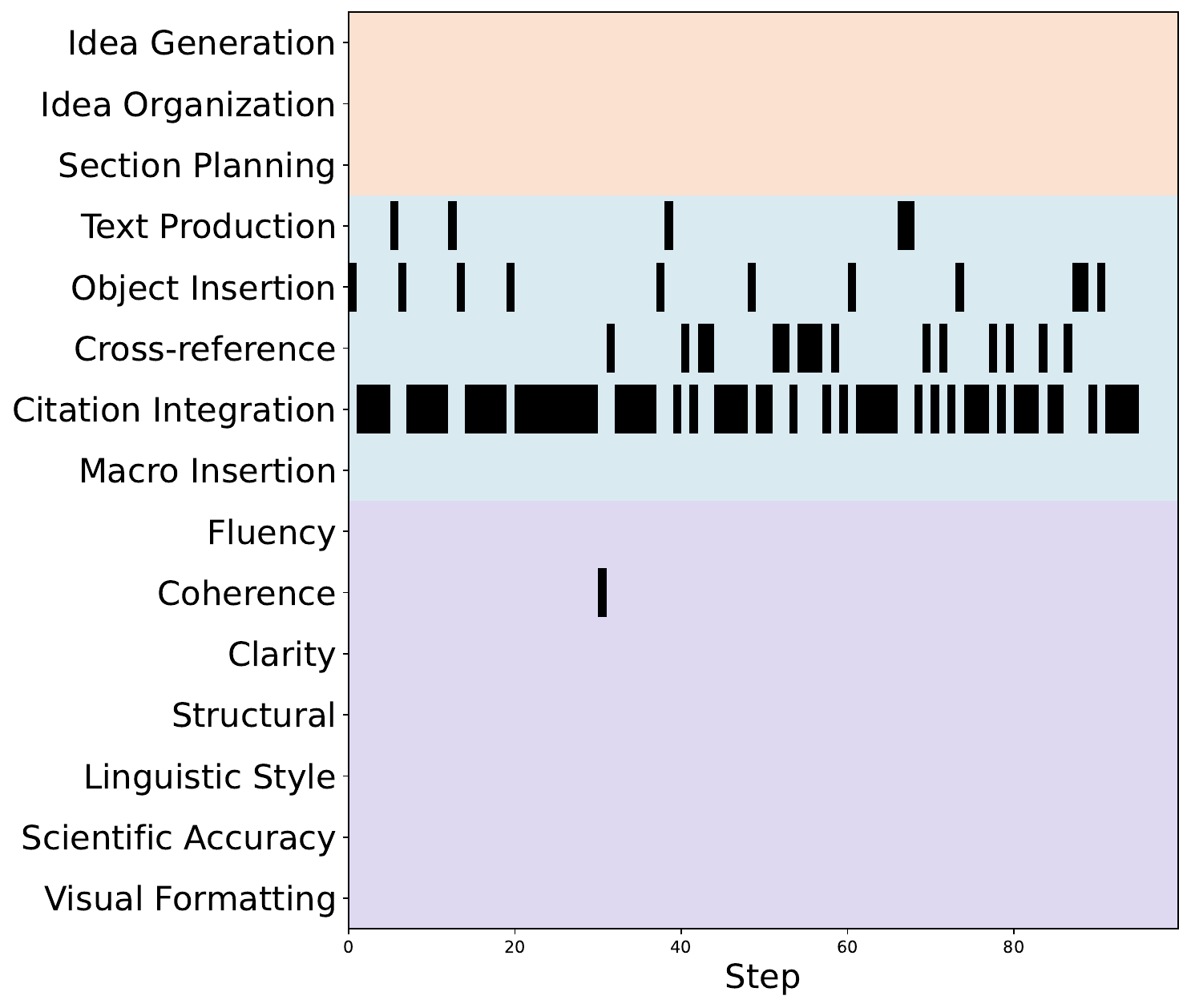}
        \\[2pt]
        \makebox[\textwidth]{\small (d) Seed 4 } 
    \\[0.5em]
    \caption{Distribution of Per-intention writing activities over time by models - \textcolor{magenta}{Llama-8B-ScholaWrite} (left); \textcolor{teal}{Llama-8B-Zero} (middle); \textcolor{blue}{GPT-4o} (right). Orange, Blue, and Purple represent Planning, Implementation, and Revision writing actions respectively. We observe different writing patterns by model during the entire 100 iterations. }
    \label{fig:writing-step-intention-all-model}
\end{figure*}

\subsubsection{Results} 
\begin{figure*}[ht!]
\vspace{-2em}
    \centering
    \begin{subfigure}[b]{0.32\textwidth}
        \centering
        \includegraphics[width=\textwidth]{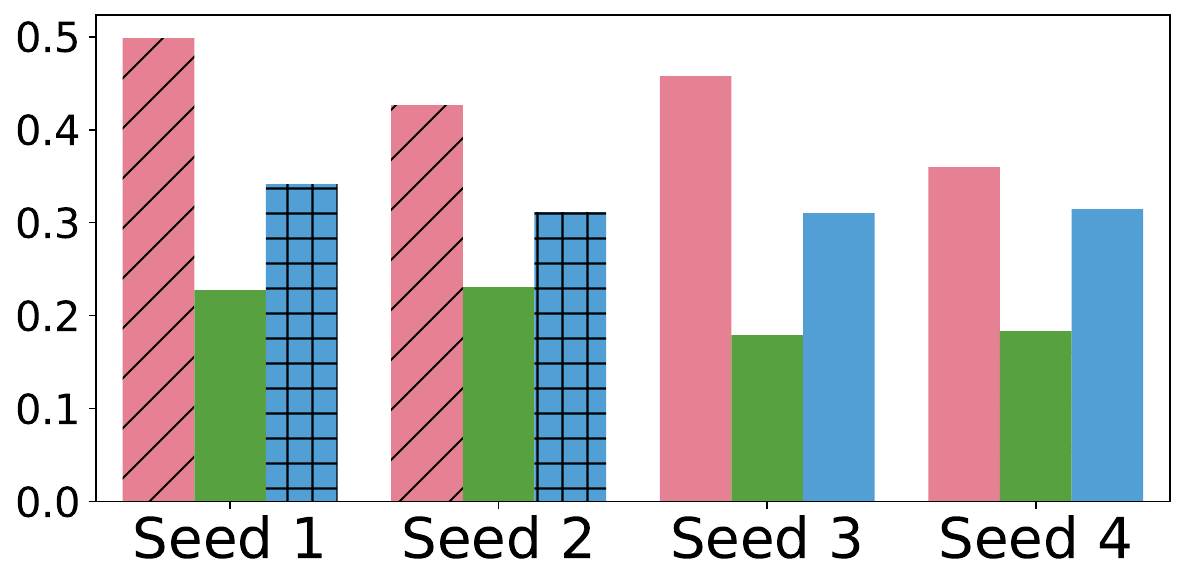}
        \vspace{-1em}
        \caption{Lexical Diversity}
        \label{fig:lexical-all}
    \end{subfigure}
    \hspace{-0.5em}
    \begin{subfigure}[b]{0.35\textwidth}
        \centering
        \includegraphics[width=\textwidth]{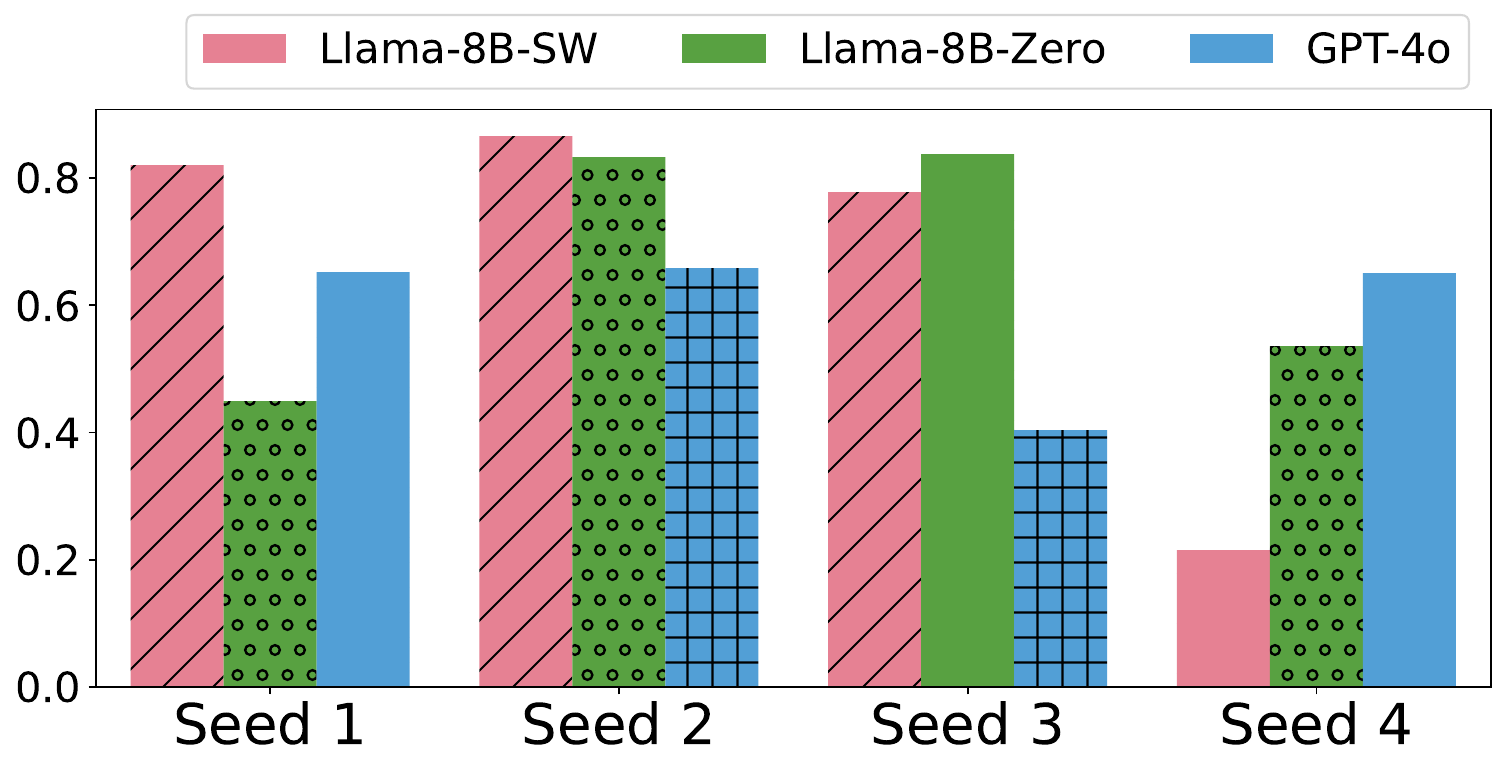}
        \vspace{-1em}
        \caption{Topic Consistency}
        \label{fig:topic-all}
    \end{subfigure}
    \begin{subfigure}[b]{0.32\textwidth}
        \centering
        \includegraphics[width=\textwidth]{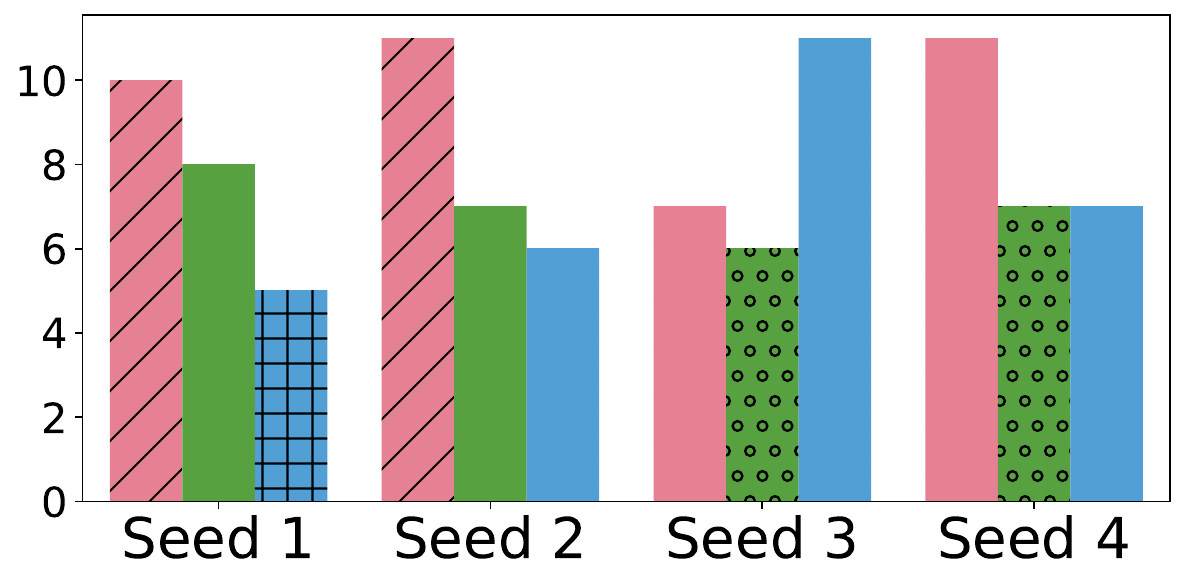}
        \vspace{-1em}
        \caption{Intention Coverage}
        \label{fig:intention-all}
    \end{subfigure}
    \caption{Metric scores of the final writing output of models after 100 iterations of the iterative self-writing experiment.
    }
    
    \label{fig:sec5-auto-all}
\end{figure*}

\begin{figure*}[t!]
\vspace{-2em}
    \centering
    \begin{subfigure}[b]{0.32\textwidth}
        \centering
        \includegraphics[width=\textwidth]{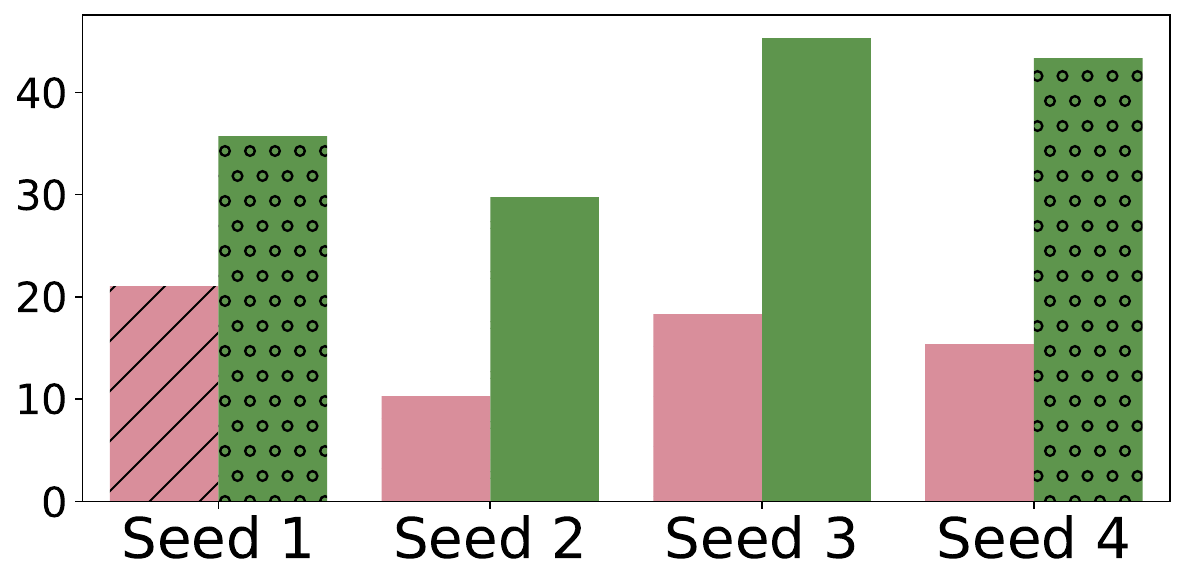}
        \vspace{-1em}
        \caption{Accuracy}
        \label{fig:Accuracy-all}
    \end{subfigure}
    \hspace{-0.5em}
    \begin{subfigure}[b]{0.30\textwidth}
        \centering
        \includegraphics[width=\textwidth]{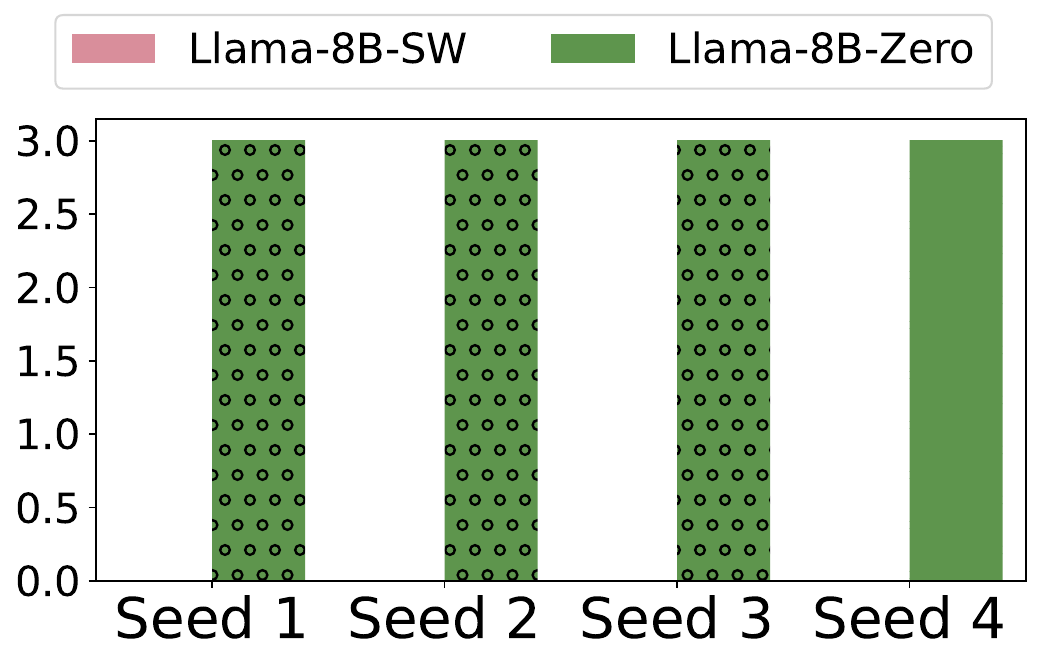}
        \vspace{-1em}
        \caption{Coherence}
        \label{fig:Coherence-all}
    \end{subfigure}
    \begin{subfigure}[b]{0.30\textwidth}
        \centering
        \includegraphics[width=\textwidth]{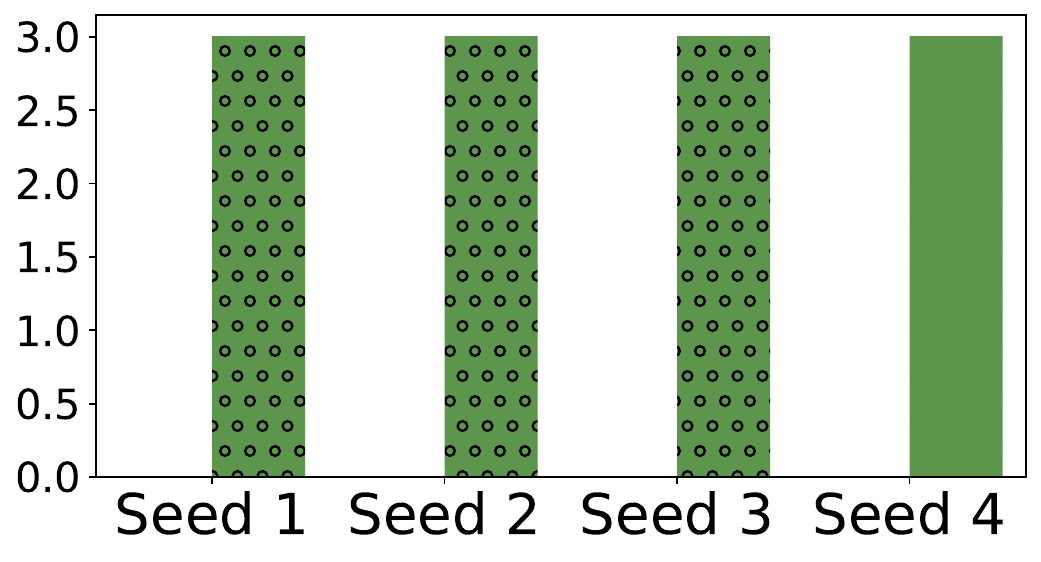}
        \vspace{-1em}
        \caption{Fluency}
        \label{fig:Fluency-all}
    \end{subfigure}
    \begin{subfigure}[b]{0.30\textwidth}
        \centering
        \includegraphics[width=\textwidth]{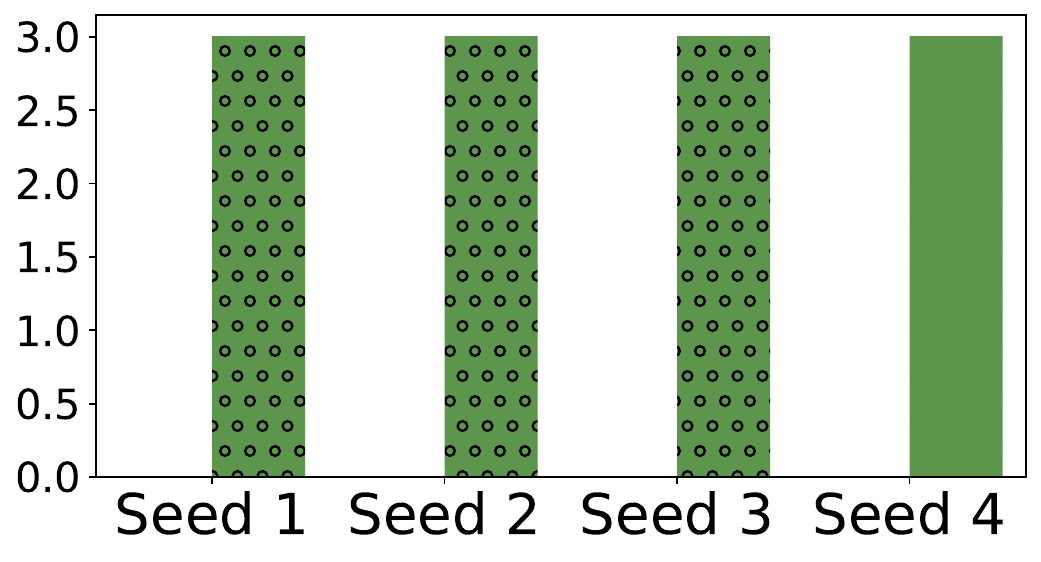}
        \vspace{-1em}
        \caption{Alignment}
        \label{fig:Alignment-all}
    \end{subfigure}
    \begin{subfigure}[b]{0.30\textwidth}
        \centering
        \includegraphics[width=\textwidth]{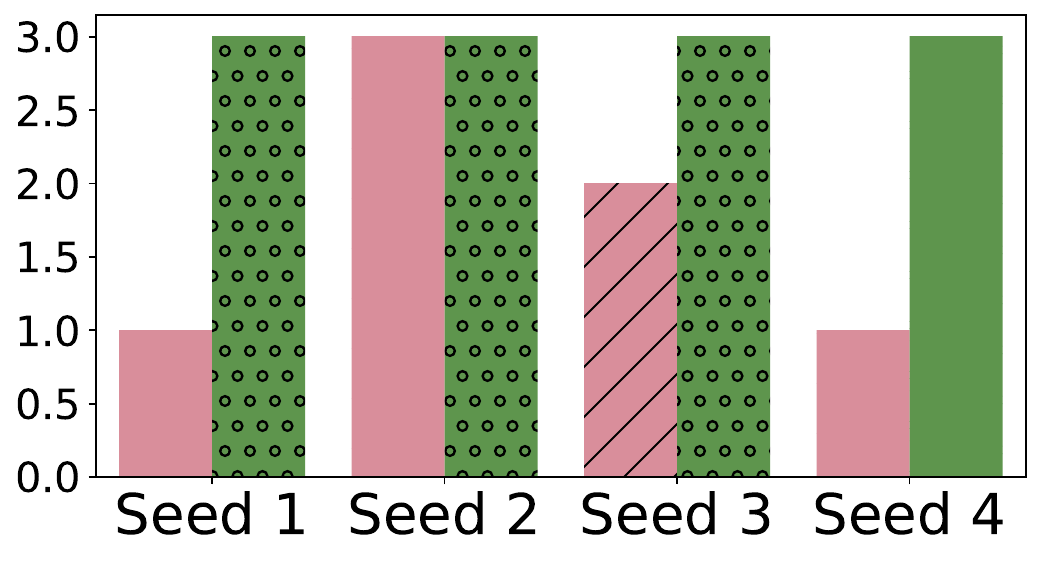}
        \vspace{-1em}
        \caption{Relevance}
        \label{fig:Relevance-all}
    \end{subfigure}
    \caption{Human evaluation results of iterative writing outputs of models
    }
    \label{fig:sec5-human-all}
\end{figure*}

\begin{figure*}
    \vspace{1em}
    \centering
    \makebox[0.32\textwidth]{\small \textcolor{magenta}{Llama-8B-SW}}
    \makebox[0.32\textwidth]{\small \textcolor{teal}{Llama-8B-Zero}}
    \makebox[0.32\textwidth]{\small \textcolor{blue}{GPT-4o}} 
    \\[0.5em]
        \includegraphics[width=0.32\textwidth]{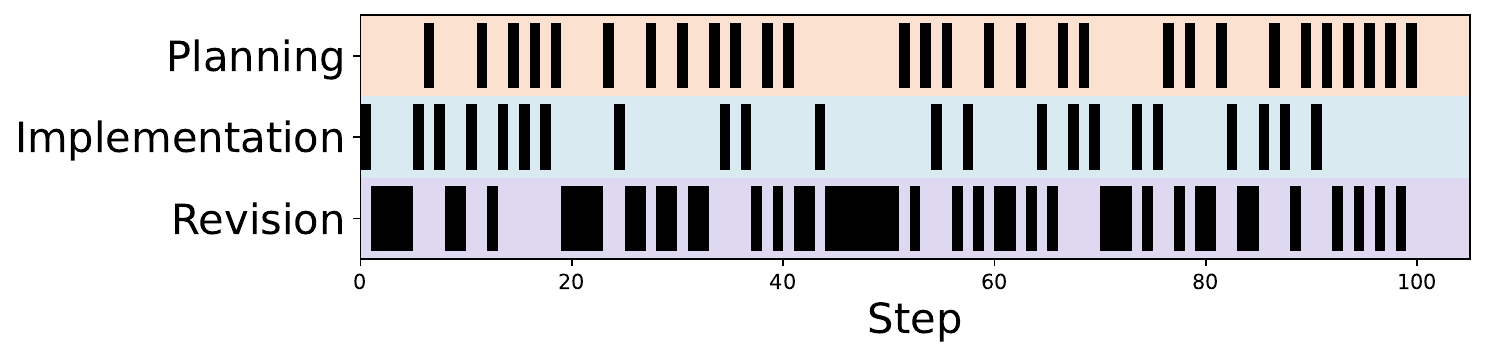}
        \includegraphics[width=0.32\textwidth]{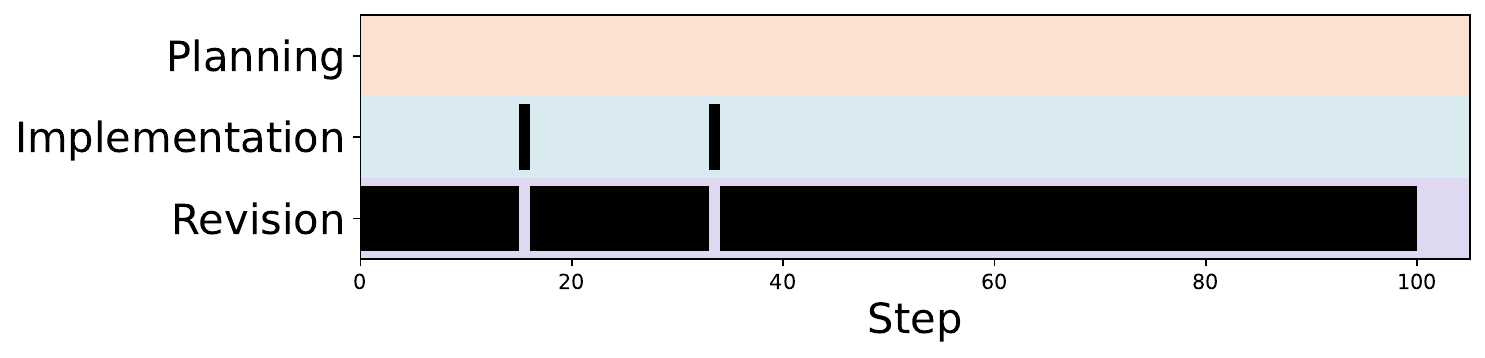}
        \includegraphics[width=0.32\textwidth]{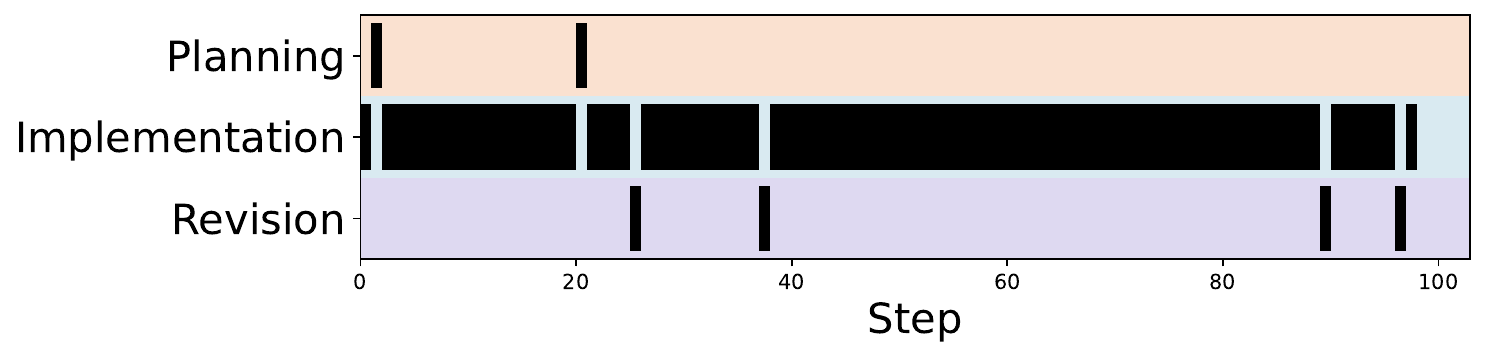}
    \caption{Distribution of high-level writing activities on seed 3 over time by models.}
    \label{fig:writing-step-broad-all-model-seed3}
\end{figure*}
Figure \ref{fig:sec5-auto-all} shows that \textcolor{magenta}{Llama-8B-SW} consistently produced the most lexically diverse words, generated the most semantically aligned topics (Seeds 1 \& 2), and covered the most writing intentions (except Seed 3). These results underscore the value of \textsc{ScholaWrite} in improving scholarly writing quality generated by language models.

However, our human evaluation (Figure \ref{fig:sec5-human-all}) revealed that \textcolor{magenta}{Llama-8B-SW} generated less human-like writing, in terms of fluency and logical claims. It also struggled with generating texts aligned with the predicted intentions. See Appendix Tables \ref{table:human-eval-seed1} to \ref{table:human-eval-seed4} for more details. Despite the weaknesses, \textcolor{magenta}{Llama-8B-SW} still produced more relevant content (Seed 2), which aligns with topic consistency trends in Figure \ref{fig:sec5-auto-all}, highlighting the usefulness of \textsc{ScholaWrite} dataset in certain contexts. 

Moreover, \textcolor{magenta}{Llama-8B-SW} exhibited the most human-like writing activity patterns over time (Figure \ref{fig:writing-step-broad-all-model-seed3}), which frequently switches between implementation and revision and covers all three high-level processes. \textcolor{teal}{Llama-8B-Zero} and \textcolor{blue}{GPT-4o} tend to remain in a single high-level stage throughout all 100 iterations of self-writing (see Appendix \ref{fig:writing-step-broad-all-model} and \ref{fig:writing-step-intention-all-model} for details). Compared to Appendix Figure \ref{fig:writing-step-detailed-all}, which depicts frequent transitions across all three stages in an early draft (e.g., the first 100 steps), \textcolor{magenta}{Llama-8B-SW} most closely replicates human writing behaviors in iterative writing tasks. These findings reinforce the potential of \textsc{ScholaWrite} in helping LLMs emulate human scholarly writing processes.

\end{document}